\documentclass[prb,twocolumn,showpacs,aps,superscriptaddress,floatfix]{revtex4}
\usepackage{amsmath}
\usepackage{amssymb}
\usepackage{amsfonts}
\usepackage{bm}
\usepackage{graphicx}
\usepackage{color}
\usepackage[svgnames]{xcolor}
\usepackage{soul}
\usepackage{hyperref}
\usepackage{verbatim}

\DeclareMathOperator{\Real}{Re}
\DeclareMathOperator{\Imag}{Im}
\graphicspath{{FIGS/}}

\begin{document}

\title{Spin density wave and electron nematicity in magic-angle twisted
bilayer graphene}

\author{A.O. Sboychakov}
\affiliation{Institute for Theoretical and Applied Electrodynamics, Russian
Academy of Sciences, Moscow, 125412 Russia}

\author{A.V. Rozhkov}
\affiliation{Institute for Theoretical and Applied Electrodynamics, Russian
Academy of Sciences, Moscow, 125412 Russia}
\affiliation{Skolkovo Institute of Science and Technology, Skolkovo Innovation Center, Moscow 143026, Russia}

\author{A.L. Rakhmanov}
\affiliation{Institute for Theoretical and Applied Electrodynamics, Russian Academy of Sciences, Moscow, 125412 Russia}

\author{Franco Nori}
\affiliation{Theoretical Quantum Physics Laboratory, RIKEN, Wako-shi,
Saitama, 351-0198, Japan}
\affiliation{Department of Physics, University of Michigan, Ann Arbor, MI
48109-1040, USA}

\begin{abstract}
We study theoretically many-body properties of magic-angle twisted
bilayer graphene for different doping levels. Our investigation is focused
on the emergence, stability, and manifestations of nematicity of the
ordered low-temperature electronic state. It is known that, at vanishing
interactions, the low-energy spectrum of the system studied consists of
four almost-flat almost-degenerate bands. Electron-electron repulsion lifts
this degeneracy. To account for such an interaction effect, a numerical
mean-field theory is used. Assuming that the ground state has
spin-density-wave-like order, we introduce a multicomponent order parameter
describing spin magnetization. Our simulations show that the order
parameter structure depends on the doping level. In particular, doping away
from the charge neutrality point reduces the rotational symmetry of the
ordered state, indicating the appearance of an electron nematic state.
Manifestations of the nematicity can be observed in the spatial
distribution of the spin magnetization within a moir{\'e} cell, as well as
in the single-electron band structure. The nematicity is the strongest at
half-filling (two extra electron or holes per supercell). We argue that
nematic symmetry breaking is a robust feature of the system ground
state, stable against model parameters variations. Specifically, it is
shown that, away from the charge neutrality point, it persists for all
three parametrizations of the interlayer hopping amplitudes discussed in
the paper. Obtained theoretical results are consistent with the available
experimental data.
\end{abstract}


%
%
%
%
%
%
%
%
%
%

\date{\today}

\maketitle

\section{Introduction}
Discovery of many-body insulating
states~\cite{NatureMott2018} and
superconductivity~\cite{NatureSC2018}
in the so-called magic-angle twisted bilayer
graphene~\cite{ourBLGreview2016}
(MAtBLG) has triggered an avalanche of both
theoretical~\cite{PhononSCMcDonald2018,ChiralSDW_SC2018,Philips2018,
PhysRevB.97.235453,PhysRevB.98.081102,OurtBLGPRB2019,KLSC2019,
AFMMottSC2019,PhononSCPRL2019,PhysRevB.99.121407,FMorderPRL2019,
fern2019nematicity, cea2020PRB2019,cea2020band}
and
experimental~\cite{PhysRevB.98.235402,MottSCNature2019,
MottNematicNature2019,KerelskyNematicNature2019,PhysRevB.99.201408,
FlatBandsARPES2019,STMNature2019,STM2019,CompressibilityPRL2019,
cao2020nematicity,Jiang_twisted_stm2019nature}
studies of this material. The MAtBLG has a twist angle
$\theta_c\sim1^{\circ}$
and it is characterized by a superstructure with a large supercell
containing several thousand carbon atoms. Single-electron states of MAtBLG
form four weakly dispersive (almost flat) low-energy
bands~\cite{dSPRB,NonAbelianGaugePot,Morell1,ourTBLG}
(these flat bands were recently visualized by ARPES in
Ref.~\onlinecite{FlatBandsARPES2019}).
Measurements~\cite{NatureMott2018,NatureSC2018}
of the conductivity $\sigma$ of MAtBLG versus doping $n$ reveal several
conductivity minima at doping values
$n/(n_s/4)\equiv\nu =0,\,\pm2,\,\pm3,\,\pm4$,
where the concentration
$n_s$
corresponds~\cite{NatureMott2018,NatureSC2018}
to four electrons per supercell. Observation of the ``missing" conductivity
minima at
$\nu=\pm1$
was later reported in
Ref.~\onlinecite{MottSCNature2019}.
Besides these findings,
Ref.~\onlinecite{NatureSC2018}
reported superconductivity domes near
$\nu=-2$.
Superconductivity domes near
$\nu=-2$,
$\nu=0$,
and
$\nu=\pm1$
were also
found~\cite{MottSCNature2019}.

Theoretically, conductivity minima at
$\nu=\pm4$
can be understood in terms of single-electron
physics~\cite{NatureMott2018}.
However, the minima at
$\nu=\pm1,\,\pm2,\,\pm3$
cannot be explained within single-particle theory, and the effects of
interactions should be taken into account. The nature of the insulating
states in MAtBLG was considered in several
papers~\cite{Philips2018,PhysRevB.98.081102,ChiralSDW_SC2018,
AFMMottSC2019,FMorderPRL2019,OurtBLGPRB2019}.
Different types of spin density wave (SDW)
states~\cite{ChiralSDW_SC2018,AFMMottSC2019,OurtBLGPRB2019},
ferromagnetic state~\cite{FMorderPRL2019}, and other symmetry broken phases~\cite{cea2020band}
have been proposed to be the ground state of the system. Potential
mechanisms of the superconductivity
(phonons~\cite{PhononSCMcDonald2018,PhononSCPRL2019},
electronic
correlations~\cite{PhysRevB.97.235453,ChiralSDW_SC2018,KLSC2019,
AFMMottSC2019,PhysRevB.99.121407})
as well as various symmetries of the superconducting order parameters
have been considered.

Neglecting the possibility of superconducting ordering, in a previous
work~\cite{OurtBLGPRB2019},
we assumed the multicomponent SDW to be the ground state of MAtBLG in
the doping range
$-4< \nu <4$.
The structure of the SDW order parameter, as well as the form of the
renormalized low-energy spectrum, was
calculated~\cite{OurtBLGPRB2019}
for different doping levels within the framework of a numerical mean-field
approach. This allowed us to explain the appearance of conductivity minima
at integer valued ratio $\nu$, consistent with
experiments~\cite{NatureMott2018,NatureSC2018,MottSCNature2019}.

Since doping affects the mean-field band structure, the dependence of the
density of states (DOS) $\rho$ versus the single-electron energy $E$ is
sensitive to the doping level of the MAtBLG sample. This theoretical
observation~\cite{OurtBLGPRB2019}
is supported by recent STM
measurements~\cite{STM2019,STMNature2019,MottNematicNature2019,
KerelskyNematicNature2019}.

Further, we observed
numerically~\cite{OurtBLGPRB2019}
that, at sufficiently strong doping, the point symmetry of the electronic
state reduces from
$C_6$
(full hexagonal symmetry) down to
$C_2$,
giving rise to electron nematic state.
References~\onlinecite{MottNematicNature2019,KerelskyNematicNature2019,
cao2020nematicity,Jiang_twisted_stm2019nature}
published claims of the experimental nematicity observations in MAtBLG
samples.

The striking agreement between our conclusions~\cite{OurtBLGPRB2019} and several independent
experimental measurements testifies in favor of the developed theoretical
approach. To build up upon this success, here we extend the study of
Ref.~\onlinecite{OurtBLGPRB2019}.
In this paper we focus on the emergence, stability and manifestations of
the electronic nematicity, demonstrating that nematic symmetry breaking
is a robust feature of the MAtBLG, stable against model modifications, such
as alterations of the interlayer hopping amplitudes. We will also argue
that the nematicity affects not only the spatial distribution of the spin
magnetization, but the single-electron spectrum as well. Experimental
implications of these findings are discussed.

The paper is organized as follows. In
Section~\ref{Geometry}
the geometry of the twisted bilayer graphene (tBLG) is outlined. In
Section~\ref{Model}
we formulate our electronic model and analyze the single-particle spectrum
of the MAtBLG for three different parametrizations of the interlayer
hopping amplitudes. We also present the general form of our multicomponent
SDW order parameter in this Section. In
Section~\ref{ResultsOP}
we analyze the spatial distribution of the SDW order parameter for
different doping levels, while in
Section~\ref{ResultsSpec}
we consider the properties of the renormalized low-energy spectrum.
Discussion of the results obtained and the conclusions are given in
Section~\ref{Discussion}.
Details of the numerical procedure used for the calculations of the SDW
order parameter are described in the Appendix.

\section{Geometry of twisted bilayer graphene}
\label{Geometry}

In this Section we present some basic facts about the geometry of twisted
bilayer graphene, which are important for further consideration (for more
details, see, e.g., review
papers~\onlinecite{MeleReview,ourBLGreview2016}).
Each graphene layer in tBLG has a hexagonal crystal structure consisting of
two triangular sublattices
${\cal A}$
and
${\cal B}$.
The coordinates of atoms in layer $1$ on sublattices
${\cal A}$
and
${\cal B}$
are
\begin{equation}
\mathbf{r}_{\mathbf{n}}^{1{\cal A} }
=
\mathbf{r}_{\mathbf{n}}^{1}\equiv n\mathbf{a}_1+m\mathbf{a}_2\,,
\;\;
\mathbf{r}_{\mathbf{n}}^{1{\cal B} }
=
\mathbf{r}_{\mathbf{n}}^{1}+\bm{\delta}\,,
\end{equation}
where
$\mathbf{n}=(n,\,m)$
is an integer-valued vector,
\begin{equation}
\label{eq::a12}
\mathbf{a}_{1,2}=a(\sqrt{3},\mp1)/2
\end{equation}
are the primitive vectors,
$\bm{\delta}=(\mathbf{a}_1+\mathbf{a}_2)/3=a(1/\sqrt{3},0)$,
and
$a=2.46$\,\AA\
is the lattice constant of graphene. Atoms in layer $2$ are located at
\begin{equation}
\mathbf{r}_{\mathbf{n}}^{2{\cal B} }
=
\mathbf{r}_{\mathbf{n}}^{2}
\equiv
d\mathbf{e}_z+n\mathbf{a}_1'+m\mathbf{a}_2'\,,
\quad
\mathbf{r}_{\mathbf{n}}^{2{\cal A} }
=
\mathbf{r}_{\mathbf{n}}^{2}-\bm{\delta}'\,,
\end{equation}
where
$\mathbf{a}_{1,2}'$
and
$\bm{\delta}'$
are the vectors
$\mathbf{a}_{1,2}$
and
$\bm{\delta}$,
rotated by the twist angle $\theta$. The unit vector along the $z$-axis is
$\mathbf{e}_z$,
the interlayer distance is
$d=3.35$\,\AA.
The limiting case
$\theta=0$
corresponds to the AB stacking.

\begin{figure}[t]
\centering
\includegraphics[width=0.98\columnwidth]{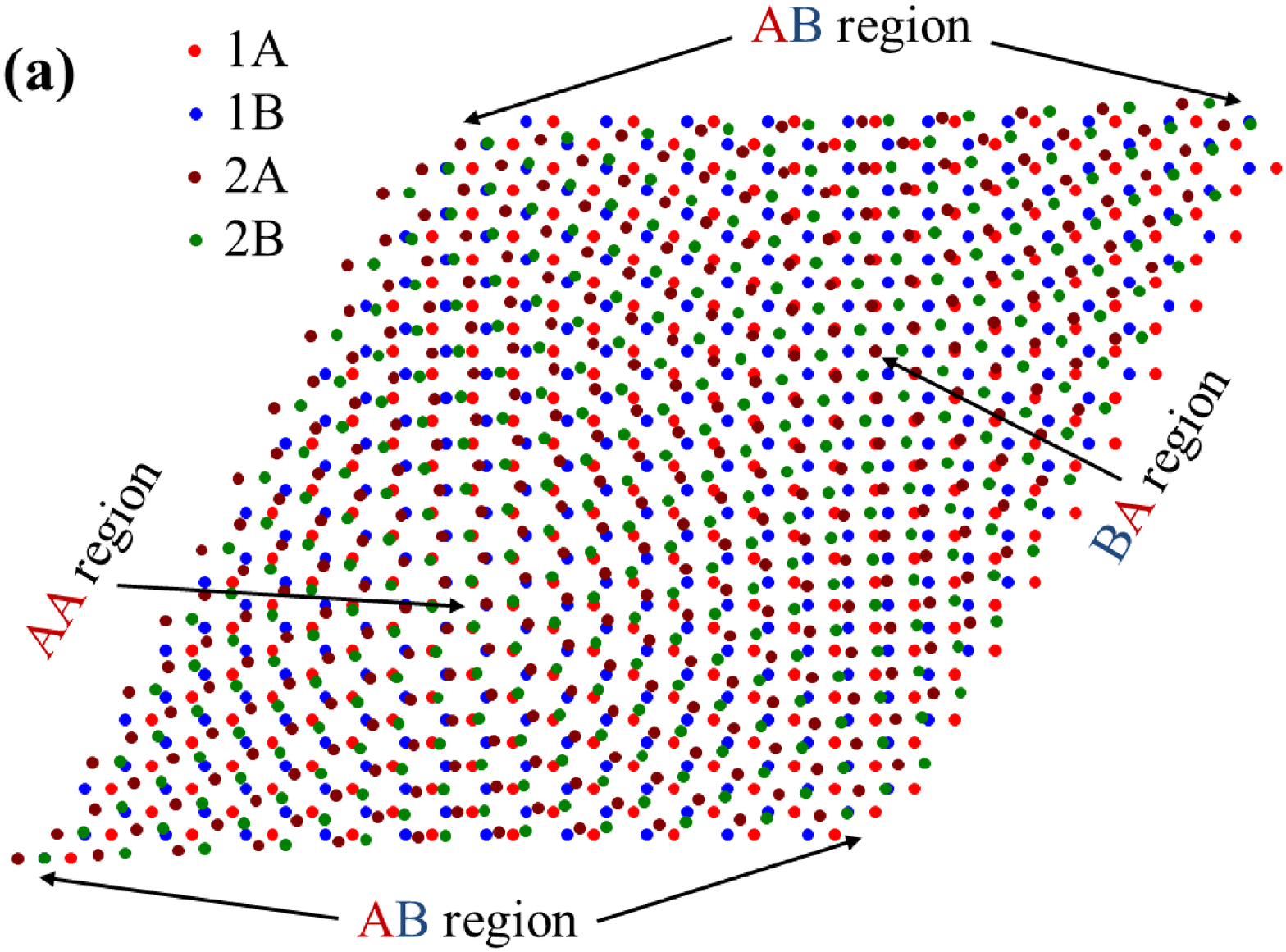}\\\vspace{7mm}
\includegraphics[width=0.8\columnwidth]{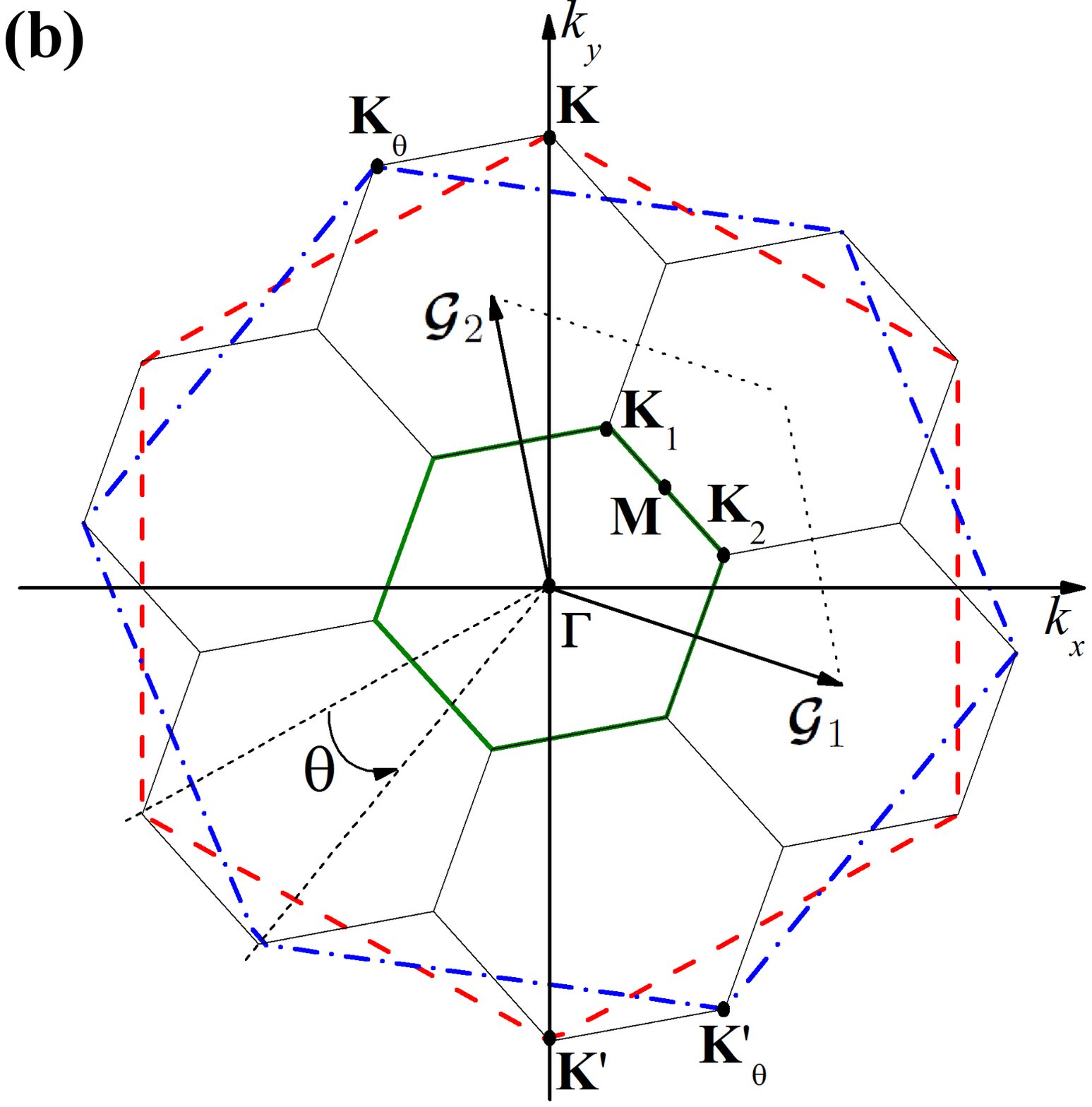}
\caption{(a) The superlattice cell of the tBLG structure with
$m_0=10$,
$r=1$
($\theta\cong3.15^{\circ}$).
Regions with almost AA, AB, and BA stackings are indicated by arrows. (b)
Brillouin zones of layers 1 and 2 (big red and blue hexagons), as well as
the Brillouin zone of the superlattice (small thick green hexagon) of the
structure
$m_0=1$,
$r=1$
($\theta\cong21.79^{\circ}$).
Reciprocal vectors of the superlattice,
$\bm{{\cal G}}_{1,2}$,
Dirac points of layers 1 and 2,
$\mathbf{K}$,
$\mathbf{K}'$,
$\mathbf{K}_{\theta}$,
and
$\mathbf{K}'_{\theta}$,
as well as symmetrical points of the reduced Brillouin zone
($\bm{\Gamma}$,
$\mathbf{M}$,
$\mathbf{K}_{1,2}$)
are also shown.
\label{fig::FigTBLGSCBZ}
}
\end{figure}

Twisting produces moir{\'{e}} patterns~\cite{ourBLGreview2016}, which can be seen as alternating
dark and bright regions in STM
images.
Measuring the moir{\'{e}} period $L$, one can extract the twist angle
according to the formula
$L=a/[2\sin(\theta/2)]$.
Moir{\'{e}} patterns exist for arbitrary twist angles. However, if the twist
angle satisfies the relationship
\begin{equation}
\label{comtheta}
\cos\theta=\frac{3m_0^2+3m_0r+r^2/2}{3m_0^2+3m_0r+r^2}\,,
\end{equation}
where
$m_0$
and $r$ are co-prime positive integers, a superstructure emerges, and a
tBLG sample splits into a periodic lattice of finite supercells. Many theoretical papers assume the twist angle to be the commensurate one, since
only in this case one can work with Bloch waves and introduce the
quasimomentum. For the commensurate structure described by
$m_0$
and $r$, the superlattice vectors are
\begin{equation}
\label{R12}
\mathbf{R}_1=m_0\mathbf{a}_1+(m_0+r)\mathbf{a}_2,
\,
\mathbf{R}_2=-(m_0+r)\mathbf{a}_1+(2m_0+r)\mathbf{a}_2,
\end{equation}
if
$r\neq3n$
($n$ is an integer), or
\begin{equation}
\mathbf{R}_1=(m_0+n)\mathbf{a}_1+n\mathbf{a}_2,
\,
\mathbf{R}_2=-n\mathbf{a}_1+(m_0+2n)\mathbf{a}_2,
\end{equation}
if
$r=3n$.
The number of graphene unit cells inside a supercell is
\begin{equation}
\label{Nsc}
N_{\rm sc}=(3m_0^2+3m_0r+r^2)/g
\end{equation}
per layer. The parameter $g$ in the latter expression is equal to unity when
$r\neq3n$.
Otherwise, it is
$g=3$.

Note that, in the general case, the superlattice cell is greater in size than
the moir{\'{e}}
cell~\cite{dSPRB,ourBLGreview2016}.
More precisely, the superlattice cell of the structure with
$m_0$
and $r$ contains
$r^2$
moir{\'{e}} cells if
$r\neq3n$,
or
$r^2/3$
moir{\'{e}} cells otherwise. The arrangements of atoms in moir{\'{e}} cells
constituting the superlattice cell are slightly different from each other.
Only when
$r=1$,
the superlattice cell coincides with the moir{\'{e}} cell. In the present
paper we consider only such structures. When $\theta$ is small enough, the
superlattice cell can be approximately described as consisting of regions
with almost AA, AB, and BA
stackings~\cite{dSPRB,ourBLGreview2016}.
To illustrate this fact, in
Fig.~\ref{fig::FigTBLGSCBZ}(a)
we present the supercell of the tBLG structure with
$m_0=10$,
$r=1$
(these values of
$m_0$
and $r$ correspond to
$\theta\cong3.15^{\circ}$).

Let us now consider what happens in momentum space. The reciprocal
lattice primitive vectors for layer~1 (layer~2) are denoted by
$\mathbf{b}_{1,2}$
($\mathbf{b}_{1,2}'$).
For layer~1 one has
\begin{equation}
\mathbf{b}_{1,2}=(2\pi/\sqrt{3},\mp 2\pi )/a\,,
\end{equation}
while
$\mathbf{b}_{1,2}'$
are connected to
$\mathbf{b}_{1,2}$
by a rotation of an angle $\theta$. Using the notation
$\bm{{\cal G}}_{1,2}$
for the primitive reciprocal vectors for the superlattice, the following
identities in reciprocal space are valid:
\begin{equation}
\mathbf{b}_1'=\mathbf{b}_1+r(\bm{{\cal G}}_{1}+\bm{{\cal G}}_{2})\,,\;
\mathbf{b}_2'=\mathbf{b}_2-r\bm{{\cal G}}_{1}\,,
\end{equation}
if
$r\neq3n$,
or
\begin{equation}
\mathbf{b}_1'=\mathbf{b}_1+n(\bm{{\cal G}}_{1}+2\bm{{\cal G}}_{2})\,,\;
\mathbf{b}_2'=\mathbf{b}_2-n(2\bm{{\cal G}}_{1}+\bm{{\cal G}}_{2})\,,
\end{equation}
if
$r=3n$.

Each graphene layer in tBLG has a hexagonal-shaped Brillouin zone. The
Brillouin zone of the layer $2$ is rotated in momentum space with respect
to the Brillouin zone of layer $1$ by the twist angle $\theta$. The
Brillouin zone of the superlattice (reduced Brillouin zone, RBZ) is also
hexagonal-shaped, but smaller in size. It can be obtained by
$N_{\rm sc}$-times
folding of the Brillouin zone of the layer $1$ or $2$. Two non-equivalent
Dirac points of the layer~$1$ are
\begin{eqnarray}
\mathbf{K}=\frac{4\pi}{3a}(0,1),
\quad
\mathbf{K}'=\frac{4\pi}{3a}(0,-1).
\nonumber
\end{eqnarray}
The Dirac points of the layer~$2$ are
\begin{eqnarray}
\mathbf{K}_{\theta}=\frac{4\pi}{3a}(-\sin\theta,\cos\theta),
\quad
\mathbf{K}'_{\theta}=\frac{4\pi}{3a}(\sin\theta,-\cos\theta).
\nonumber
\end{eqnarray}
Band folding translates these four Dirac points to the two Dirac points of
the superlattice,
$\mathbf{K}_{1,2}$.
Thus, one can say that Dirac points of the superlattice are doubly
degenerate. Points
$\mathbf{K}_1$
and
$\mathbf{K}_2$
can be expressed via vectors
$\bm{{\cal G}}_{1,2}$
as
\begin{equation}
\mathbf{K}_1=\frac{1}{3}(\bm{{\cal G}}_{1}+2\bm{{\cal G}}_{2})\,,
\quad
\mathbf{K}_2=\frac{1}{3}(2\bm{{\cal G}}_{1}+\bm{{\cal G}}_{2})\,.
\end{equation}
A typical picture illustrating these three Brillouin zones, the vectors
$\bm{{\cal G}}_{1,2}$,
as well as main symmetrical points is shown in
Fig.~\ref{fig::FigTBLGSCBZ}(b).

\section{Model Hamiltonian and its mean-field treatment}
\label{Model}

We start from the following electronic Hamiltonian of the tBLG:
\begin{eqnarray}
\label{H}
H&=&\!\!\!\sum_{{\mathbf{nm}ij\atop sr\sigma}}\!t(\mathbf{r}_{\mathbf{n}}^{is};\mathbf{r}_{\mathbf{m}}^{jr})d^{\dag}_{\mathbf{n}is\sigma}d^{\phantom{\dag}}_{\mathbf{m}jr\sigma}+
U\!\sum_{{\mathbf{n}is}}\!n_{\mathbf{n}is\uparrow}n_{\mathbf{n}is\downarrow}+\nonumber\\
&&\frac12\!\mathop{{\sum}'}_{{\mathbf{nm}ij\atop sr\sigma\sigma'}}\!V(\mathbf{r}_{\mathbf{n}}^{is}-\mathbf{r}_{\mathbf{m}}^{jr})n_{\mathbf{n}is\sigma}n_{\mathbf{m}jr\sigma'}\,.
\end{eqnarray}
In this expression
$d^{\dag}_{\mathbf{n}is\sigma}$
($d^{\phantom{\dag}}_{\mathbf{n}is\sigma}$)
are the creation (annihilation) operators of the electron with spin
$\sigma$\,($=\uparrow,\downarrow$)
at the unit cell
$\mathbf{n}$
in the layer
$i$\,($=1,2$)
in the sublattice
$s$\,($={\cal A,B}$),
while
$n_{\mathbf{n}is\sigma}=d^{\dag}_{\mathbf{n}is\sigma}d^{\phantom{\dag}}_{\mathbf{n}is\sigma}$.
The first term in
Eq.~\eqref{H}
is the single-particle tight-binding Hamiltonian with
$t(\mathbf{r}_{\mathbf{n}}^{is};\mathbf{r}_{\mathbf{m}}^{jr})$
being the amplitude of the electron hopping from site in the position
$\mathbf{r}_{\mathbf{m}}^{jr}$
to the site in the position
$\mathbf{r}_{\mathbf{n}}^{is}$.
The second term in
Eq.~\eqref{H}
describes the on-site (Hubbard) interaction of electrons with opposite
spins, while the last term corresponds to the intersite Coulomb
interaction (the prime near the last sum in Eq.~\eqref{H} means that elements with $\mathbf{r}_{\mathbf{n}}^{is}=\mathbf{r}_{\mathbf{m}}^{jr}$ should be excluded).

\begin{figure*}[t]
\centering
\includegraphics[width=0.32\textwidth]{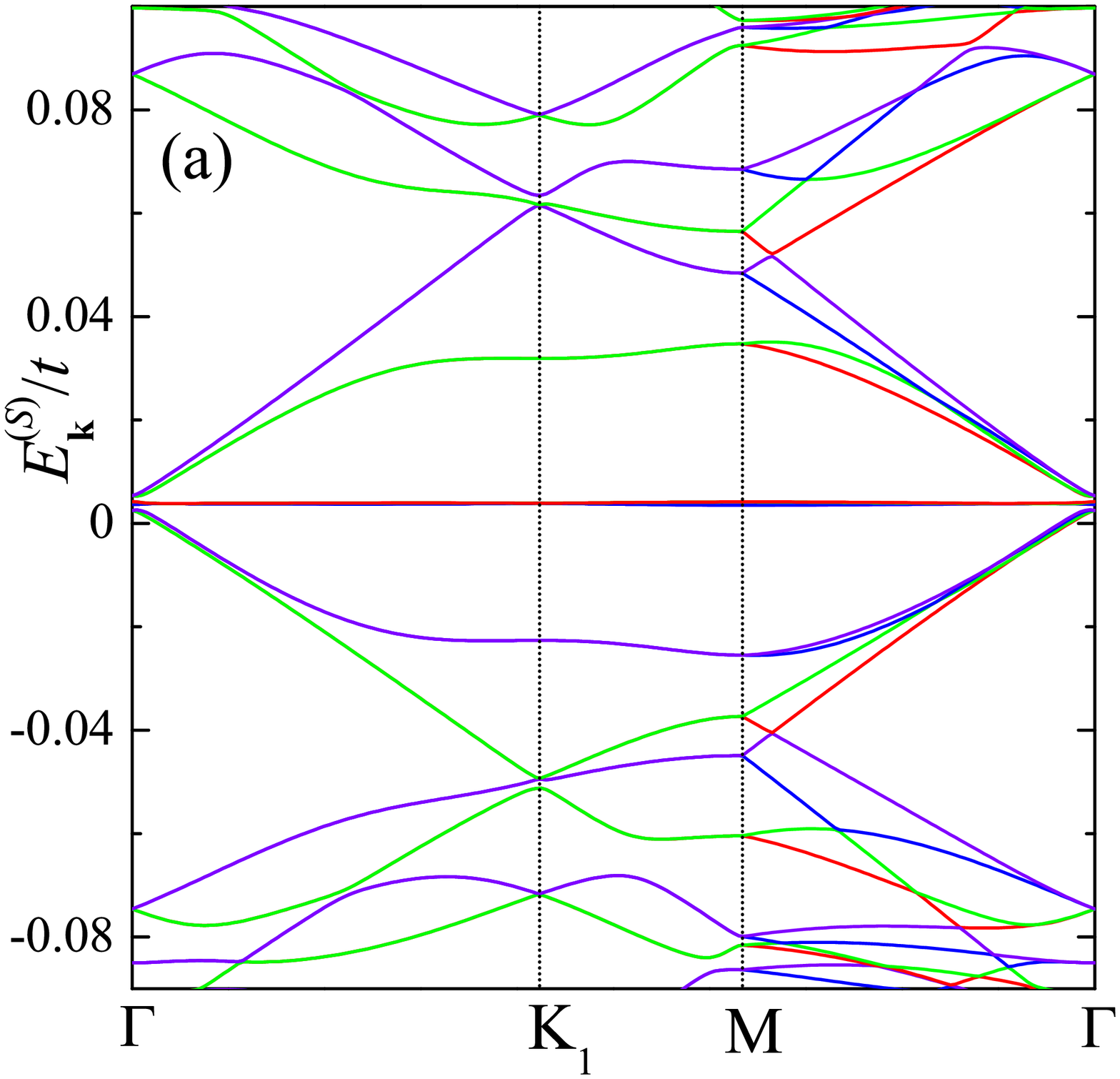}
\includegraphics[width=0.32\textwidth]{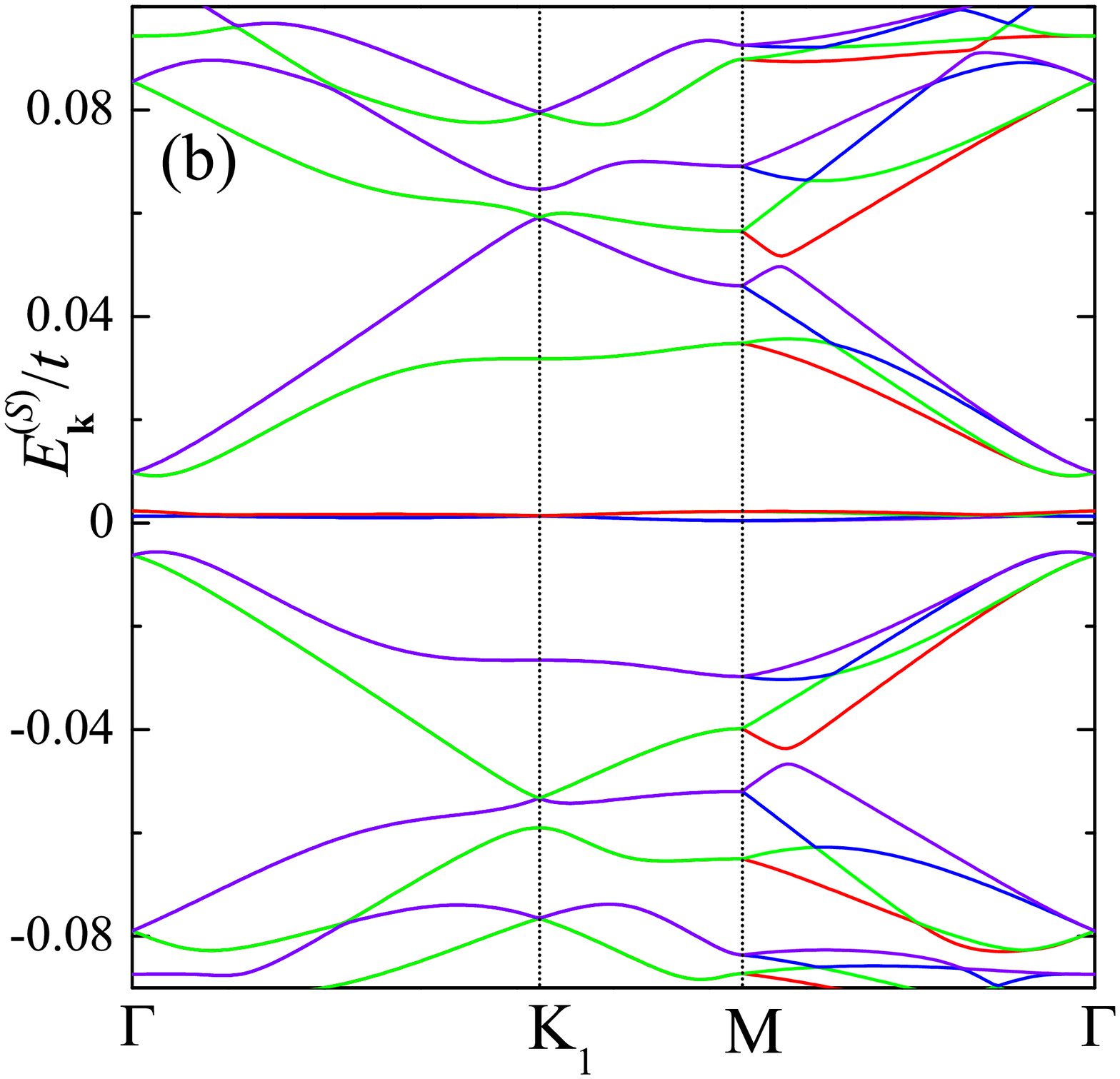}
\includegraphics[width=0.32\textwidth]{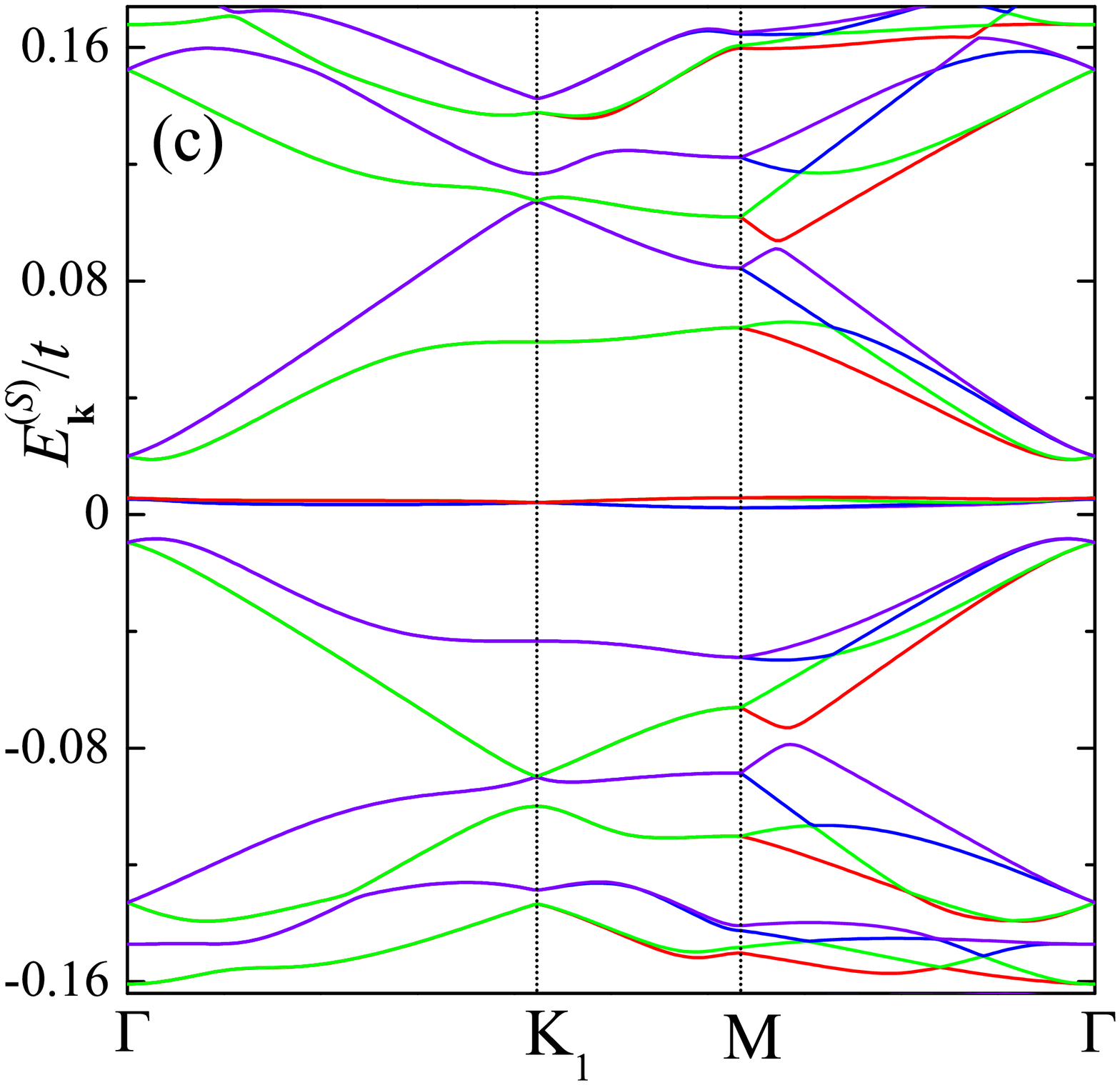}\\
\includegraphics[width=0.32\textwidth]{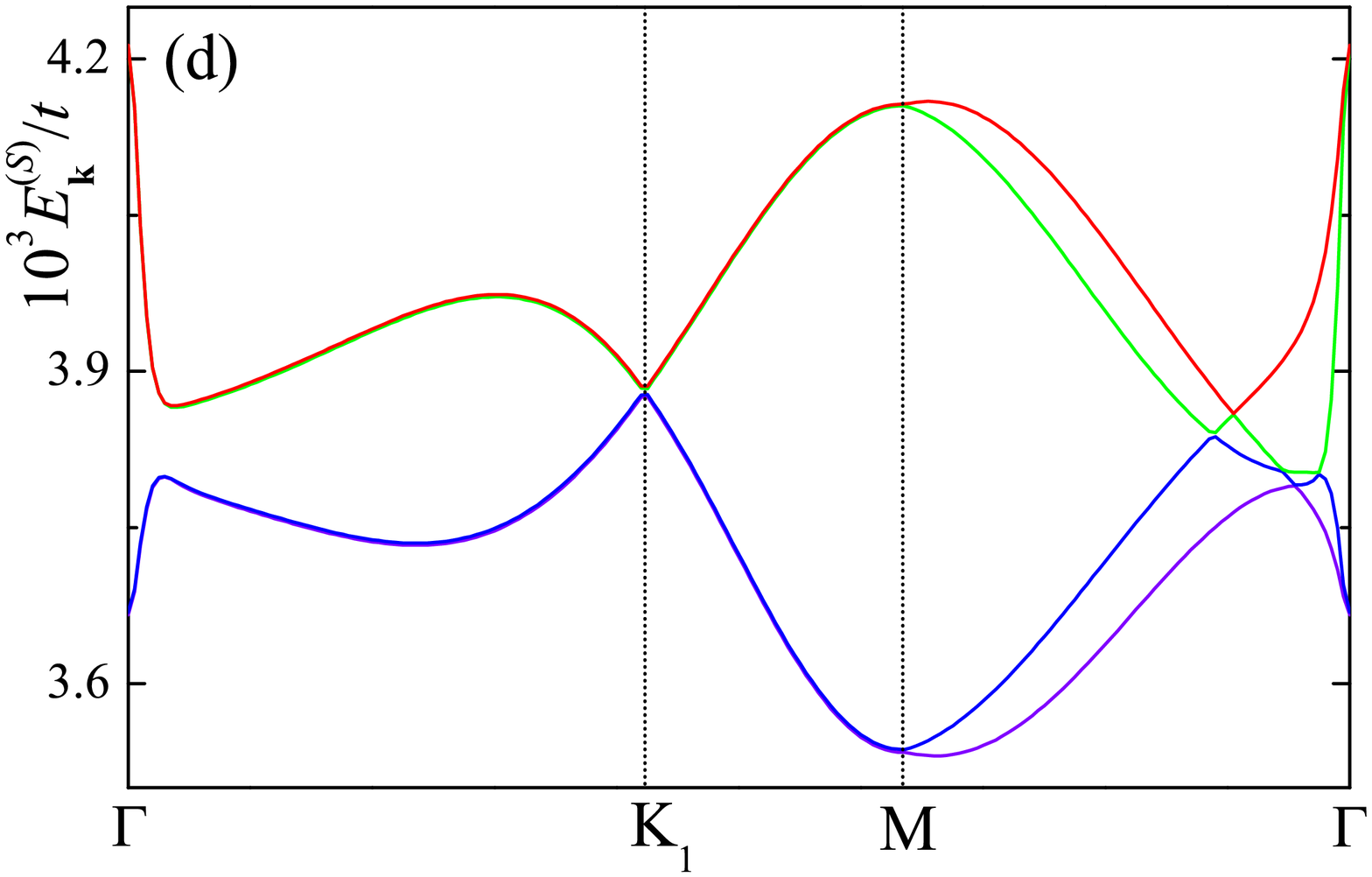}
\includegraphics[width=0.32\textwidth]{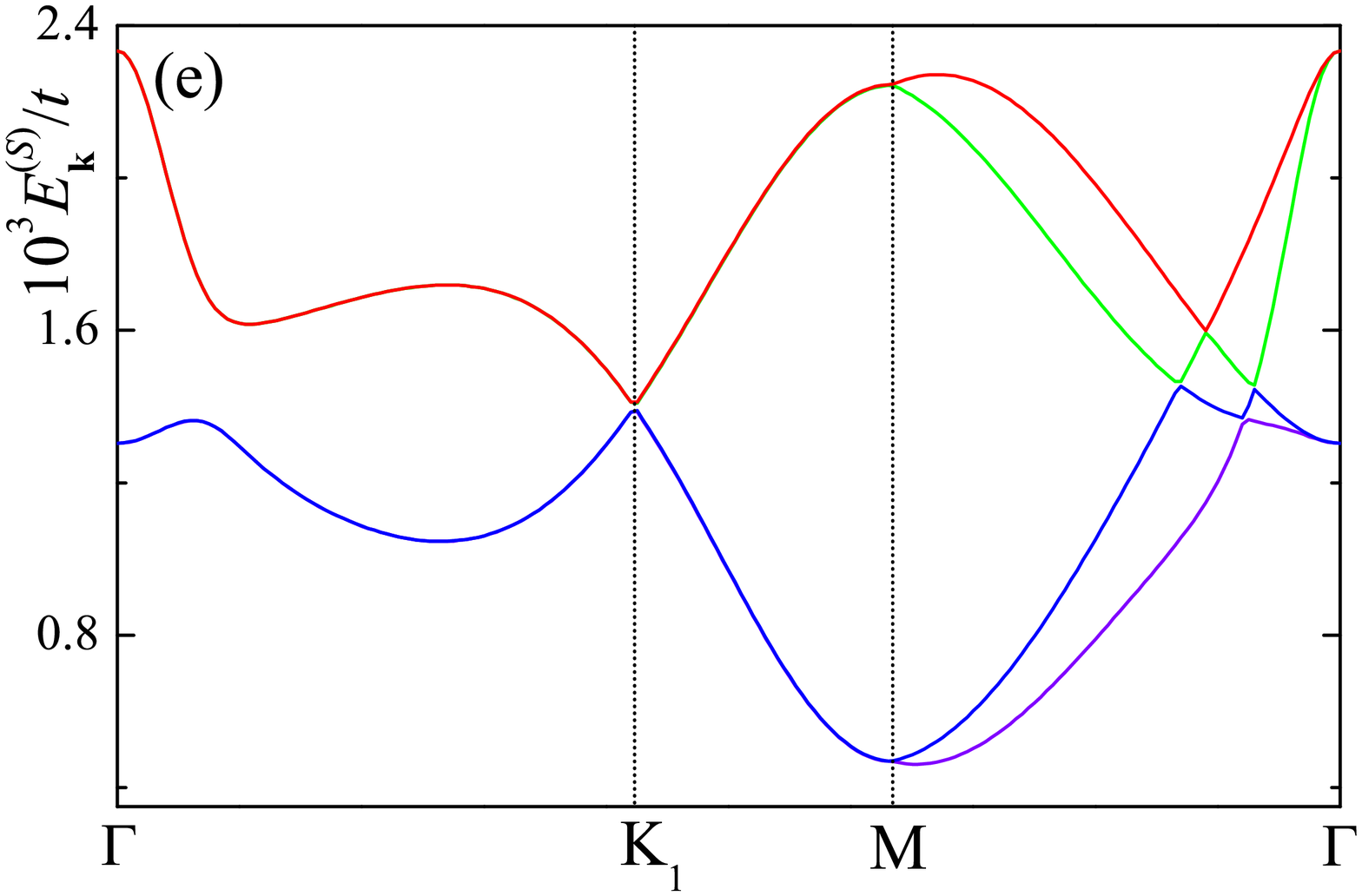}
\includegraphics[width=0.32\textwidth]{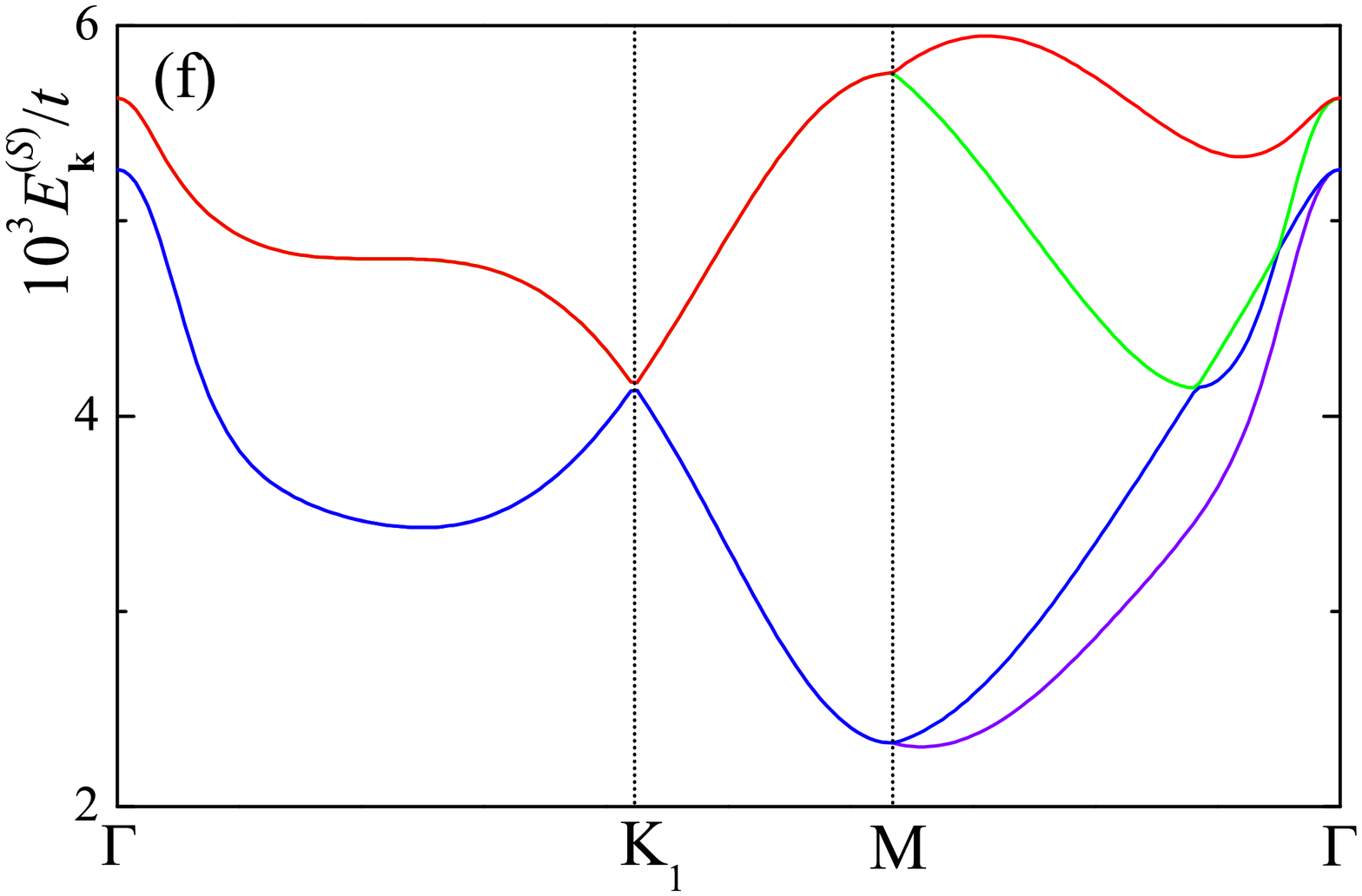}\\
\includegraphics[width=0.25\textwidth]{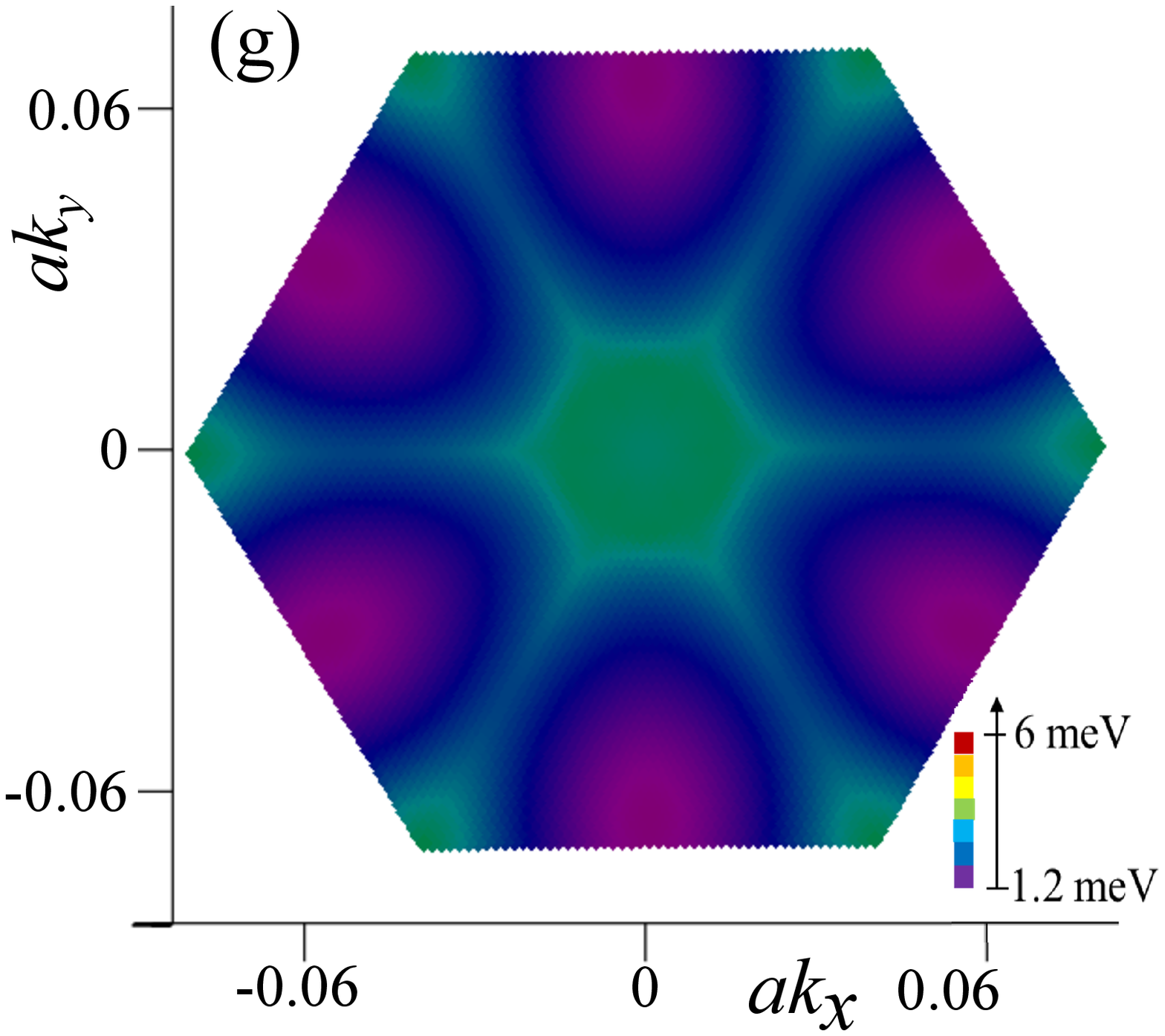}
\includegraphics[width=0.24\textwidth]{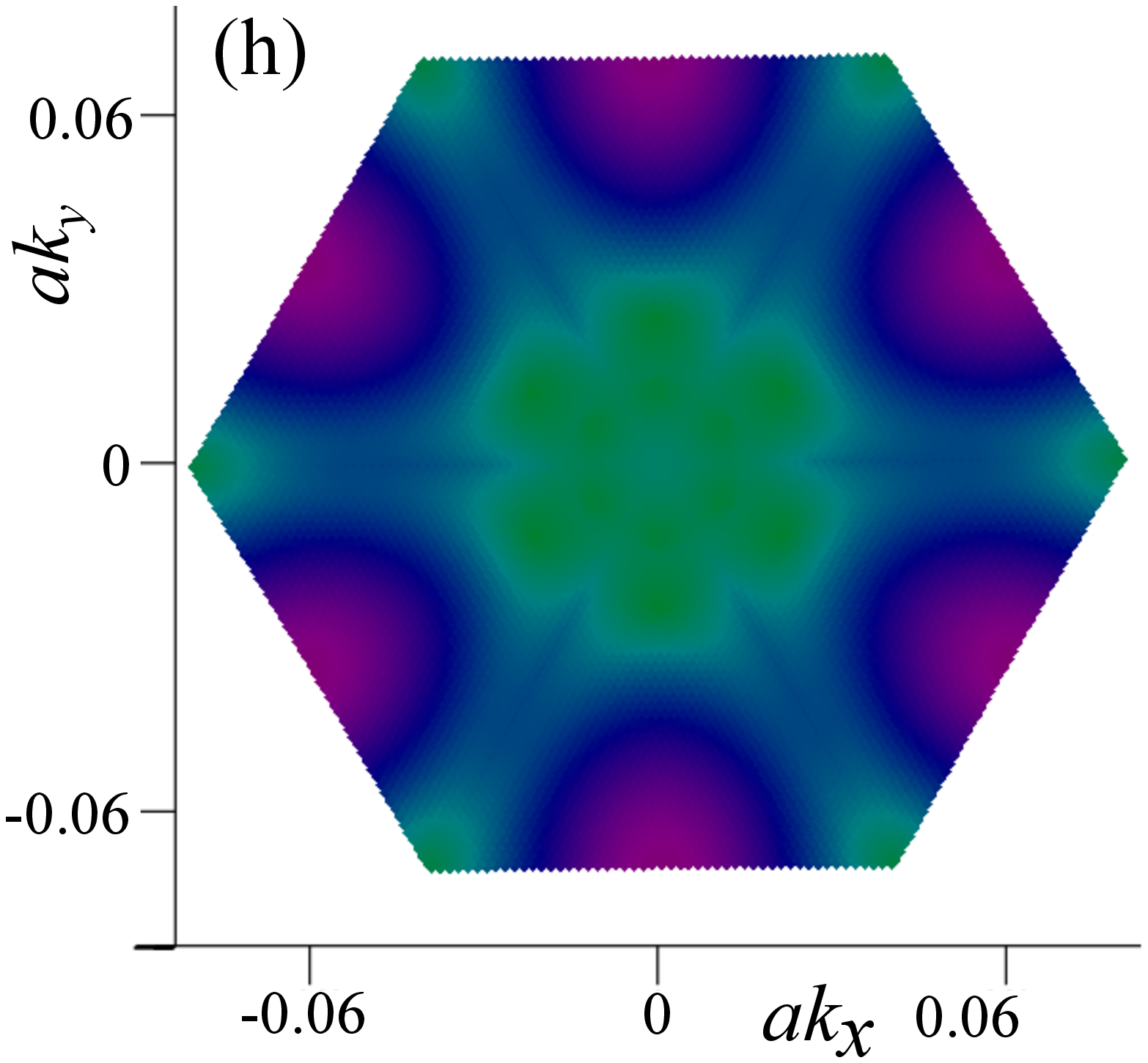}
\includegraphics[width=0.24\textwidth]{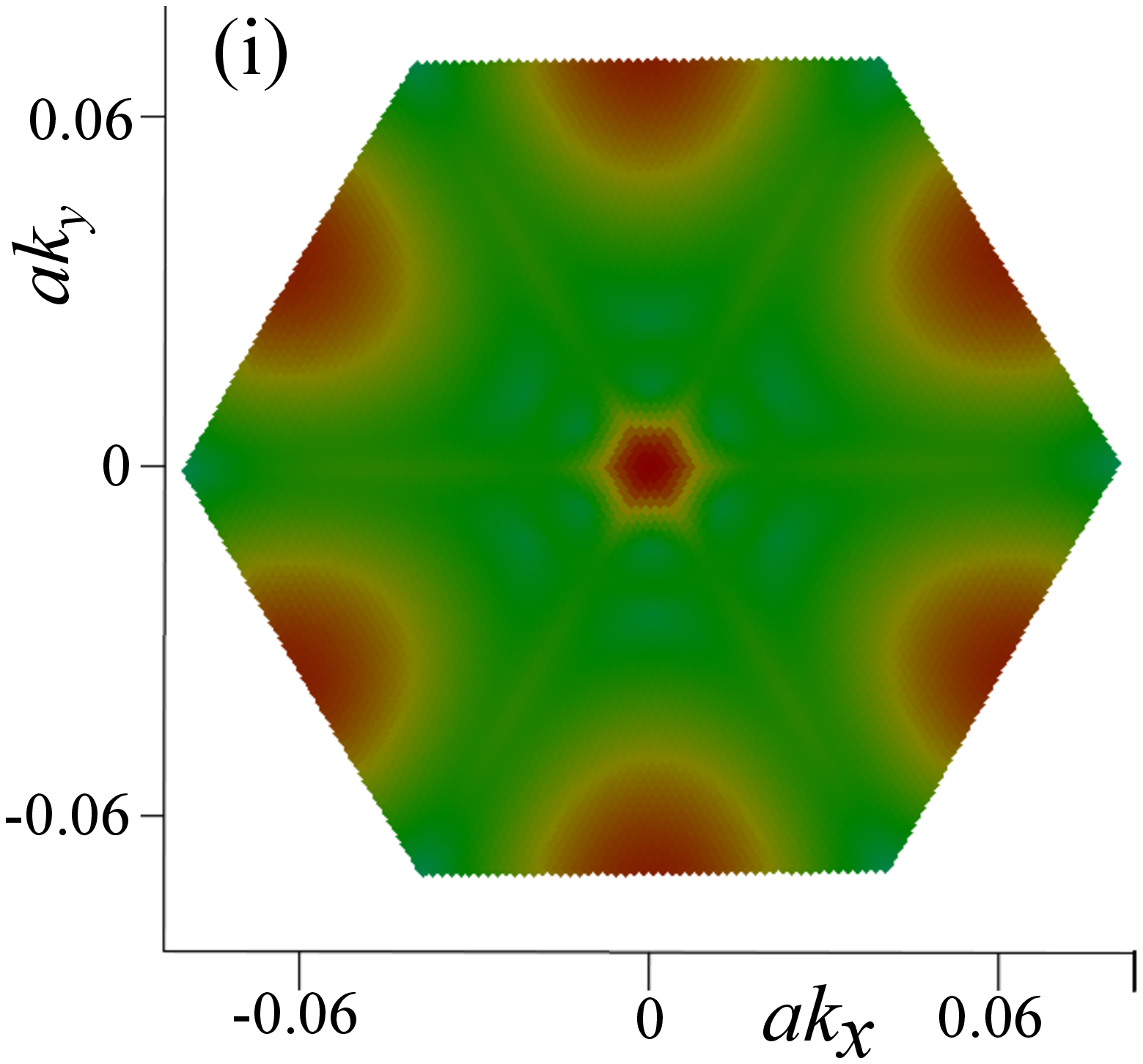}
\includegraphics[width=0.24\textwidth]{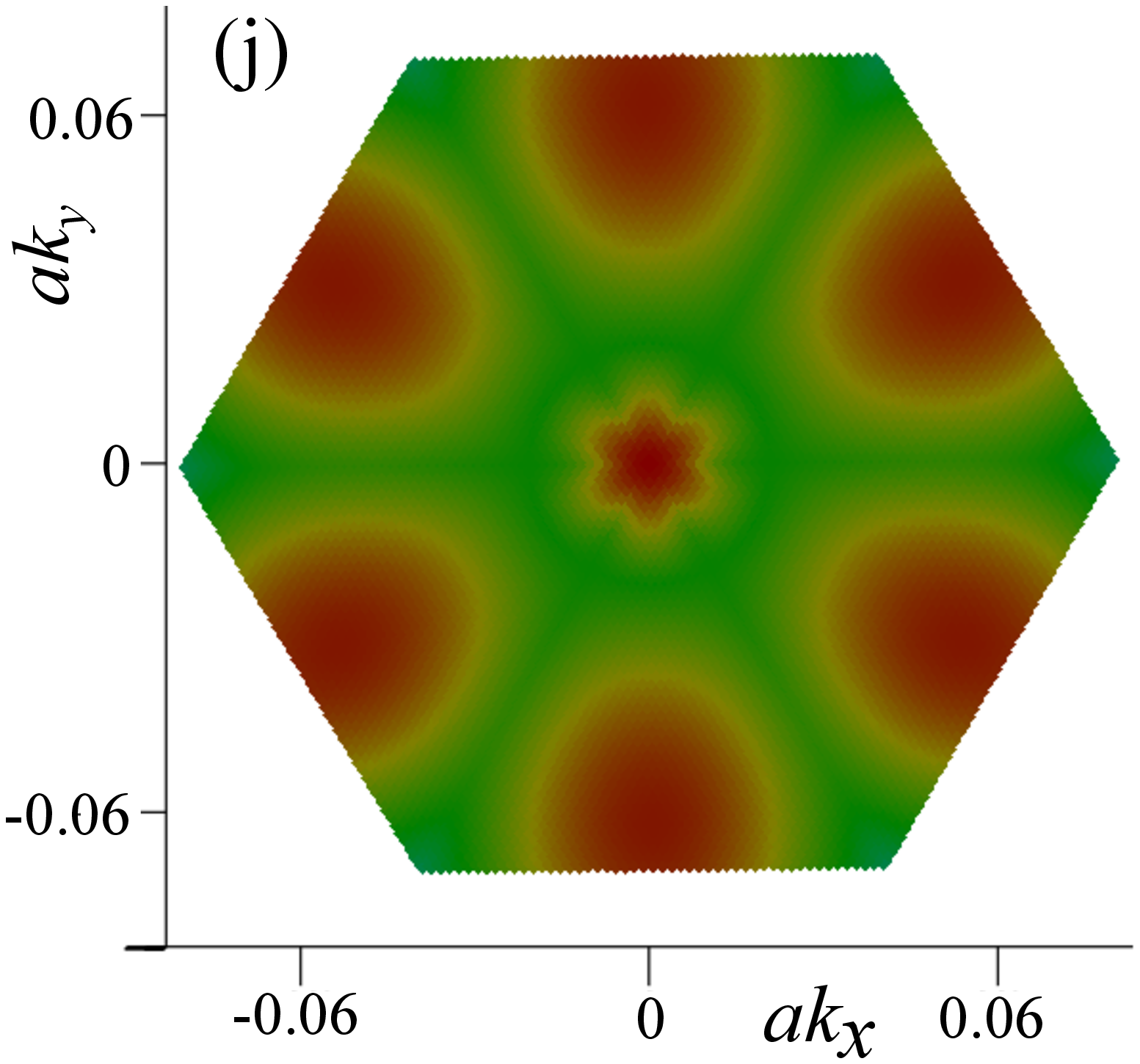}
\caption{(a)\,--\,(c) Energy spectra for
$\theta=\theta_c$
calculated for parametrization~I (a), II.A (b), and~II.B (c). The energy
window for panel (c) (parametrization~II.B) is about two times larger than
that for panels (a) and (b). (d)\,--\,(f) Fine structures of the flat
bands corresponding to panels (a)\,--\,(c), respectively. (g)\,--\,(j) Four
flat low-energy bands calculated inside RBZ for parametrization~II.A.
The color bar is the same for all four plots [see panel (g)].
\label{FigSpec}
}
\end{figure*}

\subsection{Hopping amplitude parametrization schemes}

Let us consider first the single-particle properties of MAtBLG. If we
neglect interactions, the electronic spectrum of the system is obtained by
diagonalization of the first term of the
Hamiltonian~\eqref{H}.
The result depends on the parametrization of the hopping amplitudes
$t(\mathbf{r}_{\mathbf{n}}^{is};\mathbf{r}_{\mathbf{m}}^{jr})$.
In this paper we keep only nearest-neighbor terms for the intralayer
hopping. The corresponding amplitude is
$t=-2.57$\,eV.

Unlike the intralayer hopping, there is no universally accepted
parametrization scheme for the interlayer hopping amplitudes. They are much
weaker than the intralayer amplitude $t$, and may be significantly affected
by numerous non-universal poorly controlled factors (elastic deformations,
relative layer sliding, disorder). To address this uncertainty, we will
study the
model~(\ref{H})
with three different parametrizations for the function
$t(\mathbf{r};\mathbf{r}')$.
These parametrizations, as well as the single-electron spectra
corresponding to them, are presented below.

The parametrization~I is rather simple. The function
$t(\mathbf{r};\mathbf{r}')$
is described by the following
Slater-Koster~\cite{SlaterKoster}
formula for
$p_z$
electrons (the corresponding contribution from $V_{\pi}(\mathbf{r})$ is assumed to be negligible):
\begin{equation}
\label{ParI}
t(\mathbf{r};\mathbf{r}')=\frac{\left[(\mathbf{r}-\mathbf{r}')\mathbf{e}_z\right]^2}{|\mathbf{r}-\mathbf{r}'|^2}V_{\sigma}(\mathbf{r}-\mathbf{r}')\,,
\end{equation}
where
\begin{equation}
\label{Vsigma}
V_{\sigma}(\mathbf{r})
=
t_0e^{-(|\mathbf{r}|-d)/r_0} F_c(|\mathbf{r}|)\,,\;\;
F_c(r)=\frac{1}{1+e^{(r-r_c)/l_c}}\,.
\end{equation}
The cutoff function
$F_c(r)$
is introduced to nullify the hopping amplitudes at distances larger than
$r_c$; we use
$r_c=4.92$\,\AA,
$l_c=0.2$\,\AA.
The parameter
$t_0$
defines the largest interlayer hopping amplitude. We choose
$t_0=0.37$\,eV
(this value was used to describe the AB bilayer
graphene~\cite{ourBLGreview2016}).
The parameter
$r_0$
describes how fast the hopping amplitudes decay inside the region
$r<r_c$.
We choose
$r_0=0.34$\,\AA.

The parametrization~I,
expressed by Eqs.~\eqref{ParI}\,--\,\eqref{Vsigma},
both with and without a cutoff function, is widely used in the
literature.~\cite{NanoLettTB,TramblyTB_Loc,Morell1,Morell2,OurtBLGPRB2019}
However, in the limiting case of AB bilayer
($\theta=0$),
Eqs.~\eqref{ParI}\,--\,\eqref{Vsigma}
cannot correctly reproduce the Slonczewski-Weiss-McClure (SWMc)
parametrization. Indeed, the AB bilayer has three distinct nearest-neighbor
interlayer hopping amplitudes:
$t_0$
(hopping from site
$1{\cal A}$
to the nearest site
$2{\cal B}$),
$\gamma_3$
(hopping from
$1{\cal B}$
to
$2{\cal A}$),
and
$\gamma_4$
(hopping from
$1{\cal A}$
to
$2{\cal A}$
and from
$1{\cal B}$
to
$2{\cal B}$).
The parametrization~I gives
$\gamma_3=\gamma_4$,
while experiment
shows~\cite{SWMc1exp,SWMc2exp}
that
$\gamma_3>\gamma_4$.

To comply with the SWMc scheme, we consider yet other parametrization,
designated below as `parametrization~II'. It is a more sophisticated
approach, initially proposed in
Ref.~\onlinecite{Tang}.
Parametrization~II takes into account the environment dependence of the
hopping. That is, the electron hopping amplitude connecting two atoms at
positions $\mathbf{r}$ and $\mathbf{r}'$ depends not only on the difference
$\mathbf{r}-\mathbf{r}'$,
but also on positions of other atoms in the lattice. Extra flexibility of
the formalism becomes useful when the tunneling between
$\mathbf{r}$
and
$\mathbf{r}'$
is depleted by nearby atoms, which act as obstacles to a tunneling
electron. For tBLG, the parametrization~II was used in
Refs.~\onlinecite{Pankratov1,ourTBLG,ourTBLG2017},
among other papers.

To use parametrization~II for MAtBLG, values of several fitting parameters
have to be assigned. We choose them in such a way as to correctly
describe the case of the AB bilayer, with
$\gamma_3>\gamma_4$
(details can be found in our previous paper
Ref.~\onlinecite{ourTBLG}).
One of the fitting parameters is
$t_0$: the largest interlayer hopping amplitude. It also scales all other
interlayer hopping amplitudes. We perform all calculations for two versions
of parametrization~II denoted below as II.A and II.B. They have different
values of
$t_0$.
All other fitting parameters are identical for II.A and II.B. Specifically,
for the parametrization~II.A we assign
$t_0=0.21$\,eV
to guarantee that the angle
$\theta_c$
is the same for both II.A and I (the precise definition of
$\theta_c$
will be given below, in the next subsection). For parametrization~II.B the
value of
$t_0$
is the same as for parametrization~I:
$t_0=0.37$\,eV.
In other words, the overall interlayer tunneling energy scale is the same
for both I and II.B. However, the values of
$\theta_c$
for these parametrizations deviate significantly from each other.

\begin{table}
\begin{center}
\begin{tabular}{|l|c|c|c|c|c|}
\hline
&$t_0$&$\theta_c$&$W$&$\Delta_u$&$\Delta_d$\\
\hline
Par. I&\,$0.37$\,eV\,&\,$1.08^{\circ}$\,&\,$1.8$\,meV\,&\,$2.7$\,meV\,&\,$2.3$\,meV\,\\
\hline
Par. II.A&\,$0.21$\,eV\,&\,$1.08^{\circ}$\,&\,$4.8$\,meV\,&\,$17.5$\,meV\,&\,$15.5$\,meV\,\\
\hline
Par. II.B&\,$0.37$\,eV\,&\,$1.89^{\circ}$\,&\,$9.4$\,meV\,&\,$33.2$\,meV\,&\,$27.1$\,meV\,\\
\hline
\end{tabular}
\end{center}
\caption{Various single-electron characteristics of the MAtBLG spectrum for
three parametrizations of the interlayer hopping.
\label{table1}
}
\end{table}

\subsection{Single-particle spectrum of MAtBLG}

Once a specific parametrization is chosen, the single-electron part of our
model may be diagonalized, and its single-electron spectrum may be found.
Regardless of the type of the parametrization, the tBLG spectrum has
common features. For each superstructure
$(m_0,\,r)$,
the tBLG spectrum consists of
$4N_{\text{\rm sc}}$
energy bands
$E_{0\mathbf{k}}^{(S)}$
with quasimomentum $\mathbf{k}$ lying inside the reduced Brillouin zone,
and
$1\leq S\leq4N_{\text{\rm sc}}$.
For given $\mathbf{k}$,the energies
$E_{0\mathbf{k}}^{(S)}$
are arranged in ascending order.

When the twist angle is not too small, the spectrum at low energies
consists of two doubly degenerate Dirac cones located near the RBZ Dirac points
$\mathbf{K}_1$
and
$\mathbf{K}_2$.
These Dirac cones intersect at energies above and below the cone apex
energy giving rise to the low-energy van Hove singularities.

The interlayer hybridization renormalizes the Fermi velocity of the Dirac
cones, making it smaller than the Fermi velocity of the single-layer
graphene. At not-too-small $\theta$, the renormalized velocity decreases
when $\theta$
decreases~\cite{dSPRB,PNAS}.
The energies of the van Hove singularities demonstrate a similar dependence
on $\theta$.

The Dirac cones inherited from two graphene sheets are hosted by four bands
$E_{0\mathbf{k}}^{({\cal S})}$,
with
${\cal S}
=
2N_{\text{\rm sc}}-1,\,2N_{\text{\rm sc}},\,2N_{\text{\rm sc}}+1,\,2N_{\text{\rm sc}}+2$.
Since in a pristine or weakly doped sample these are the single-electron
states closest to the Fermi energy, the low-temperature properties of the
MAtBLG are controlled by these bands. Consequently,
their total width $W$ defined as
\begin{equation}
W=\max_{\mathbf{k}}(E_{0\mathbf{k}}^{(2N_{\text{\rm sc}}+2)})-\min_{\mathbf{k}}(E_{0\mathbf{k}}^{(2N_{\text{\rm sc}}-1)})
\end{equation}
is an important characteristics of the MAtBLG spectrum. As long as the
twist angle is not too small, $W$ decreases with decreasing $\theta$.

Both numerical and analytical studies demonstrate that both the Fermi
velocity and the width $W$ experience substantial reduction as $\theta_c$ decreases. Yet, in a wide range of $\theta$, the tBLG formally remains a semimetal at the charge neutrality point.
However, at some critical twist angle
$\theta_c$
the system acquires a Fermi surface even at zero doping. For
$\theta < \theta_c$
the tBLG remains in a formally metallic state.

The value of
$\theta_c$
is not universal, and depends on particulars of the interlayer tunneling.
For parametrizations~I and~II.A one has
$\theta_c\cong1.08^{\circ}$
[$(m_0,\,r)=(30,\,1)$].
For parametrization~II.B, the Fermi surface arises at larger angle,
$\theta_c\cong1.89^{\circ}$
[$(m_0,\,r)=(17,\,1)$].
With further decrease of the twist angle, the bandwidth $W$ becomes
an oscillating function of $\theta$. For all three parametrizations under
study, the width $W$ has a minimum at
$\theta=\theta_c$.
For each parametrization, the numerical calculations presented below were
performed at
$\theta = \theta_c$
(one must remember that
$\theta_c$
is a parametrization-specific quantity).

Formally speaking, our
$\theta_c$
differs from the common definition of the first magic angle introduced in
Ref.~\onlinecite{PNAS}.
According to the latter, the first magic angle corresponds to nullification
of the Fermi velocity at the Dirac points, yet, in our study this velocity
remains non-zero when
$\theta=\theta_c$.
While both definitions give similar values of the twist angle, these values
are not identical. We choose to work in the regime of smallest $W$ since
the logic of the mean-field approximation suggests that this regime
corresponds to the largest condensation energy.

The low-energy structure of the numerically calculated spectra at
$\theta=\theta_c$,
are shown in
Figs.~\ref{FigSpec}(a)\,--\,(c)
for all three parametrizations. The finer details for the flat bands
$E_{0\mathbf{k}}^{({\cal S})}$
may be examined in
Figs.~\ref{FigSpec}(d)\,--\,(f).
We show the bands along the contour
${\bm\Gamma}\to\mathbf{K}_1\to\mathbf{M}\to{\bm\Gamma}$.
Qualitatively, the low-energy spectra for all parametrizations look very
similar. We see a Dirac cone near point
$\mathbf{K}_{1}$,
local extrema near the
$\mathbf{M}$
point, and complicated behavior on the line
$\mathbf{M}\to{\bm\Gamma}$.

On the quantitative level, however, the characteristics of the low-energy
bands are different. For example, the bandwidth $W$ for the
parametrization~II.B is about $6$ times larger than that for the
parametrization~I, and about $2$ larger than that for the
parametrization~II.A. Other important parametrization-dependent quantities
are the energy gaps separating the flat bands
$E_{0\mathbf{k}}^{({\cal S})}$
from dispersive bands at higher and lower energies.
Formally speaking, these gaps can be defined as
\begin{eqnarray}
\Delta_{d}&=&\max_{\mathbf{k}}(E_{0\mathbf{k}}^{(2N_{\text{\rm sc}}-2)})-\min_{\mathbf{k}}(E_{0\mathbf{k}}^{(2N_{\text{\rm sc}}-1)}),\nonumber\\
\Delta_{u}&=&\max_{\mathbf{k}}(E_{0\mathbf{k}}^{(2N_{\text{\rm sc}}+2)})-\min_{\mathbf{k}}(E_{0\mathbf{k}}^{(2N_{\text{\rm sc}}+3)}).
\end{eqnarray}
Our numerical data demonstrates that the values of
$\Delta_{u}$
and
$\Delta_{d}$
for parametrizations~II.A and~II.B exceed the values for the
parametrization~I by order of magnitude. The characteristics of the
low-energy spectra for all three parametrizations at
$\theta=\theta_c$
are summarized in
Table~\ref{table1}.

Finally, let us briefly discuss the symmetry properties of the flat
bands.
Figures~\ref{FigSpec}(g)\,--\,(j)
show the low-energy spectra calculated inside the reduced Brillouin zone for
parametrization~II.A. We see that the spectra have hexagonal symmetry.
Spectra are also symmetric under reflections with respect to the axes
parallel and perpendicular to
$\mathbf{G}_1$,
$\mathbf{G}_2$,
and
$\mathbf{G}_1+\mathbf{G}_2$.
All these symmetries are observed also for the other parametrizations as
well. However, below we will see that the symmetry of the low-energy
spectra can be reduced if we include interactions into account.

\subsection{Structure of the SDW order parameters}

The system having flat bands intersecting the Fermi level is very
susceptible to interactions. In our model, the interactions are described by the second
and the third terms in the total
Hamiltonian~\eqref{H}.
They represent the on-site and intersite Coulomb repulsion.
Interactions spontaneously break symmetries of the single-particle
Hamiltonian generating a finite order parameter. We assume here that this
order parameter is a spin density wave. This choice is not arbitrary. It
was shown in many papers (see, e.g.,
Refs.~\onlinecite{dSPRB,NonAbelianGaugePot,ourTBLG}),
that at small twist angles, electrons at the Fermi level occupy mainly the
regions with almost perfect AA stacking within a supercell. At the same
time, it was demonstrated
theoretically~\cite{AAPRL,AAPRB2013,aa_phasep2013,akzyanov_aa2014}
that the ground state of the AA
stacked bilayer graphene should be
antiferromagnetic.
For this reason we believe that the SDW should be a good candidate for the
ground state of the MAtBLG.

Our SDW order parameter is a multicomponent one. First, it contains terms
proportional to the on-site expectation values of electrons with opposite
spins. To be more specific, we define
\begin{eqnarray}
\label{Deltania}
\Delta_{\mathbf{n}is}=U\langle d^{\dag}_{\mathbf{n}is\uparrow}d^{\phantom{\dag}}_{\mathbf{n}is\downarrow}\rangle\,.
\end{eqnarray}
These components are controlled by the Hubbard interaction. We take
$U=2t$.
This value is somewhat smaller than the critical value for a single-layer
graphene transition into a mean-field antiferromagnetic
state~\cite{MF_Uc_sorella1992},
$U_c=2.23t$.
Thus, our Hubbard interaction is rather strong, but not too strong to open
a gap in the single layer graphene.

Next, we include the intralayer nearest-neighbor SDW order parameter. In a
graphene layer, each atom in one sublattice has three nearest-neighbors
belonging to another sublattice (for example, an atom on sublattice
${\cal B}$
has three nearest neighbors on sublattice
${\cal A}$).
For this reason we consider three types of intralayer nearest-neighbor
order parameters,
$A^{(\ell)}_{\mathbf{n}i\sigma}$
($\ell=1,\,2,\,3$),
corresponding to three different links connecting the nearest-neighbor
sites. These order parameters are defined as follows
\begin{equation}
\label{Anis}
A^{(\ell)}_{\mathbf{n}i\sigma}
=
V_{\rm nn}\langle
	d^{\dag}_{\mathbf{n}+\mathbf{n}_{\ell}i{\cal A}\sigma}
	d^{\phantom{\dag}}_{\mathbf{n}i{\cal B}\bar{\sigma}}
\rangle\,,
\end{equation}
where
$\mathbf{n}_{1}=(0,\,0)$,
$\mathbf{n}_{2}=(1,\,0)$,
$\mathbf{n}_{3}=(0,\,1)$,
$\bar{\sigma}=-\sigma$,
and
$V_{\rm nn}=V (|\bm{\delta}|)$
is the in-plane nearest-neighbor Coulomb repulsion energy. We take
$V_{\text{nn}}/U=0.59$,
in agreement with
Ref.~\onlinecite{Wehling}.

Finally, we consider the interlayer SDW order parameter. It is defined as
follows
\begin{equation}
\label{Mmn}
B^{rs}_{\mathbf{m};\mathbf{n}\sigma}
=
V(\mathbf{r}^{1r}_{\mathbf{m}}-\mathbf{r}^{2s}_{\mathbf{n}})
\langle
	d^{\dag}_{\mathbf{m}1r\sigma}
	d^{\phantom{\dag}}_{\mathbf{n}2s\bar{\sigma}}
\rangle\,.
\end{equation}
For calculations we assume that
$B^{rs}_{\mathbf{m};\mathbf{n}\sigma}$
is non-zero only when sites
$\mathbf{r}_{\mathbf{m}}^{1s}$
and
$\mathbf{r}_{\mathbf{n}}^{2r}$
are sufficiently close. Namely, if the hopping amplitude connecting
$\mathbf{r}_{\mathbf{m}}^{1s}$
and
$\mathbf{r}_{\mathbf{n}}^{2r}$
vanishes, then the parameter
$B^{rs}_{\mathbf{m};\mathbf{n}\sigma}$
is zero. The number of non-zero
$B^{rs}_{\mathbf{m};\mathbf{n}\sigma}$
depends on the type of the hopping amplitude parametrization. For
parametrizations~II.A and~II.B we have up to three non-zero
$B^{rs}_{\mathbf{m};\mathbf{n}\sigma}$
for a given $\mathbf{n}$, $r$, $s$, and $\sigma$. For parametrization~I we
have up to $9$ such
$B^{rs}_{\mathbf{m};\mathbf{n}\sigma}$.
Assuming screening is small at short distances we model the function
$V(\mathbf{r})$
in
Eq.~\eqref{Mmn}
as
$V(\mathbf{r})\propto1/|\mathbf{r}|$
with
$V(d)=V_{\text{nn}}|\bm{\delta}|/d=0.25U$.

\begin{figure*}[t]
\centering
\includegraphics[width=0.32\textwidth]{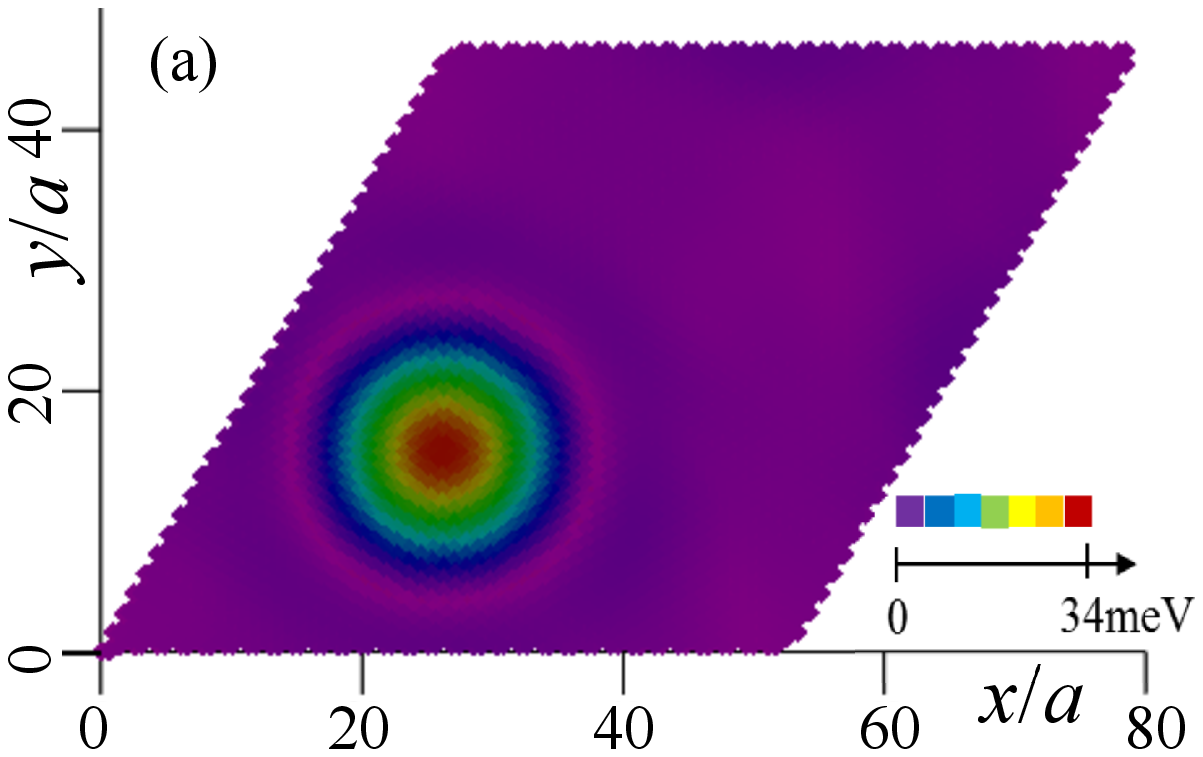}
\includegraphics[width=0.32\textwidth]{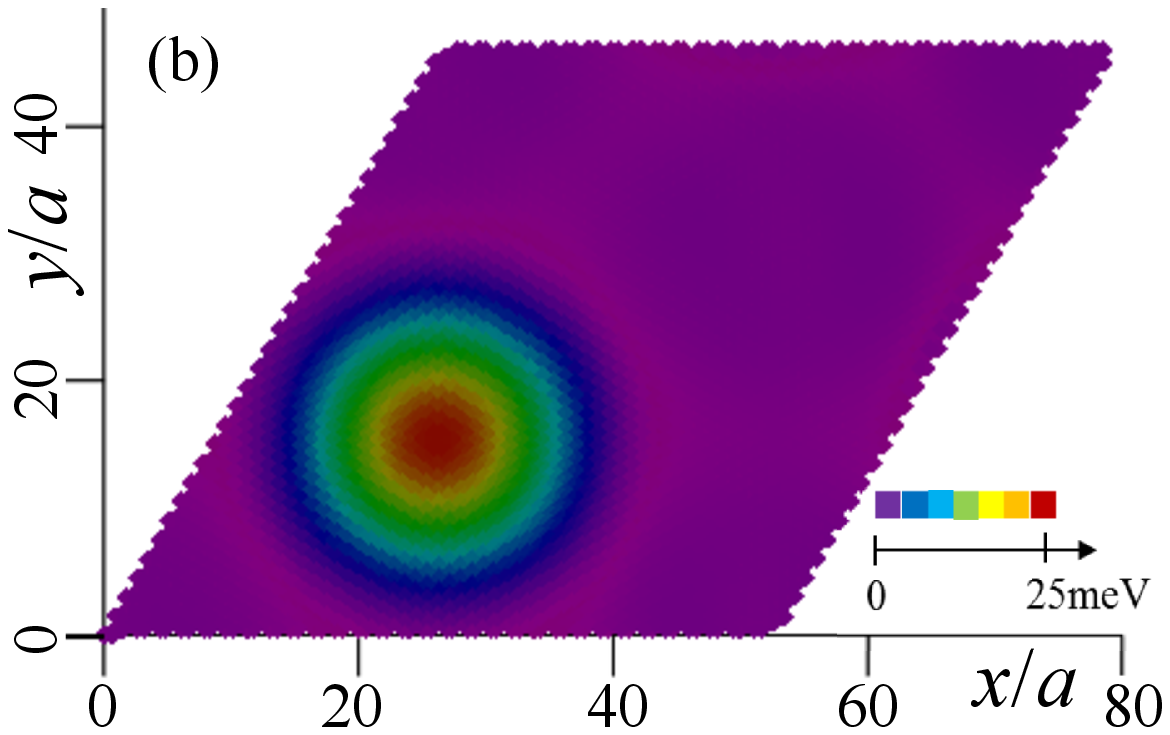}
\includegraphics[width=0.32\textwidth]{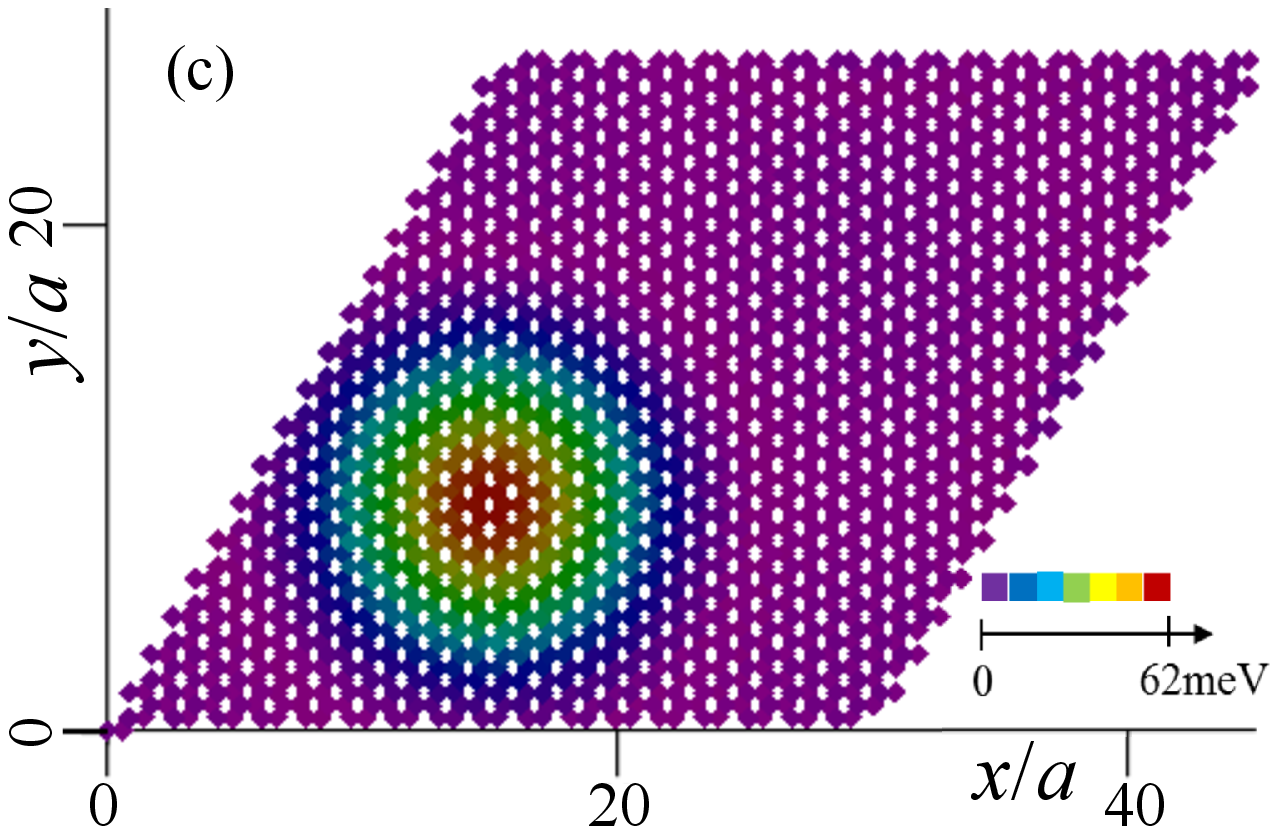}\vspace{5mm}\\
\includegraphics[width=0.32\textwidth]{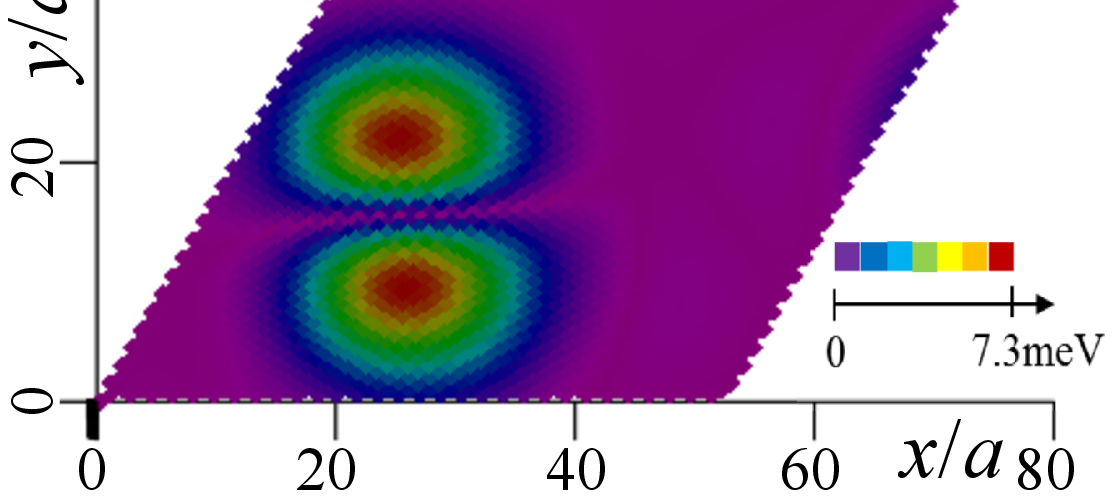}
\includegraphics[width=0.32\textwidth]{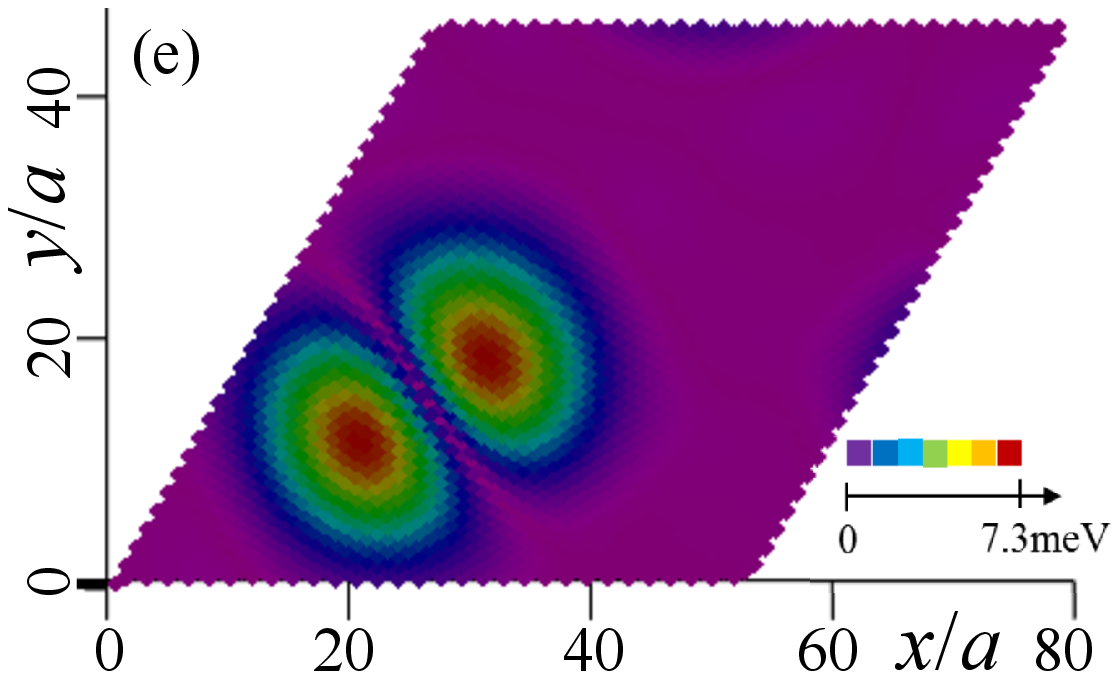}
\includegraphics[width=0.32\textwidth]{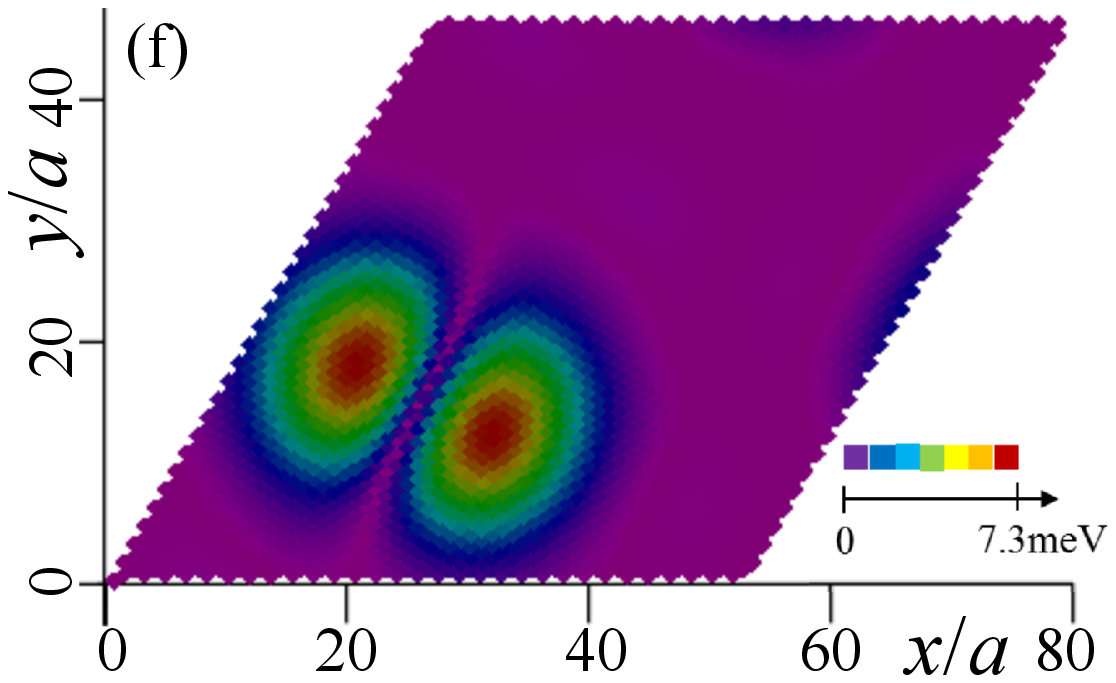}\vspace{5mm}\\
\includegraphics[width=0.28\textwidth]{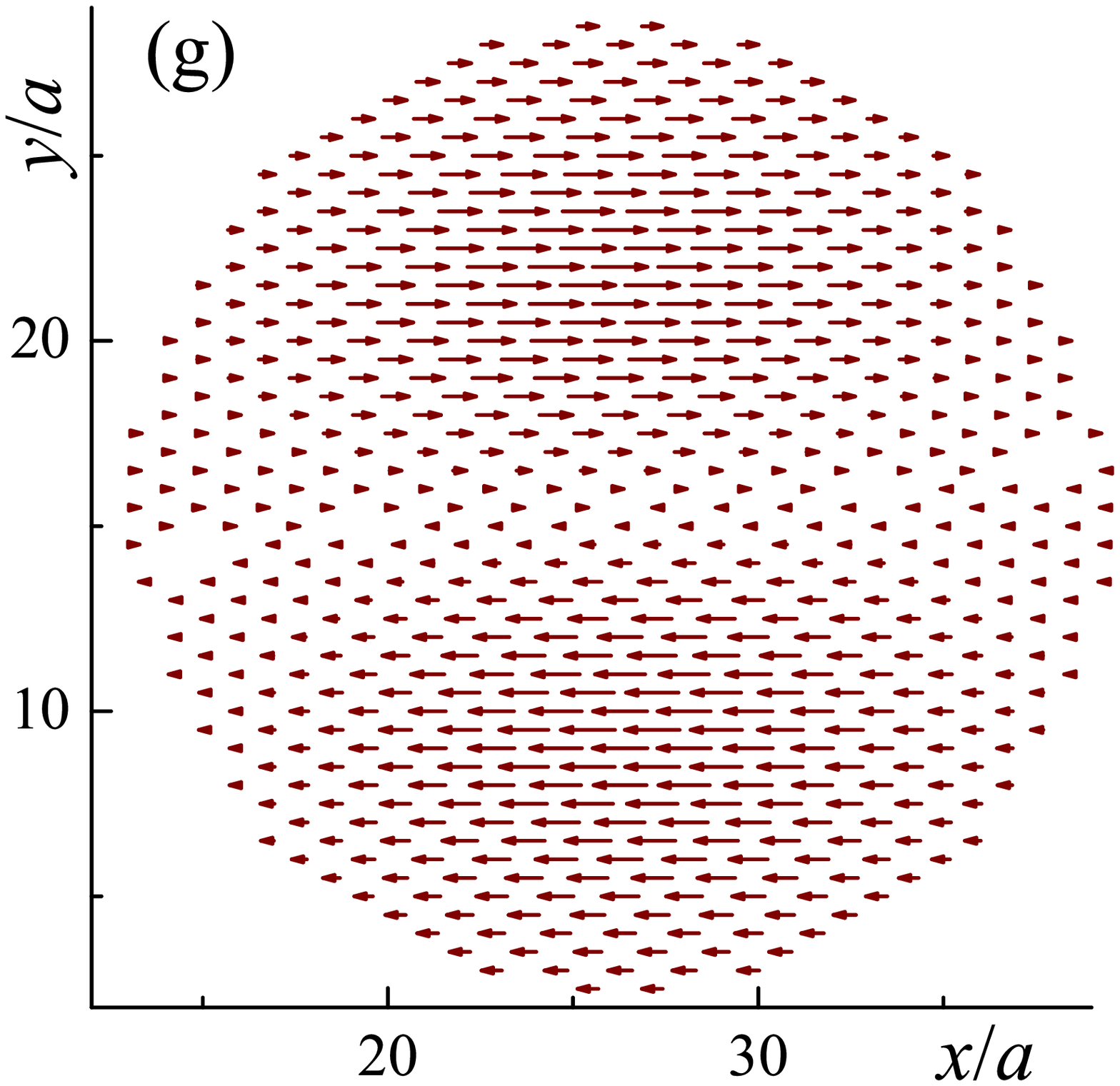}\hspace{7mm}
\includegraphics[width=0.28\textwidth]{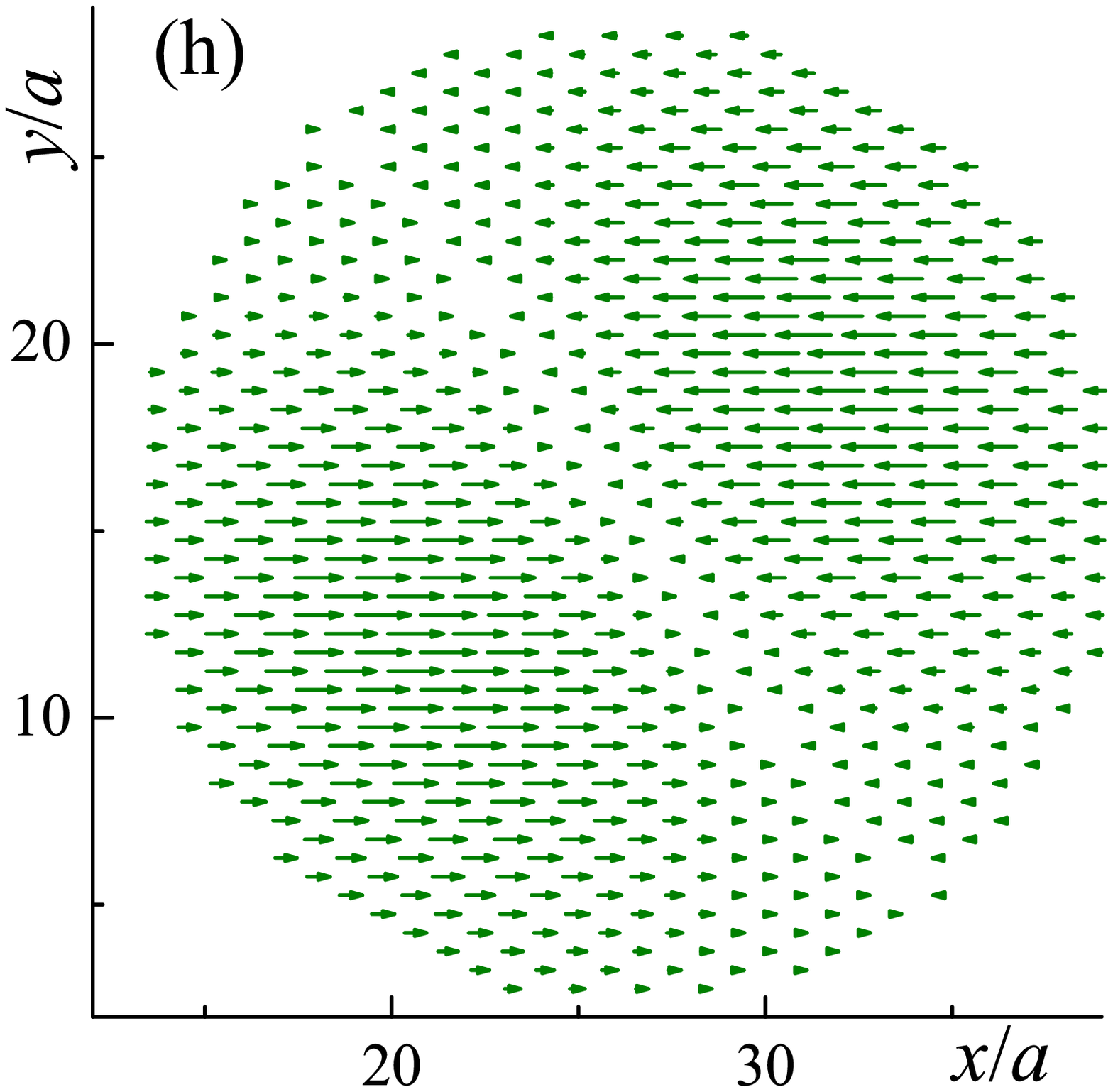}\hspace{7mm}
\includegraphics[width=0.28\textwidth]{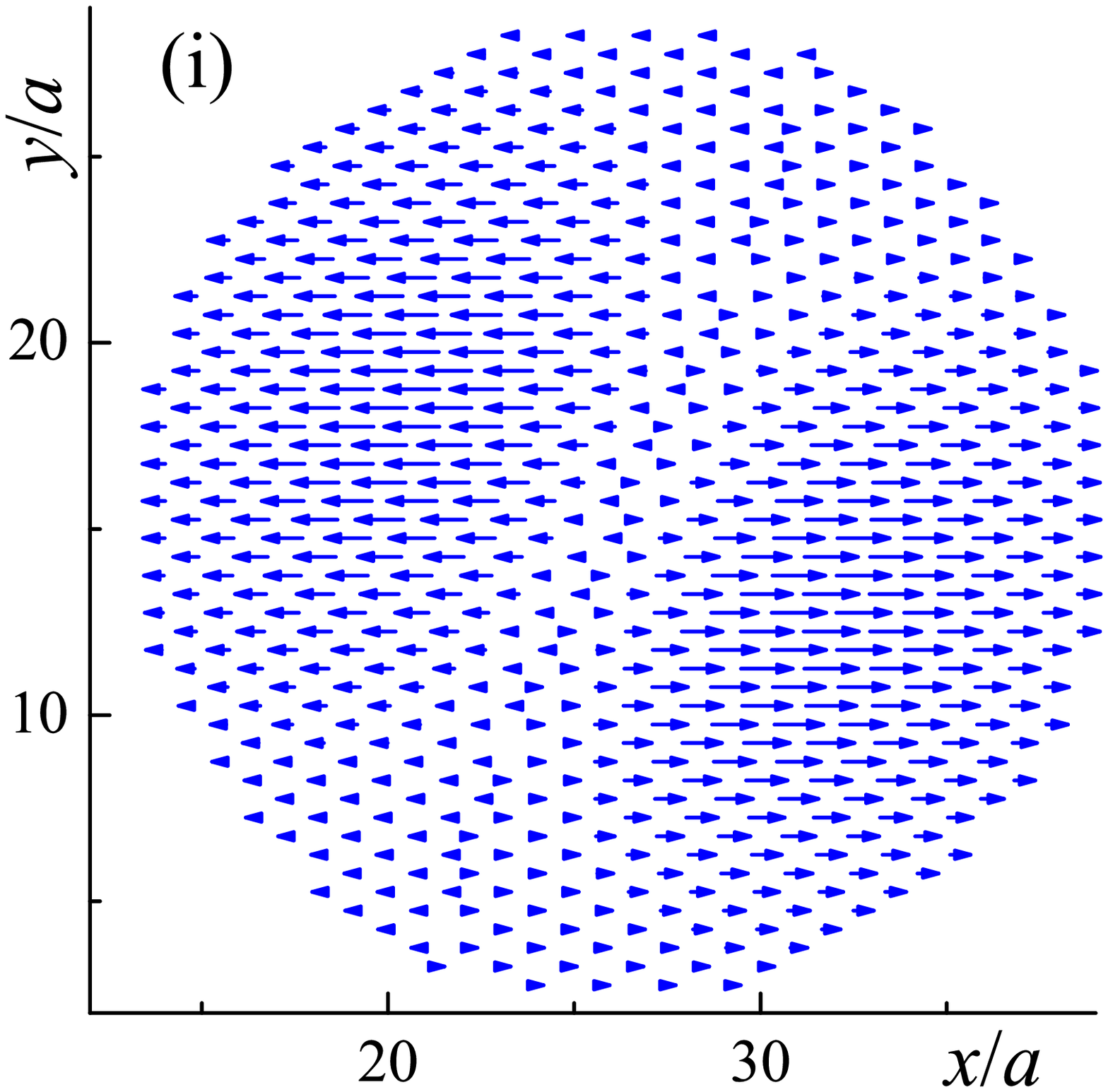}
\caption{Characteristics of the SDW order at the charge neutrality point.
(a)\,--\,(c) Spatial distributions of the absolute values of the on-site
order parameter $\Delta_{\mathbf{n}1s}$ in layer $1$,
Eq.~\eqref{Deltania},
within a single supercell [the supercell position relative to the MAtBLG
lattice is the same as in
Fig.~\ref{fig::FigTBLGSCBZ}\,(a)].
The order parameter is
calculated for parametrizations~I (a), II.A (b),
and~II.B (c). Since different parametrizations have different values of
$\theta_c$, the superlattice cell area for parametrization~II.B is about four times
smaller than that for parametrizations~I and~II.A.
(d)\,--\,(f) Spatial distributions of the in-plane (layer 1) order
parameter $\mathfrak{A}^{(\ell)}_{\mathbf{n}1}=|A^{(\ell)}_{\mathbf{n}1\uparrow}+A^{(\ell)*}_{\mathbf{n}1\downarrow}|$
for
$\ell=1$~(d),
$\ell=2$~(e),
and
$\ell=3$~(f),
calculated for parametrization~I.
(g)\,--\,(i)
Spins on the links,
Eq.~\eqref{Slink},
corresponding to order parameters
$A^{(\ell)}_{\mathbf{n}i\sigma}$
with
$\ell=1$~(g),
$\ell=2$~(h),
and
$\ell=3$~(i),
calculated for parametrization~I. Only the AA region of the superlattice cell
is shown.
\label{FigOPN0}
}
\end{figure*}

We assume superlattice periodicity for all three types of SDW order
parameters. A superlattice translation preserves the SDW texture.
With this constraint we write down the system of mean-field equations for the
functions
$\Delta_{\mathbf{n}is}$,
$A^{(\ell)}_{\mathbf{n}i\sigma}$,
and
$B^{rs}_{\mathbf{m};\mathbf{n}\sigma}$,
and solve it numerically for different doping levels, $n$, varying from $-4$
to $+4$ extra electrons per supercell. Details of the calculation procedure
are given in the Appendix.

\subsection{Approximation quality}
\label{subsec::approx_quality}

There are several circumstances which we must keep in mind assessing the
reliability of the approximations utilized in this study. Our approach is
based on the mean-field framework. It is well-known that the mean-field approximation is reliable for: (i)~a three-dimensional model; (ii)~in the limit of weak coupling; (iii)~with a single order parameter. If either of these three conditions is violated, more care is necessary interpreting the obtained results.

As our system is two-dimensional, finite-temperature long-range order in
the MAtBLG is impossible, as postulated by the Hohenberg-Mermin-Wagner
theorem. Yet, it is believed that, despite the absence of the true order,
the mean-field energy scale remains an observable quantity: it may be
experimentally measured as a low-$T$ single-particle (pseudo)gap. Consistent with
this expectation, our calculations reproduce energy scales observed in
experiment, see the Discussion for details.

Further, many real-life systems violate condition~(ii). To address this
issue for the MAtBLG, let us evaluate the effective coupling constant for
our model using the following argument. The main contribution to the
formation of SDW order comes from the flat bands. Thus, the effective
Hubbard interaction can be estimated as
$$U_{\rm eff}= U \sum_{n} |\Phi_{n}|^4,$$
where
$\Phi_{n}$
is the wave function of the flat band in real space representation, and the
summation is performed over all sites within a single supercell. Since the
electrons at the Fermi level are localized inside the AA~region of the
superlattice, occupying about 1/3 of the superlattice’s area, one can write
that inside this region
$$|\Phi_{n}| \sim \sqrt{3/N_{\rm sc}},$$
and we obtain the estimate
$$U_{\rm eff} \sim 3U/N_{\rm sc}.$$
Substituting specific numbers, we obtain
$U_{\rm eff} = 5.5$\,meV
for parametrizations~I and~II.A, and
$U_{\rm eff} = 16.8$\,meV
for parametrizations~II.B. These values must be compared against the
width of the flat bands $W$, see
Table~\ref{table1}.
Since
$U_{\rm eff}/W$
for all parametrizations is of the order of unity, the studied model is in
the limit of intermediate coupling. As the ratio
$U_{\rm eff}/W$
grows beyond unity, the mean-field approximation becomes progressively less
controlled, but we expect that our results remain qualitatively valid at
not too strong interaction. As an example of a successful application of
the mean-field calculations in the intermediate-coupling regime see Ref.~\onlinecite{zaanen1989stripes}.

Another complication would be the violation of the condition~(iii) above: for the
MAtBLG, several (metastable) order parameters compete against each other to
become the true ground state. This situation is not unique, and similar
competitions were discussed in the contexts of other
models~\cite{q1d2009,phasep_pnics2013,igoshev2010,ourPRL_half-met2017}.
Since there is no known procedure which allows one to compile an exhaustive
list of metastable phases for a given Hamiltonian, a compromise based on
general physical arguments, input from experiments, and other factors is
unavoidable. It is not surprising, therefore, that our numerical search
for the most optimal order parameter is constrained in several respects. We
already pointed out that the order parameters violating superlattice
translations are not considered as they incur unacceptable computational
costs.

Also, the numerical implementation of our mean-field procedure does not
account for non-coplanar spin textures, whose relevance for the studied
system is an open question. Non-coplanar textures naturally
appear~\cite{lu2020chiral} in recently introduced effective
models~\cite{lu2020chiral,ChiralSDW_SC2018,nesting_multiple_OP2018PRX,
you_twisted_supercond2019npj}, where they stabilize due to Fermi surface nesting and a significantly enhanced symmetry group. To which extent the weak-coupling nesting-based
argument of Refs.~\onlinecite{lu2020chiral,ChiralSDW_SC2018,nesting_multiple_OP2018PRX}
is applicable to the MAtBLG (a system in intermediate-coupling
regime~\cite{you_twisted_supercond2019npj}, with very complex Fermi
surface~\cite{ourTBLG}) remains an interesting issue for future studies.

Finally, our procedure, as it is described above, does not attempt to
obtain self-consistency for the charge density distribution within a
supercell. Indeed, one must remember that, since the atoms locations within
a supercell are not equivalent to each other, the average charge on a given
atom depends on its position (the same is true for the local density of
states). In principle, the interaction attempts to suppress spatial
variations of charge through ``Hartree" terms; however, we neglect them in
our numerical code.

\section{Results: Spatial distributions of SDW order parameters.}
\label{ResultsOP}

\begin{figure*}[t]
\centering
\includegraphics[width=0.32\textwidth]{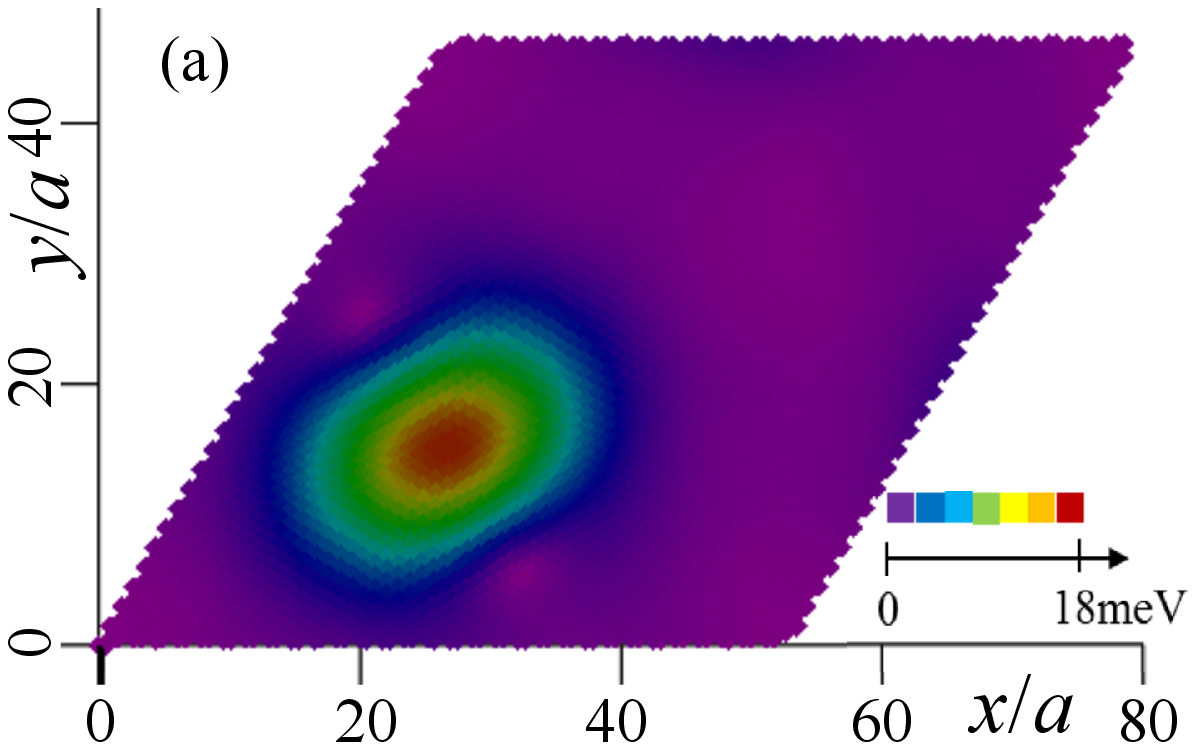}\hspace{10mm}
\includegraphics[width=0.22\textwidth]{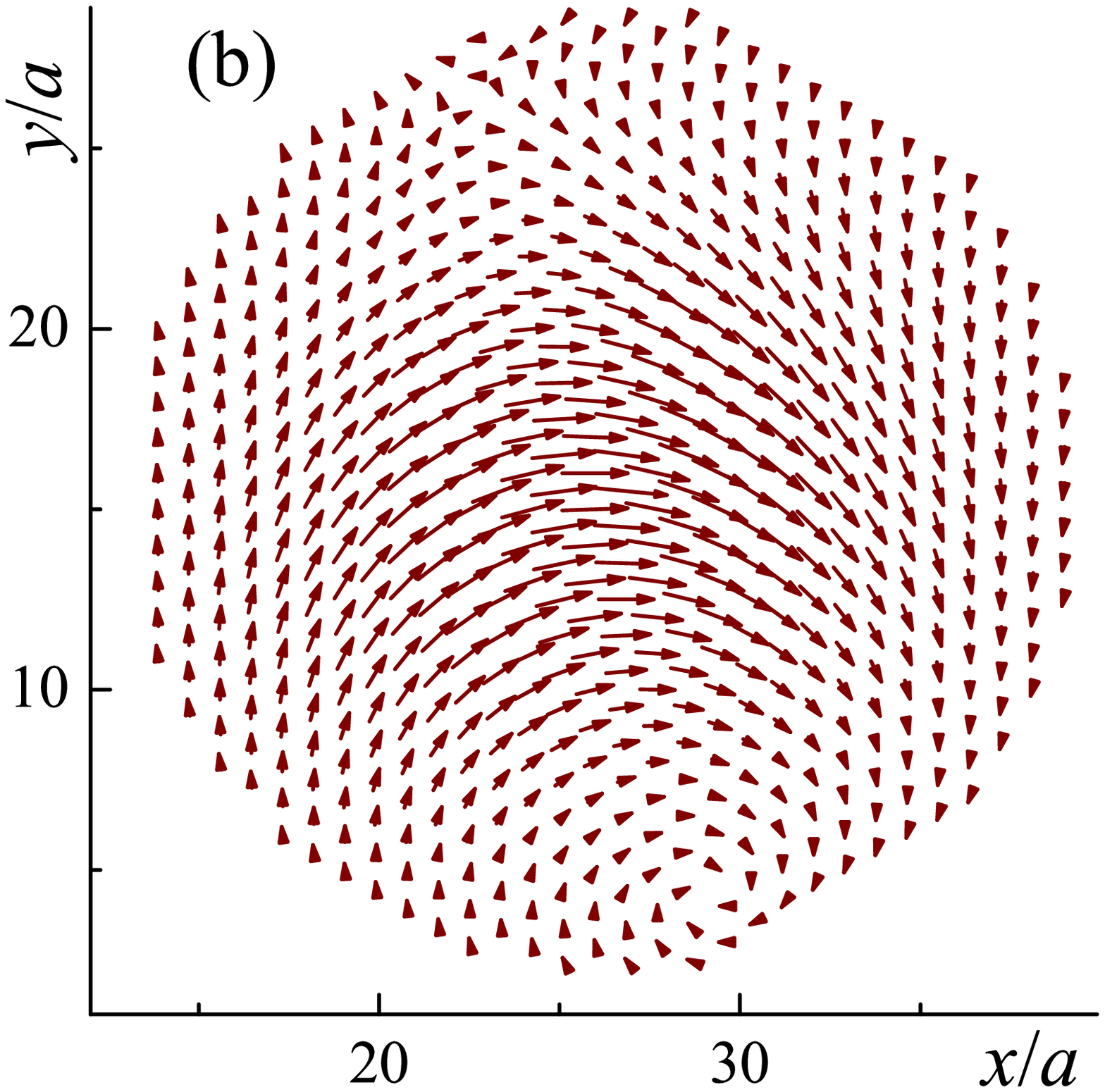}\hspace{18mm}
\includegraphics[width=0.22\textwidth]{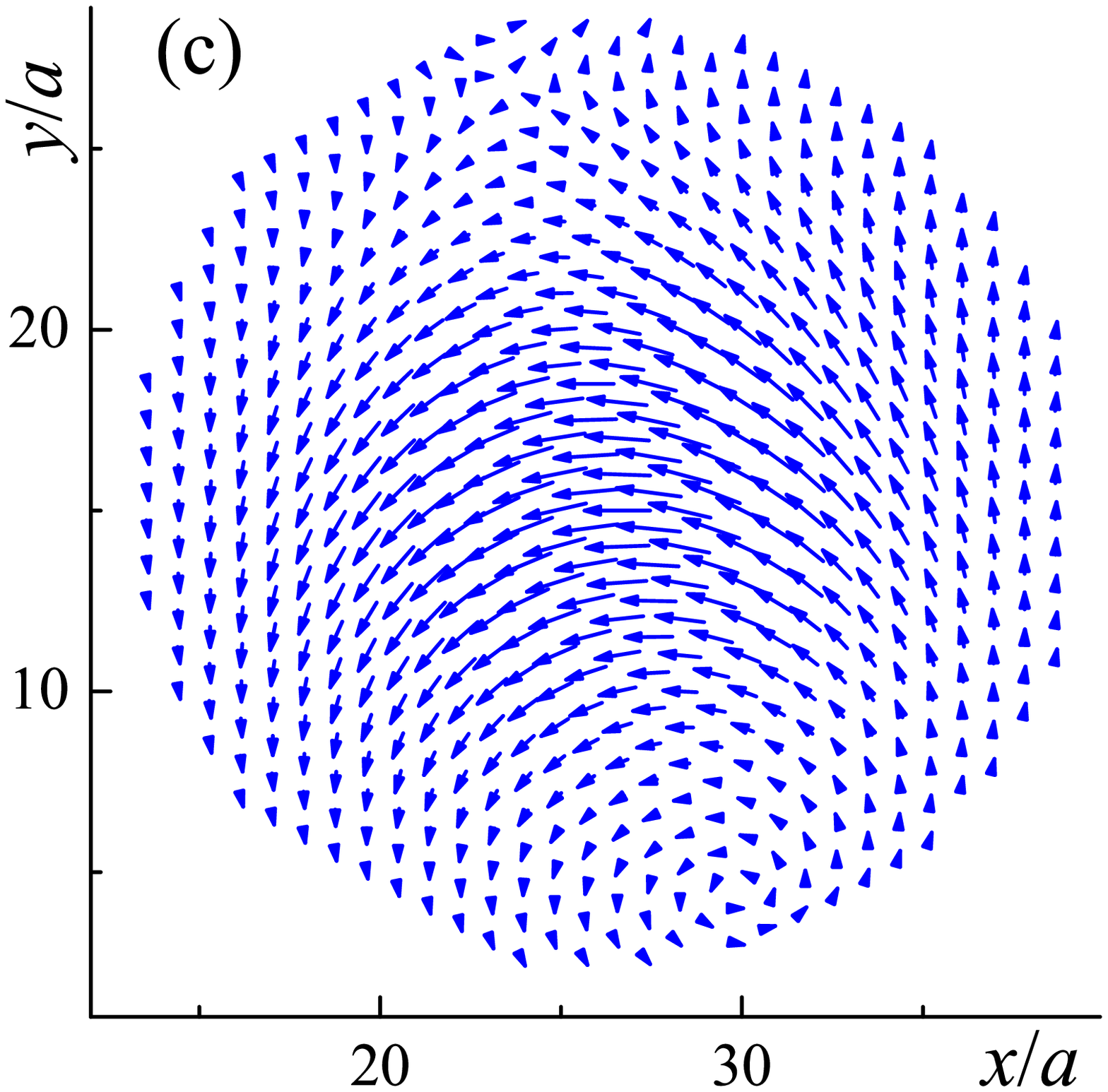}\hspace{7mm}\vspace{5mm}\\
\includegraphics[width=0.32\textwidth]{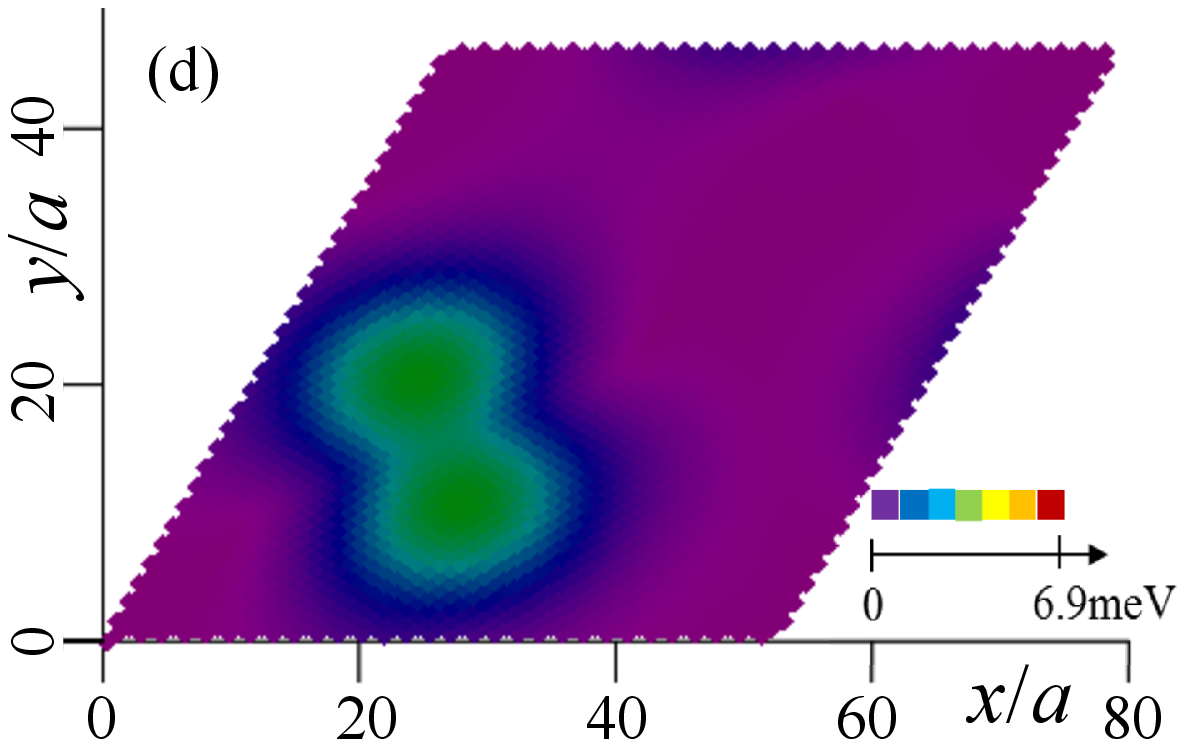}
\includegraphics[width=0.32\textwidth]{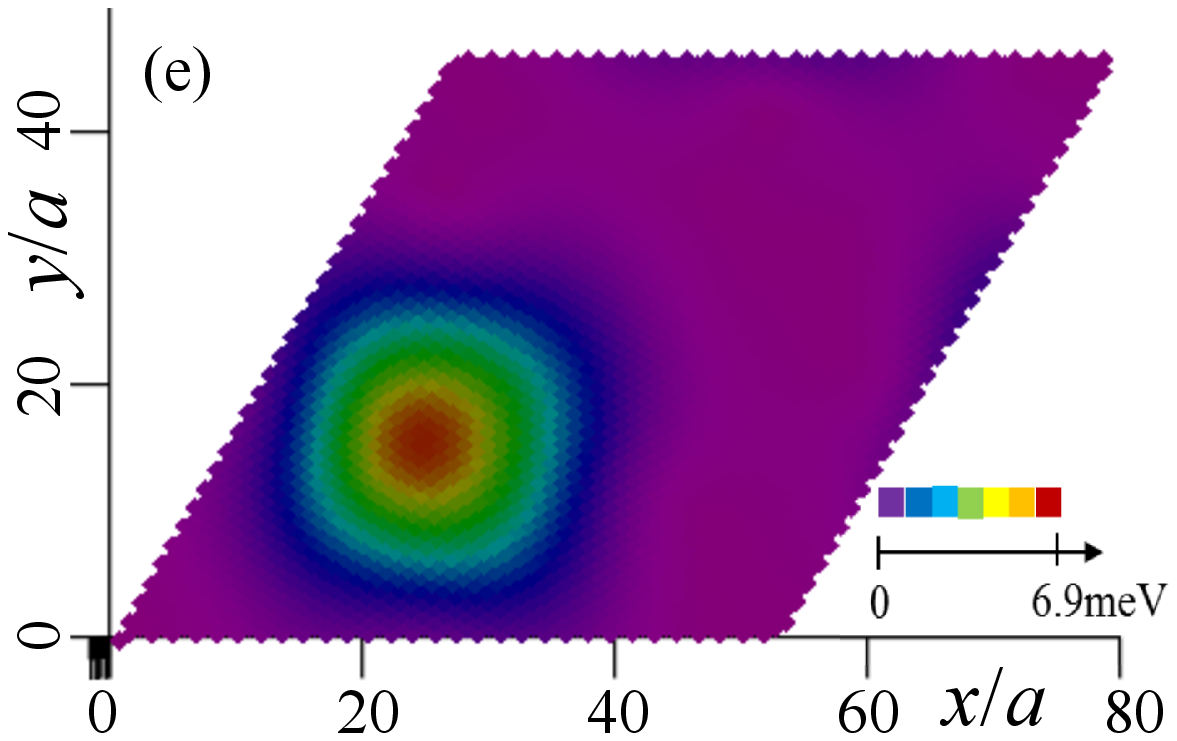}
\includegraphics[width=0.32\textwidth]{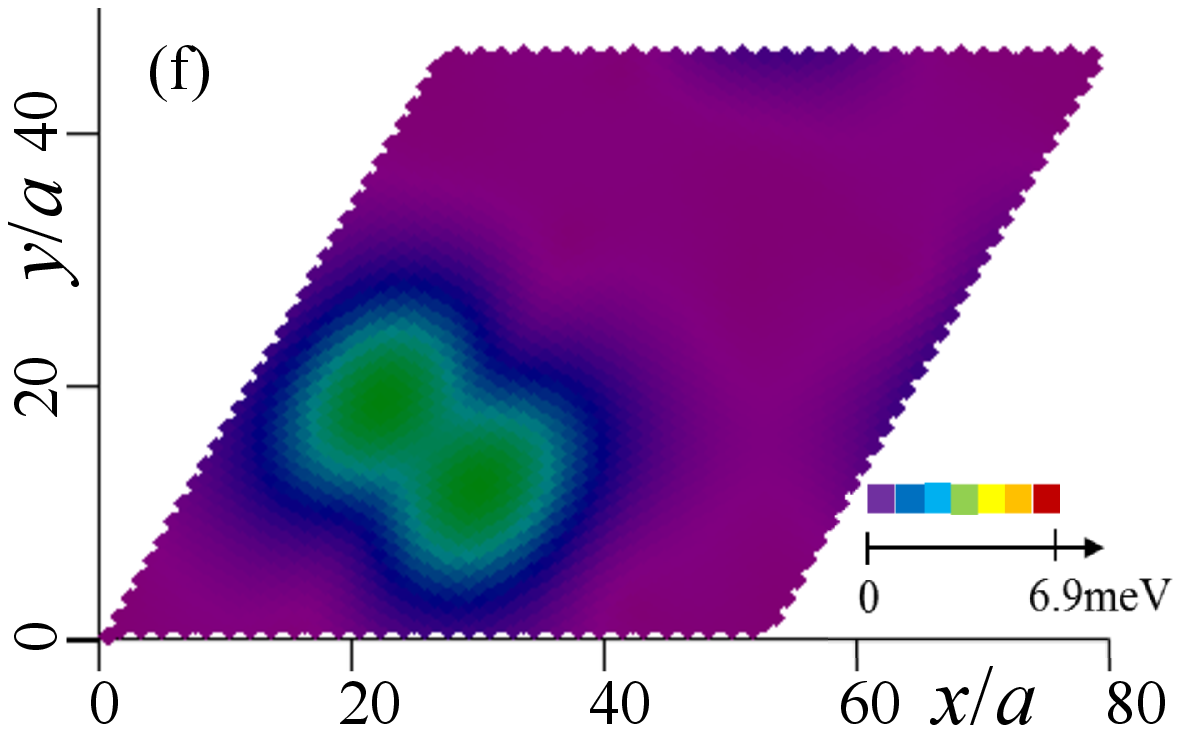}\vspace{5mm}\\
\includegraphics[width=0.28\textwidth]{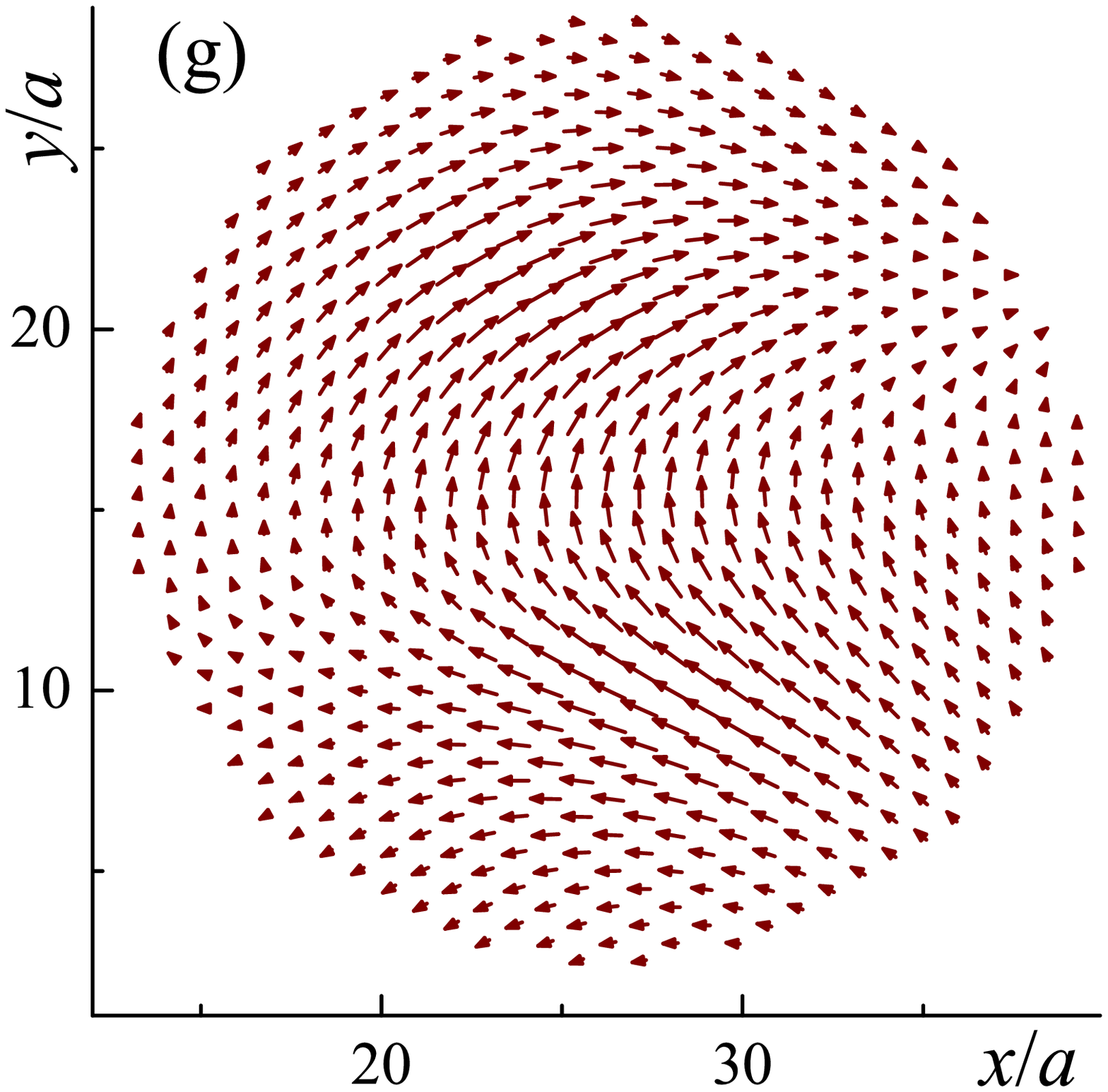}\hspace{7mm}
\includegraphics[width=0.28\textwidth]{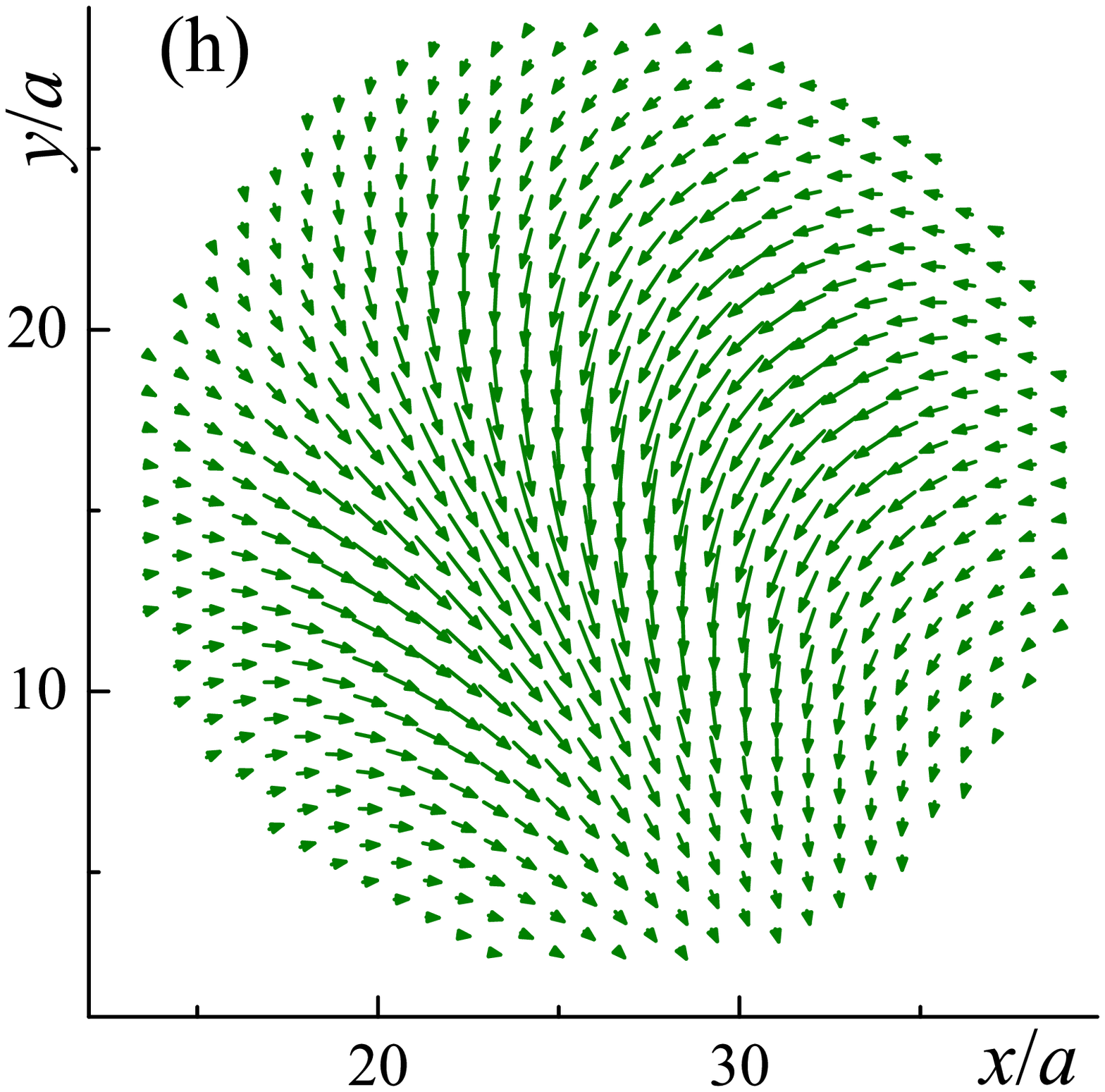}\hspace{7mm}
\includegraphics[width=0.28\textwidth]{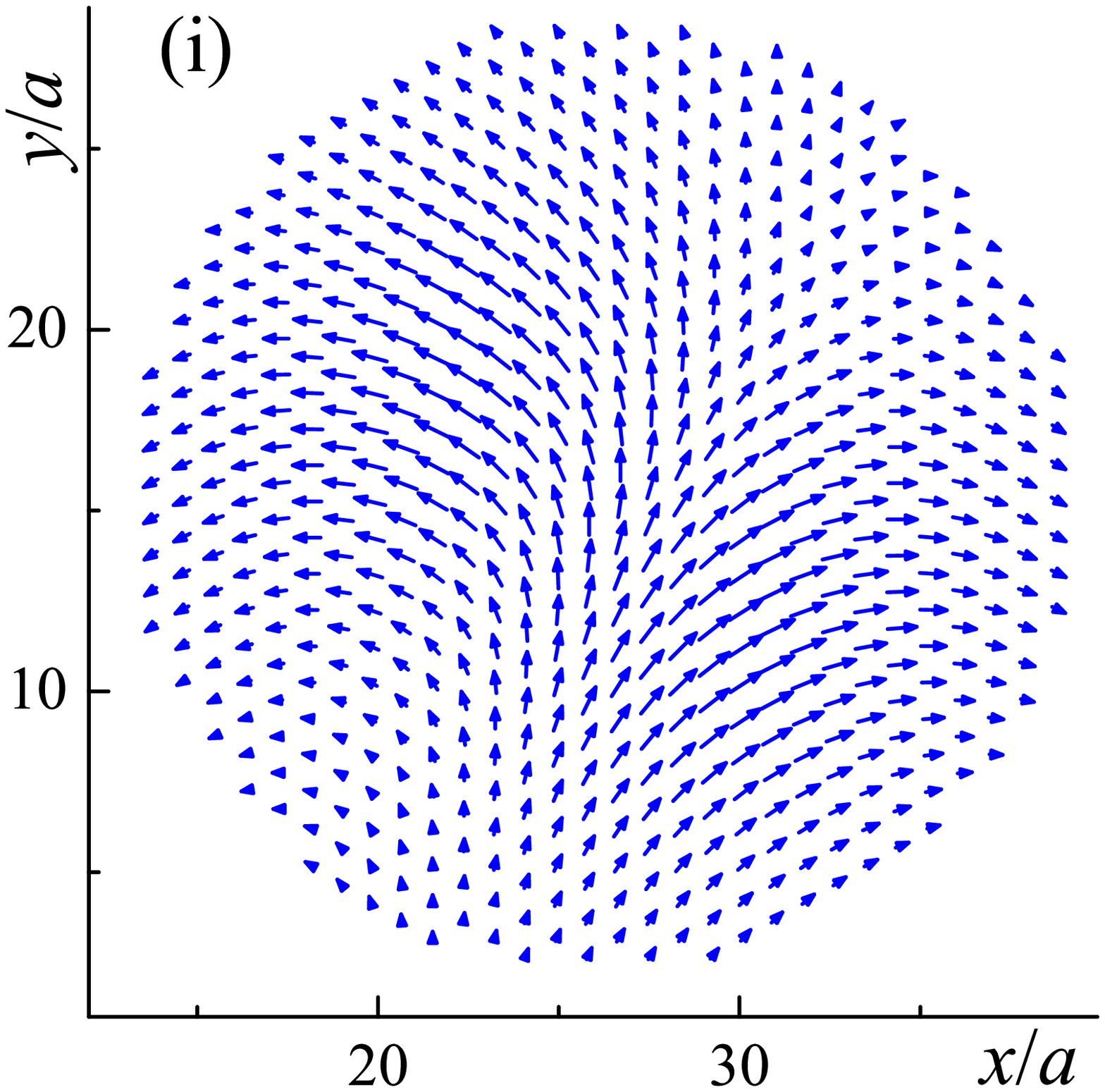}
\caption{Characteristics of the SDW order at the half-filled state
($\nu=2$).
(a) Spatial distribution of
$|\Delta_{\mathbf{n}1s}|$
within a single supercell [the supercell position relative to the MAtBLG
lattice is the same as in
Fig.~\ref{fig::FigTBLGSCBZ}\,(a)].
(b)\,--\,(c) Orientations of the on-site
spins,
Eq.~\eqref{Snia},
in the layer 1, sublattice
${\cal A}$~(b)
and
${\cal B}$~(c)
shown inside the AA region of the supercell.
(d)\,--\,(f) Spatial distributions of
$\mathfrak{A}^{(\ell)}_{\mathbf{n}1}
=
|A^{(\ell)}_{\mathbf{n}1\uparrow}+A^{(\ell)*}_{\mathbf{n}1\downarrow}|$
within a single supercell,
for
$\ell=1$~(d),
$\ell=2$~(e),
and
$\ell=3$~(f).
(g)\,--\,(i) Spins on the links corresponding to the order parameters
$A^{(\ell)}_{\mathbf{n}i\sigma}$,
with
$\ell=1$~(g),
$\ell=2$~(h),
and
$\ell=3$~(i),
shown inside the AA region of the superlattice cell.
The order parameters shown are
calculated for parametrization~I. Similar pictures are obtained for
parametrizations~II.A and~II.B.
\label{FigOPN2}
}
\end{figure*}

In this section we present the results of our calculation of the SDW order
parameters and analyze their symmetry properties. The spatial
distribution of the order parameters inside the superlattice cell is different
for different doping levels. However, it turns out that for a given doping
the properties of the order parameters are very similar for the three parametrizations used.

\subsection{Charge neutrality point}

We start from the charge neutrality point.
Figures~\ref{FigOPN0}(a)\,--\,(c)
show the color plots of the spatial distribution of the absolute values of
the on-site order parameter in layer 1,
$|\Delta_{\mathbf{n}1s}|$,
calculated for three parametrizations. Similar structures are observed for
layer 2. We see that the maximum values of the
$\Delta_{\mathbf{n}1s}$
are different for the three parametrizations, but the plots themselves look very
similar. The maxima of
$\Delta_{\mathbf{n}1s}$
are located in the center of the AA region of the superlattice cell [c.f.
with
Fig.~\ref{fig::FigTBLGSCBZ}(a)].
The order parameter
$\Delta_{\mathbf{n}is}$
defines the spin on a site in position $\mathbf{n}$, layer $i$, and
sublattice $s$, as follows:
\begin{equation}
\label{Snia}
\mathbf{S}_{\mathbf{n}is}
=
\frac{1}{U}\left(
	\Real \Delta_{\mathbf{n}is},\,\Imag \Delta_{\mathbf{n}is},\, 0
\right).
\end{equation}
A vanishing $z$~component of $\mathbf{S}_{\mathbf{n}is}$
in our definition~(\ref{Snia})
implies that only planar spin textures are allowed (this is a limitation of
our numerical code, as explained in
Sec.~\ref{subsec::approx_quality}).
However, at the charge neutrality point this constraint turns out to be
unimportant, since all spins are collinear.
If in layer~1 and sublattice
${\cal A}$
(and in layer~2 and sublattice
${\cal B}$)
they point in one direction (along the $x$ axis), then the layer 1 and sublattice
${\cal B}$
(and in layer 2 and sublattice
${\cal A}$)
they point in the opposite direction. Thus, we have antiferromagnetic ordering of spins.

Let us now visualize the intralayer nearest-neighbor order parameters
$A^{(\ell)}_{\mathbf{n}1\sigma}$.
Using these quantities one can define the spins on the link connecting
nearest-neighbor sites in each layer as
follows:
\begin{eqnarray}
\label{Slink}
\mathbf{S}^{(\ell)}_{\mathbf{n}i}
&\!\equiv\!&
\frac12\sum_{\sigma\sigma'}\!
	\bm{\sigma}_{\sigma\sigma'}
	\langle
		d^{\dag}_{\mathbf{n}+\mathbf{n}_{\ell}i{\cal A}\sigma}
		d^{\phantom{\dag}}_{\mathbf{n}i{\cal B}\sigma'}
	\rangle
	+
	{\rm c.c.}
\\
\nonumber
&\!=\!&
\frac{1}{V_{\text{nn}}}\!
\left(
	\Real[A^{(\ell)}_{\mathbf{n}i\uparrow}+
		A^{(\ell)}_{\mathbf{n}i\downarrow}],\,
	\Imag[A^{(\ell)}_{\mathbf{n}i\uparrow}-
		A^{(\ell)}_{\mathbf{n}i\downarrow}],\,
	0
\right),
\end{eqnarray}
where
$\bm{\sigma}$
is a three-component vector composed of the Pauli matrices. The quantity
$$\mathfrak{A}^{(\ell)}_{\mathbf{n}1}
=
|A^{(\ell)}_{\mathbf{n}1\uparrow}+A^{(\ell)*}_{\mathbf{n}1\downarrow}|$$
is proportional to the absolute values of
$\mathbf{S}^{(\ell)}_{\mathbf{n}i}$ calculated in layer 1 for parametrization~I. The spatial distributions
of
$\mathfrak{A}^{(\ell)}_{\mathbf{n}1}$
are shown in
Figs.~\ref{FigOPN0}(d)\,--\,(f)
for all three possible values of $\ell$. The distributions are shaped like
dumbbells localized in the AA region of the superlattice. The orientations
of these dumbbells are different for different orientations of the
carbon-carbon links. Similar figures are obtained for other two
parametrizations. The directions of the vectors
$\mathbf{S}^{(\ell)}_{\mathbf{n}i}$
inside the AA region are shown in
Figs.~\ref{FigOPN0}(g)\,--\,(i).
We see that all spins
$\mathbf{S}^{(\ell)}_{\mathbf{n}i}$
are collinear; but if in one part of a dumbbell they point in one
direction, then in another part of the dumbbell they are oriented in the
opposite direction.

Absolute values of the order parameters
$A^{(\ell)}_{\mathbf{n}i\sigma}$
are several times smaller than that for
$\Delta_{\mathbf{n}is}$.
Our calculations show that the interlayer order parameters
$B^{rs}_{\mathbf{m};\mathbf{n}\sigma}$
are one order of magnitude smaller than
$A^{(\ell)}_{\mathbf{n}i\sigma}$;
thus, we do not discuss them here in details.

We now demonstrate that the calculated SDW magnetization texture has the
same geometrical symmetries as the tBLG superstructure. We start with the
following observation about the tBLG lattice symmetry. The center of the AA
region of the superlattice cell is located at
$\mathbf{R}_{\text{AA}}=(\mathbf{R}_1+\mathbf{R}_2)/3$.
For the
$r=1$
superstructures considered here, one can prove using
Eqs.~\eqref{eq::a12},~\eqref{comtheta},
and~\eqref{R12}
that
\begin{equation}
\mathbf{R}_{\text{AA}}
=
m_0\mathbf{a}_2-\mathbf{a}_1+2\bm{\delta}
=
m_0\mathbf{a}'_2+\bm{\delta}'\,.
\end{equation}
It is easily seen from this equation that the point
$\mathbf{R}_{\text{AA}}$
is located at the center of the hexagons of both layers. This means that the tBLG
lattice is invariant under a rotation by
$60^{\circ}$
around the axis perpendicular to the layers and passing through the point
$\mathbf{R}_{\text{AA}}$.
\begin{figure}[t]
\includegraphics[width=0.9\columnwidth]{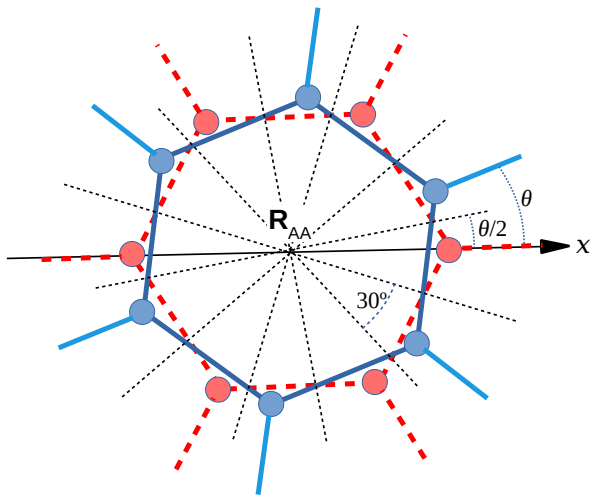}
\caption{Mirror symmetries of the tBLG. Point
$\mathbf{R}_{\text{AA}}=(\mathbf{R}_1+\mathbf{R}_2)/3$
is the center of the AA region of the superlattice cell. The solid (blue)
lines represent carbon-carbon bonds in the (rotated) top layer, dashed
(red) lines correspond to the bonds in the (immobile) bottom layer. The
circles represent carbon atoms. Thin dashed (black) lines are the
reflection axes. It is easy to see from this figure that a reflection
relative to any of these axes, accompanied by exchange of the layers,
leaves the tBLG lattice unchanged.
\label{fig::symm}
}
\end{figure}

Further, the tBLG lattice also has mirror-like symmetries. Indeed, the
lattice remains invariant if one exchanges layers and then performs a
reflection with respect to a certain axis in the $xy$-plane passing through
point
$\mathbf{R}_{\text{AA}}$,
see
Fig.~\ref{fig::symm}.
There are six such axes. They cross the $x$-axis at
$30^{\circ}p +\theta/2$,
where
$p = 0, 1, \ldots, 5$.
Since for the superstructures considered the twist angle is small, any mirror
symmetry axis is either approximately parallel to
$\mathbf{R}_1$,
$\mathbf{R}_2$,
or
$(\mathbf{R}_1-\mathbf{R}_2)$,
or approximately perpendicular to one of these vectors.

One can easily see from
Figs.~\ref{FigOPN0}(a)\,--\,(f)
that the SDW order parameters are invariant with respect to all geometrical
symmetries of the tBLG lattice. Indeed, the on-site order parameter does not
change under the action of the rotations and reflections mentioned
above, while the intersite order parameters
$\mathfrak{A}^{(\ell)}_{\mathbf{n}i}$
either remain invariant or convert into
$\mathfrak{A}^{(\ell')}_{\mathbf{n}i}$,
with
$\ell'\ne\ell$.

\subsection{Half-filled state}

Our simulations show that doping of the system away from the charge-neutrality point spontaneously reduces the symmetry of the SDW order parameters. For illustration of this fact we consider only the case of half-filling, which corresponds to $2$ extra electrons or extra holes per
supercell. Electron and hole dopings are equivalent at the qualitative
level; thus, for definiteness, we consider electron doping.
Figure~\ref{FigOPN2}(a) shows the spatial distribution of
$|\Delta_{\mathbf{n}1s}|$, calculated for parametrization~I. A similar pattern is observed in layer $2$. We see that in contrast to Fig.~\ref{FigOPN0}(a)
the spatial profile becomes uniaxially stretched. The stretching axis is
(approximately) parallel to the vector $(\mathbf{R}_1+\mathbf{R}_2)$. As a result of this distortion, the $60^{\circ}$ rotation is no longer a symmetry of the system. However, the spatial profile is still symmetric under $180^{\circ}$ rotation around
${\bf R}_{\rm AA}$. Regarding the mirror symmetry, only axes parallel and perpendicular to
$(\mathbf{R}_1-\mathbf{R}_2)$ remain mirror symmetry axes. Another difference, in comparison to the charge-neutrality point, is that the on-site spins are no longer collinear,
even though an antiferromagnetic type configuration is preserved, see
Figs.~\ref{FigOPN2}(b)\,--\,(c).

The change in the inter-site order parameters under doping is even more
dramatic: the spatial profile for
$\mathfrak{A}^{(\ell)}_{\mathbf{n}1}$
with
$\ell=2$
does not have the form of a dumbbell,
Fig.~\ref{FigOPN2}(e), and it is different from that for
$\ell=1$
and
$\ell=3$,
Figs.~\ref{FigOPN2}(d),\,(f).
However, the
$180^{\circ}$
rotation symmetry endures for all three types of intersite order
parameters. Moreover, the symmetry under the reflection with respect to the
axis parallel to
$(\mathbf{R}_1-\mathbf{R}_2)$
also remains unbroken. This reflection transforms the profiles shown in
Figs.~\ref{FigOPN2}(e),\,(f)
into each other, while the profile of
Fig.~\ref{FigOPN2}(d)
is unchanged. Likewise, one can argue that the line perpendicular to
$(\mathbf{R}_1-\mathbf{R}_2)$
is also a valid symmetry axis for the inter-site texture. Finally, our
calculations demonstrate that vectors
$\mathbf{S}^{(\ell)}_{\mathbf{n}1}$
are no longer collinear,
and their textures have a complicated structure.

Thus, at half-filling the symmetry of the order parameters is partially
reduced, indicating spontaneous formation
of the electron nematic state. The nematicity of the ordered state in doped
MAtBLG is a robust property which does not require fine-tuning. Indeed,
the rotation symmetry of the order parameter is lowered for all three
parametrizations studied in this paper.

\begin{figure}[t]
\includegraphics[width=0.9\columnwidth]{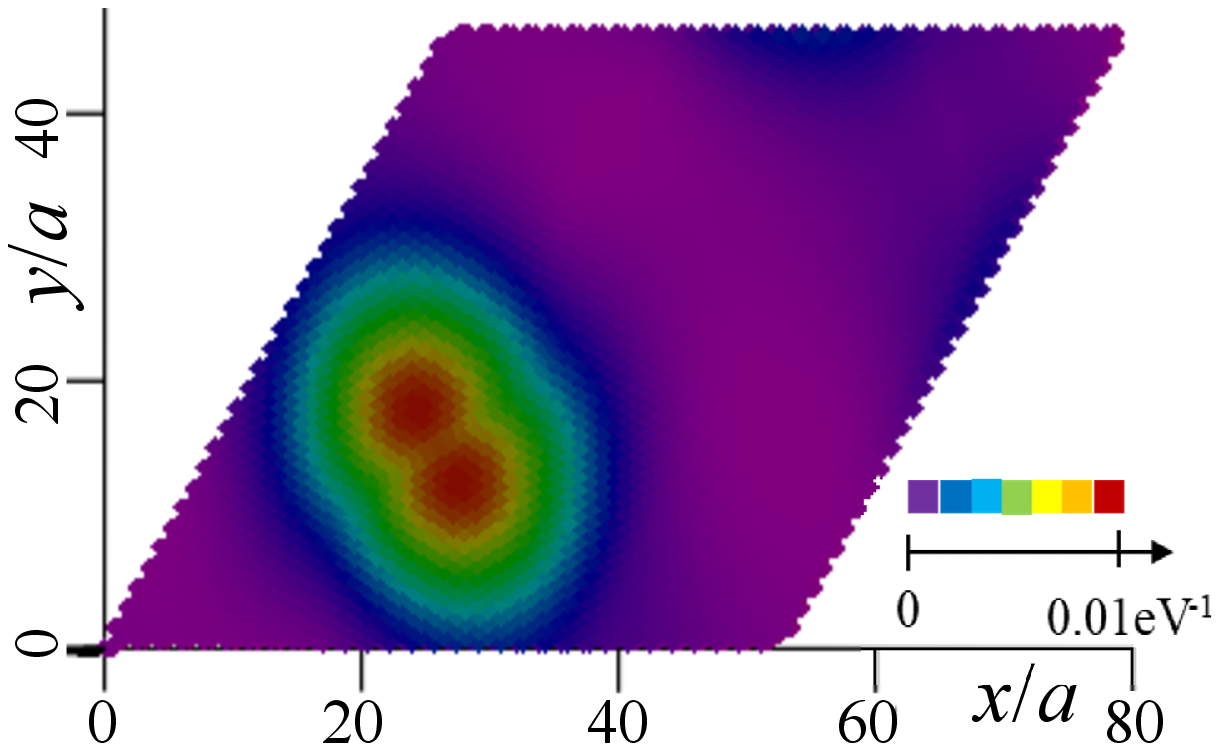}
\includegraphics[width=0.9\columnwidth]{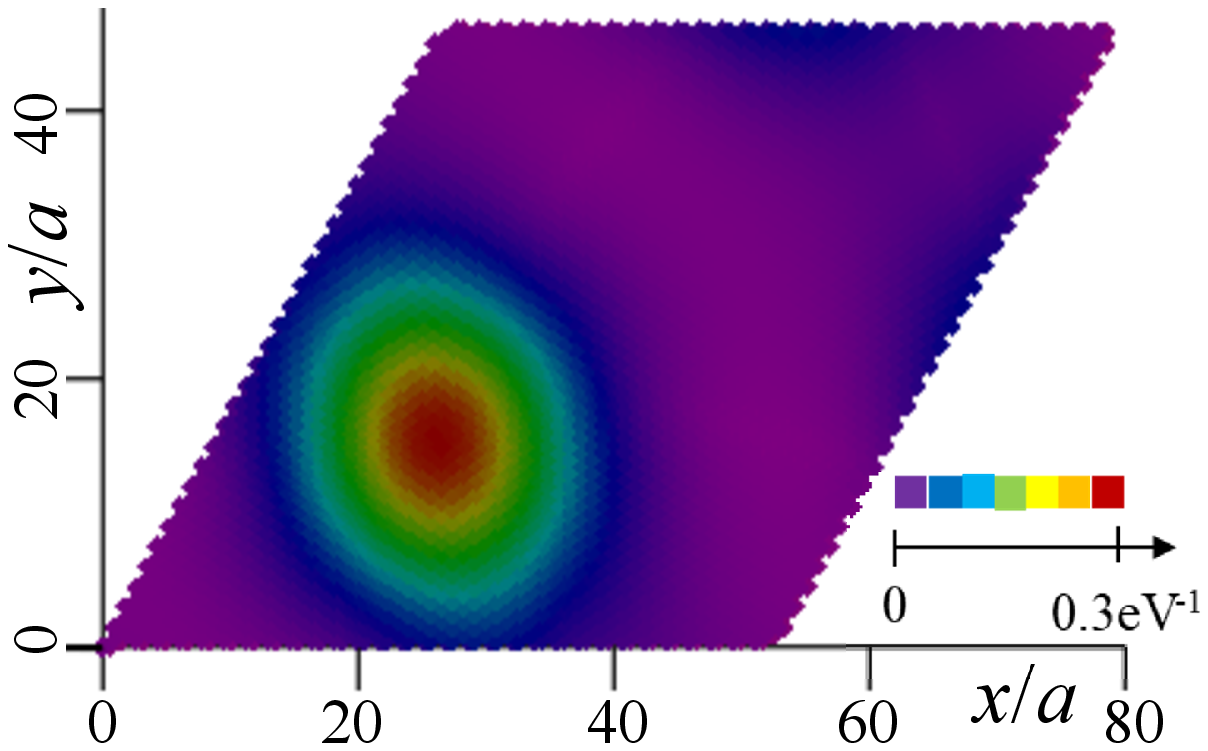}
\caption{Local density of states at the Fermi energy for the
many-body states with nematicity. Both panels present a single supercell,
whose position relative to the MAtBLG lattice is the same as in
Fig.~\ref{fig::FigTBLGSCBZ}\,(a).
The numerical data in the top panel is for
$\nu = 2$
(half filling, two extra electrons per supercell), in the bottom panel
is for
$\nu = 2.25$.
For both filling fractions, the local densities of states demonstrate a $C_2$
point symmetry group, instead of the larger
$C_6$
group of the moir\'{e} superlattice. This manifestation of nematicity may
be detected in STM experiments.
\label{fig::nematic_DOS}
}
\end{figure}

The nematicity also manifests itself in the local DOS, see
Fig.~\ref{fig::nematic_DOS}.
The two panels of this figure show the local DOS at the Fermi energy for
the half-filled
($\nu=2$)
and
$\nu=2.25$
states. In both cases, the local density of states demonstrates invariance
under the $C_2$ point symmetry group, which is smaller than
$C_6$
of the MAtBLG moir\'{e} superlattice. In this context, the symmetry
reduction can be interpreted as a type of commensurate charge-density wave,
with the modulation wave vector
${\bf Q}_{\rm cdw}$
being equal to zero in the RBZ. In real space the charge modulation has the
same period as the superlattice. To discriminate between the superlattice
and the charge-density wave in experiments, one has to rely on differences
in the point symmetry groups of the two.

\section{Results: The low-energy band structure}
\label{ResultsSpec}
\begin{figure}[t]
\centering
\includegraphics[width=0.99\columnwidth]{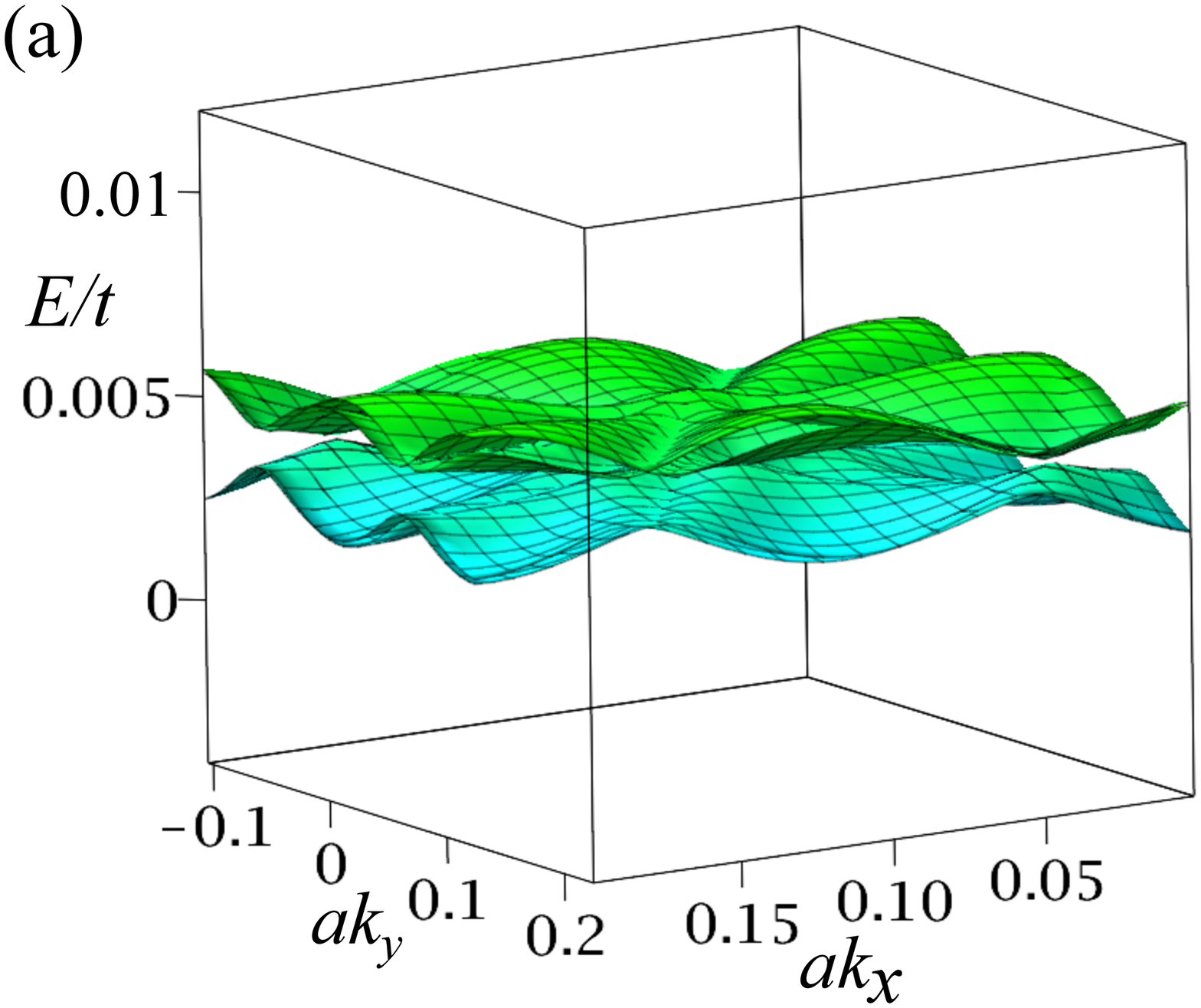}\\
\includegraphics[width=0.99\columnwidth]{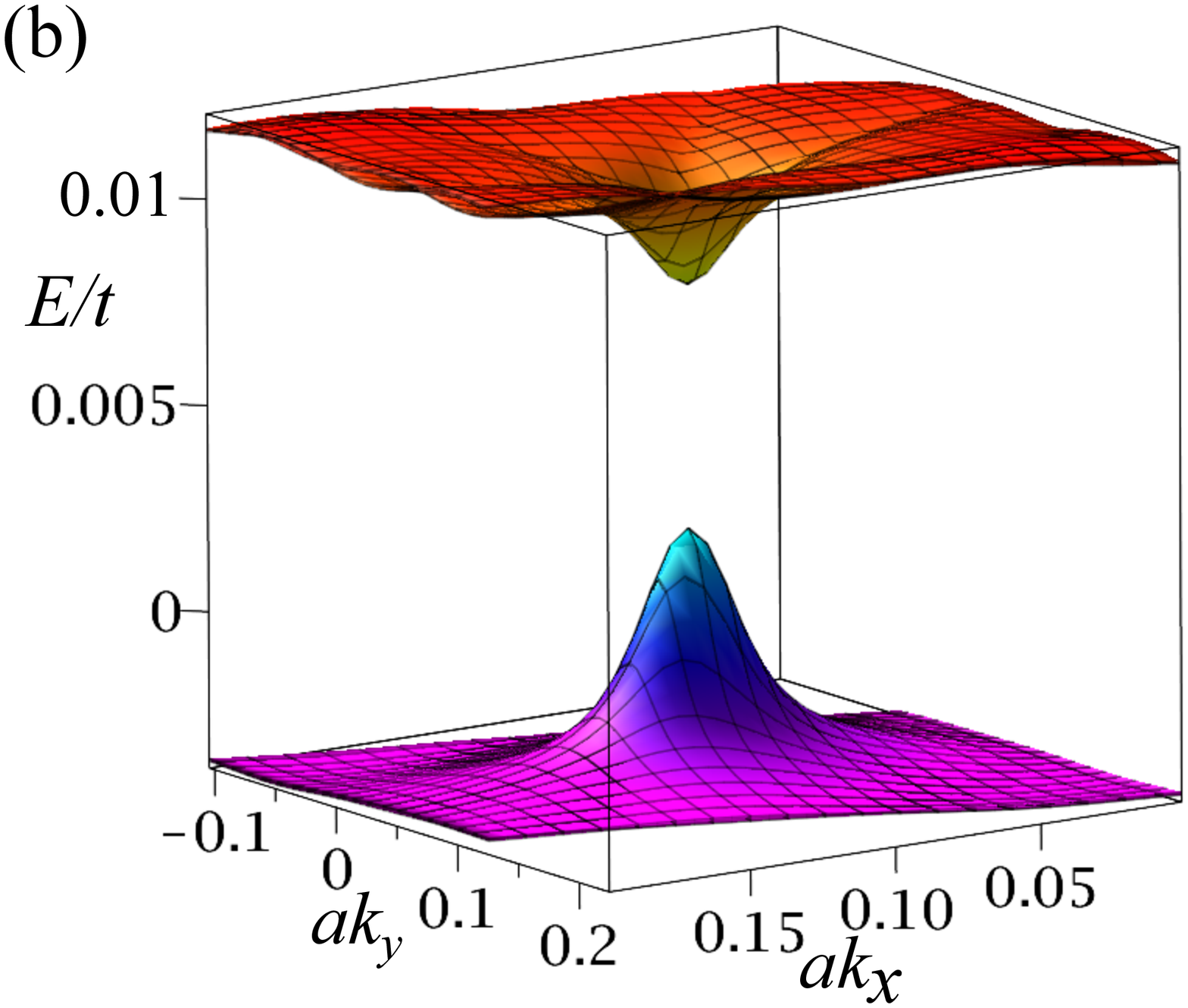}
\caption{The spectrum of the non-interacting model (a) compared with the
spectrum calculated within the mean-field approximation (b).
The spectrum in (a) consists of four warped bands, each band is spin-degenerate. In (b), two quartets of single-electron bands are visible, while the individual bands are indiscernible on this scale.
We see that the characteristic
gap-inducing splitting between the quartets exceeds the warping of the
non-interacting single-electron bands. The calculations were performed at
the charge neutrality point
$n=0$,
for parametrization II.B.
\label{fig::compare_spectra}
}
\end{figure}
\begin{figure*}[t]
\centering
\includegraphics[width=0.24\textwidth]{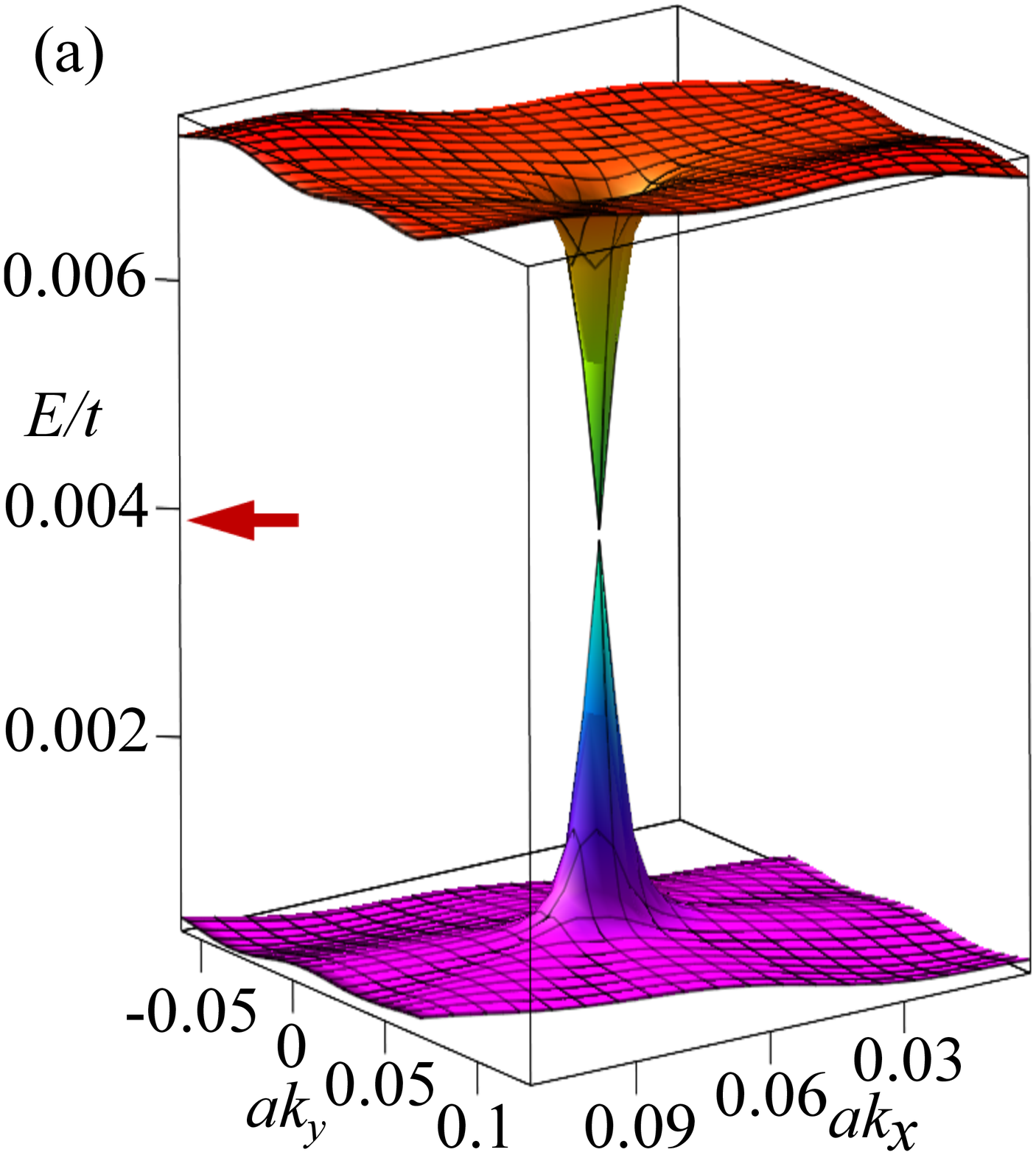}
\includegraphics[width=0.24\textwidth]{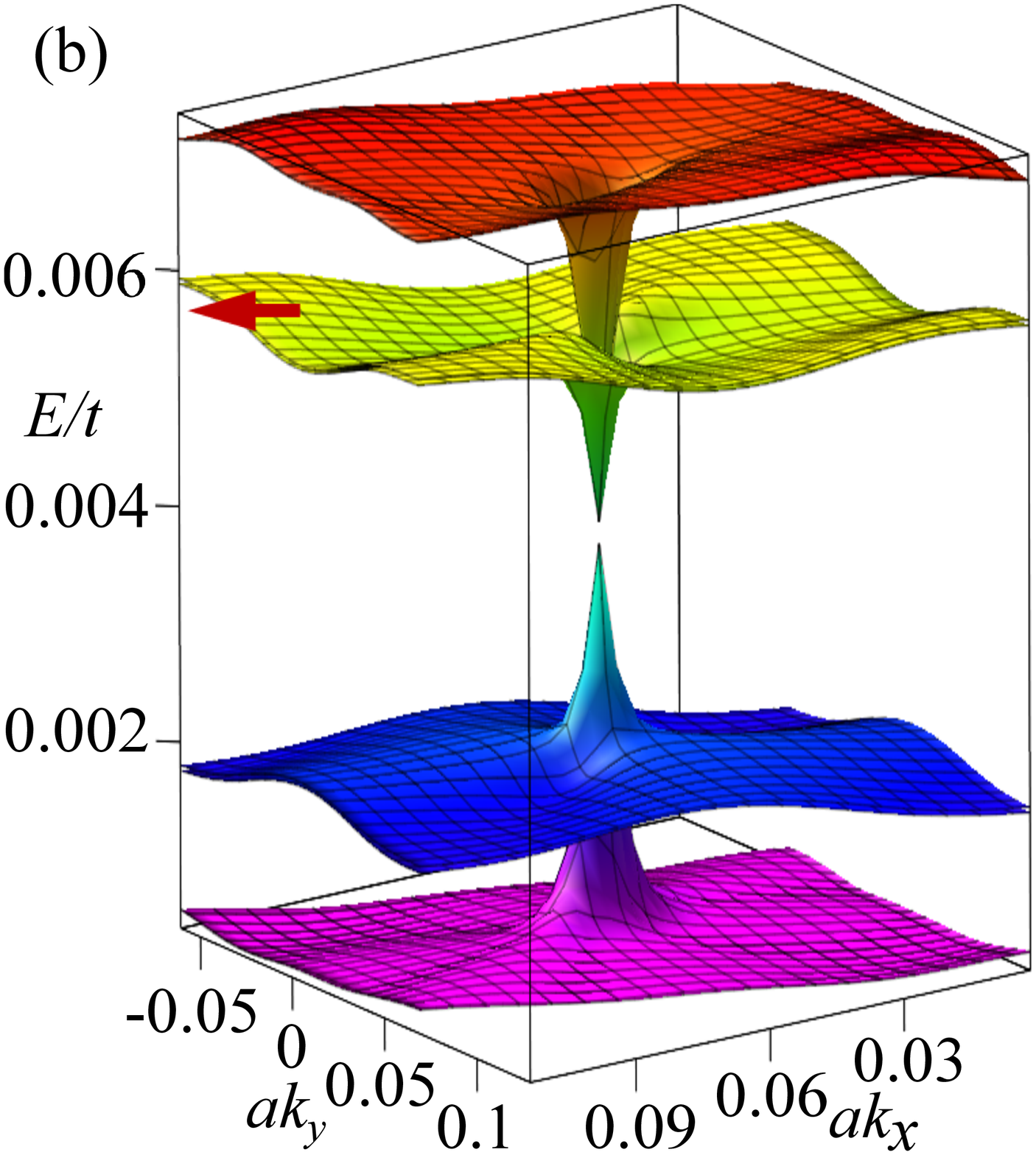}
\includegraphics[width=0.24\textwidth]{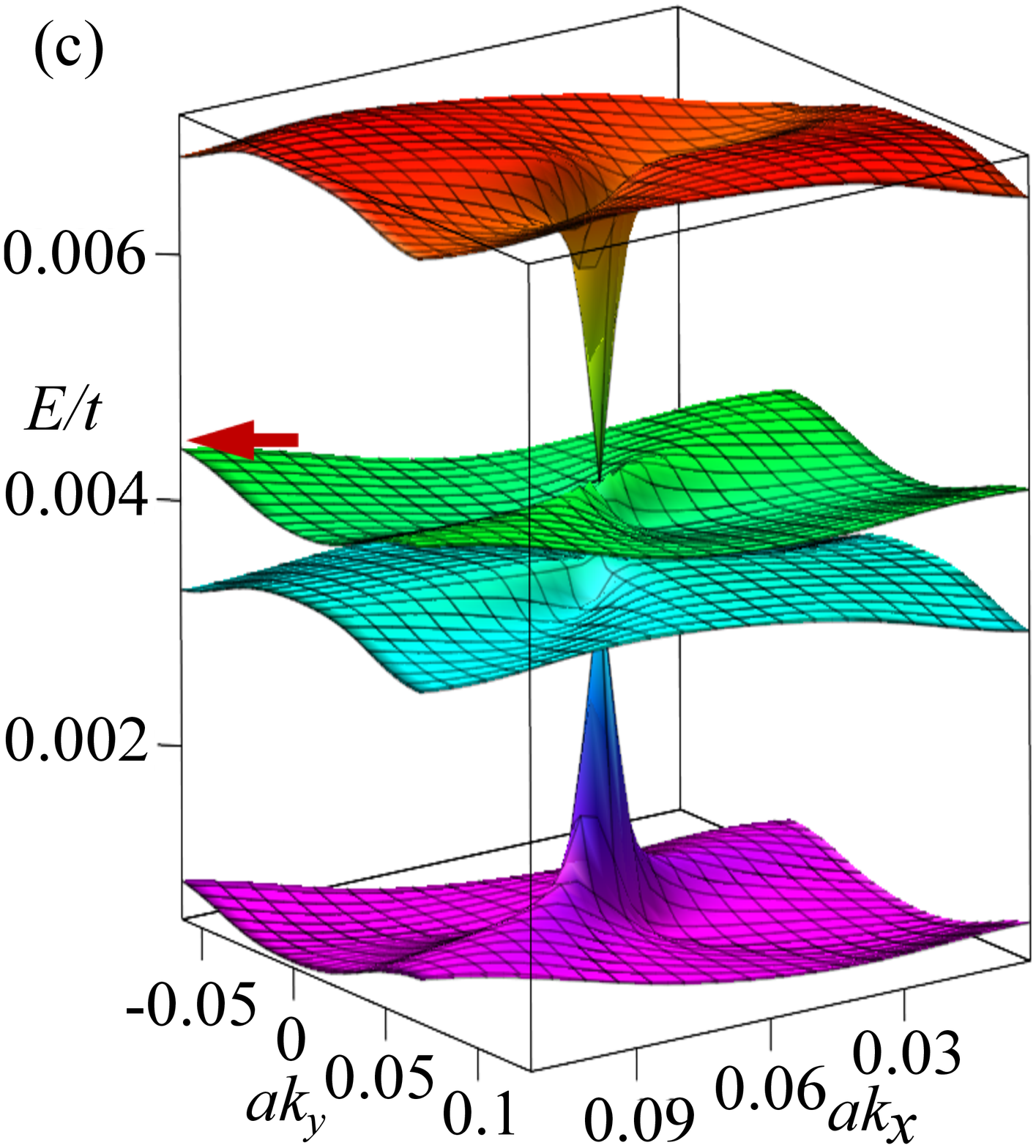}
\includegraphics[width=0.24\textwidth]{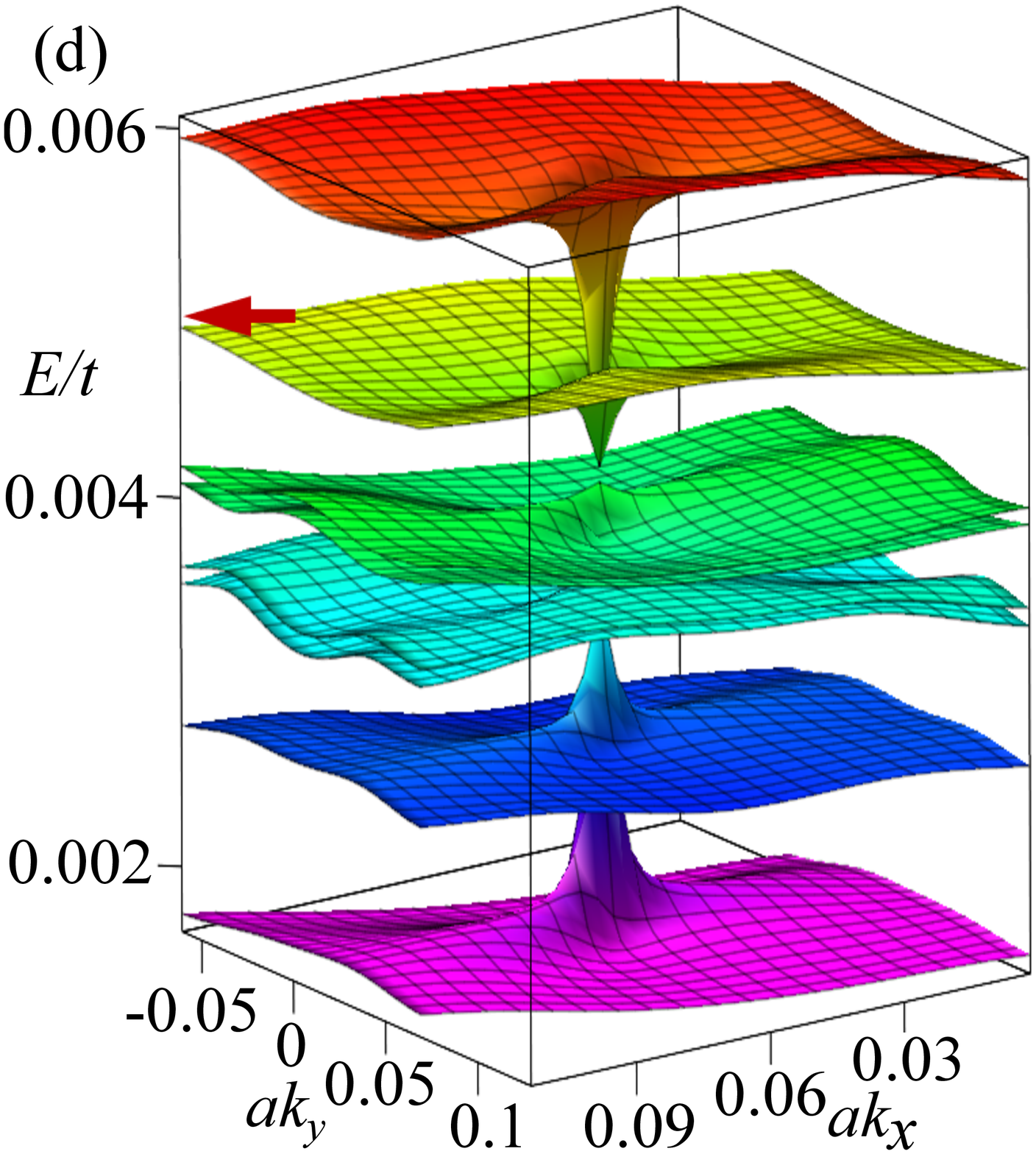}\vspace{5mm}\\
\includegraphics[width=0.24\textwidth]{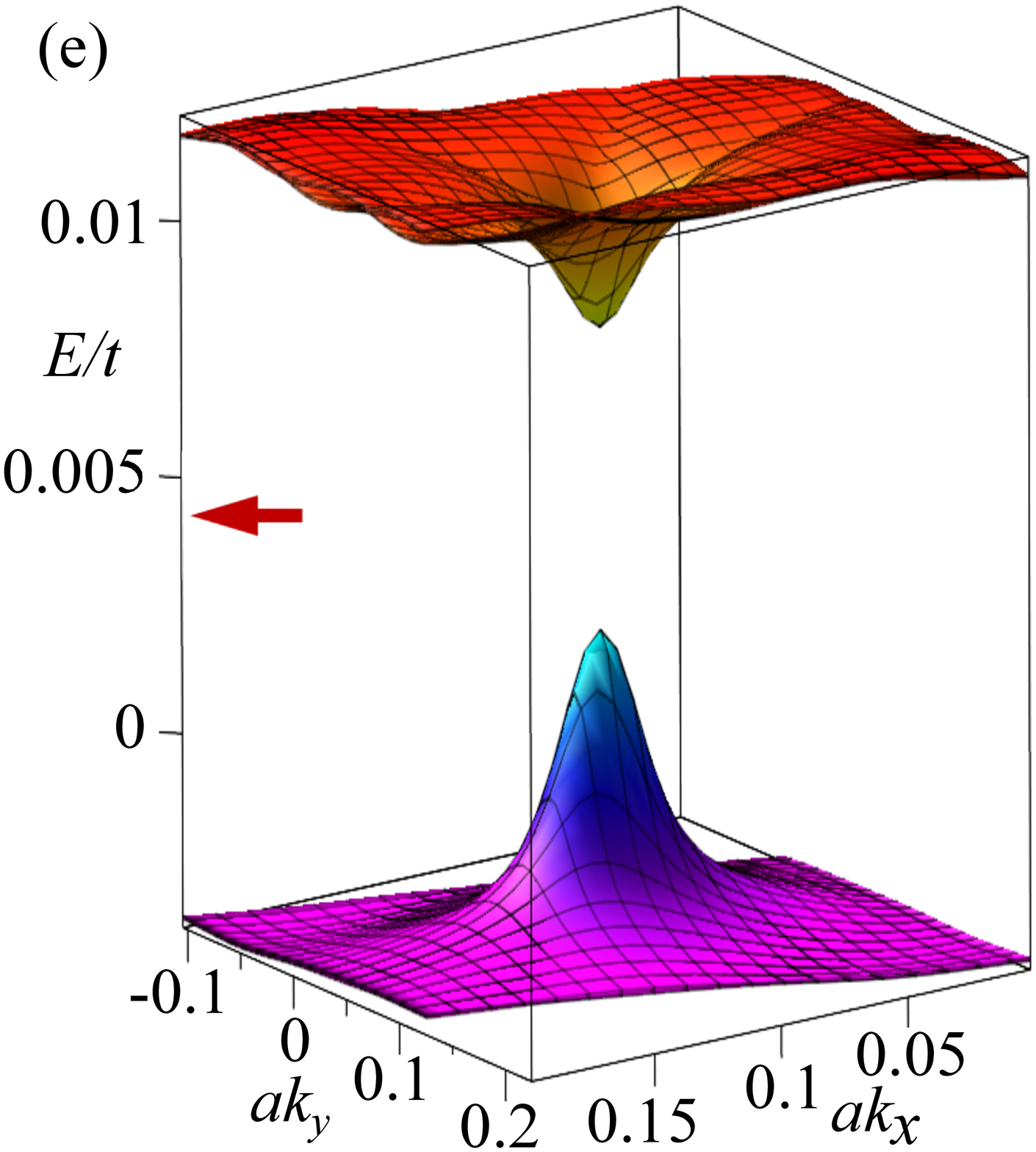}
\includegraphics[width=0.24\textwidth]{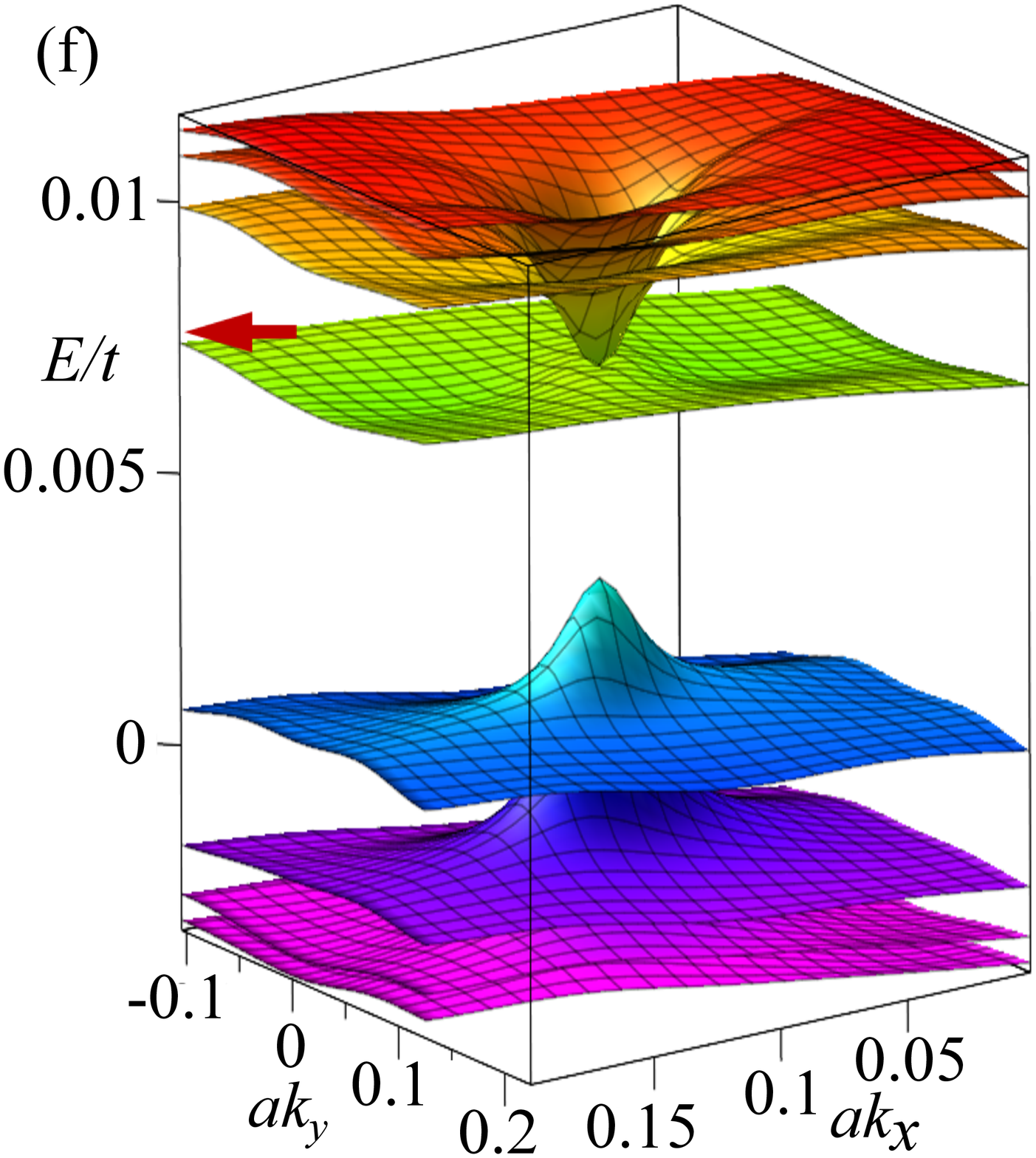}
\includegraphics[width=0.24\textwidth]{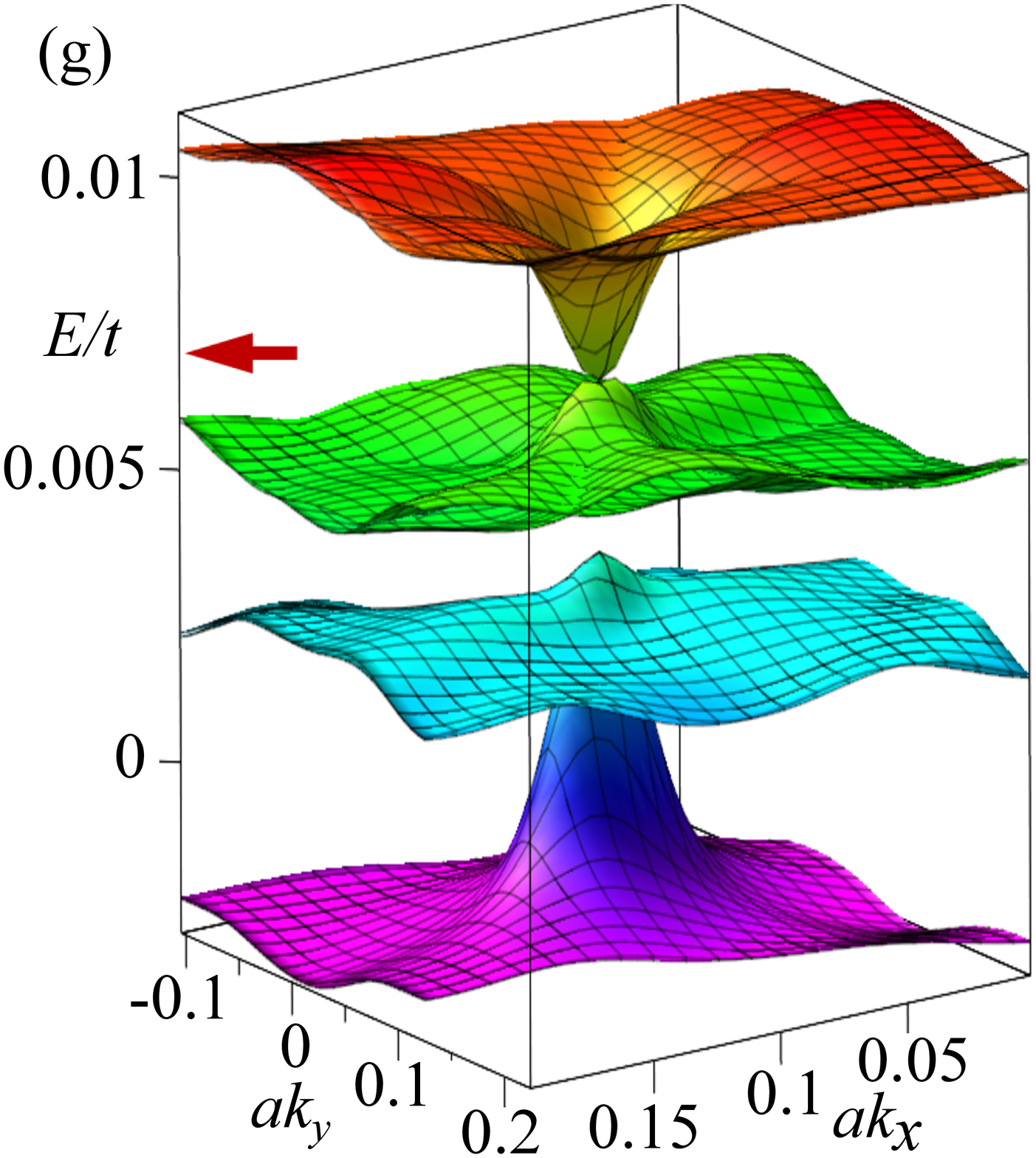}
\includegraphics[width=0.24\textwidth]{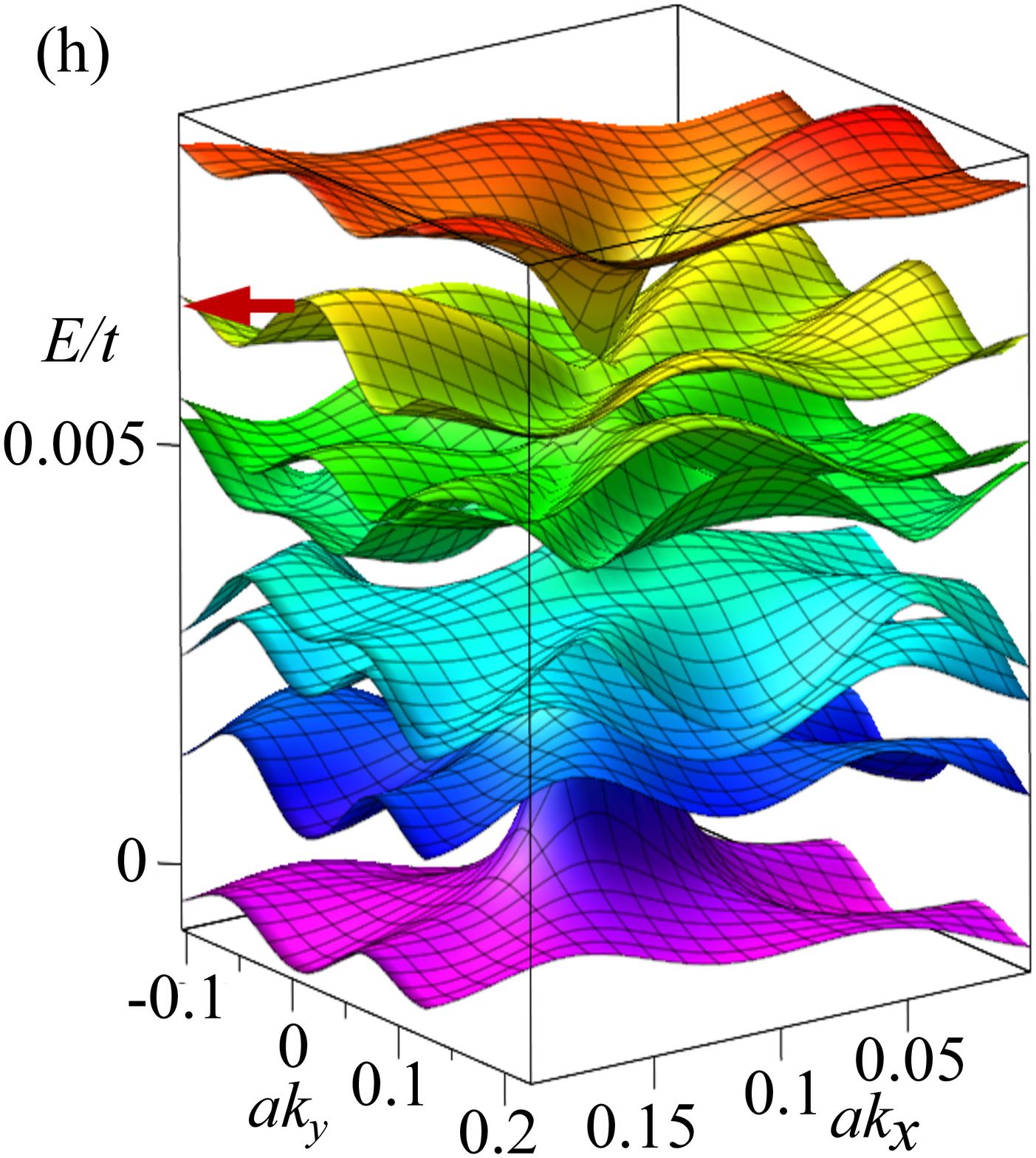}
\caption{Mean-field low-energy spectra calculated at different
integer-valued doping levels for
parametrizations~I [(a)\,--\,(d)]
and~II.B [(e)\,--\,(h)].
Doping levels are:
$\nu=0$ [(a),\,(e)],
$\nu=1$ [(b),\,(f)],
$\nu=2$ [(c),\,(g)],
and
$\nu=3$ [(d),\,(h)].
The red thick arrow near the $E/t$ axis marks the position of the Fermi level.
\label{Fig3DSpec}}
\end{figure*}

\subsection{Symmetry of the mean-field low-energy spectra}

If interactions are neglected, the tBLG single-electron states are doubly
degenerate. The SDW order parameters
lift the spin degeneracy. As a result, at low energies we have eight
non-degenerate flat bands.
Figure~\ref{fig::compare_spectra}
allows one to compare the spectrum of the non-interacting model with the mean-field spectrum at the charge neutrality point.
The degeneracy-lifting patterns for different doping levels
and parametrizations are illustrated by plots in
Figs.~\ref{Fig3DSpec}(a)\,--\,(h)
which show the low-energy spectra inside a reciprocal supercell (this data
will be discussed in detail in
subsection~\ref{subsec::spectra_doping},
see also
Ref.~\onlinecite{OurtBLGPRB2019}).

\begin{figure*}[t]
\centering
\includegraphics[width=0.9\columnwidth]{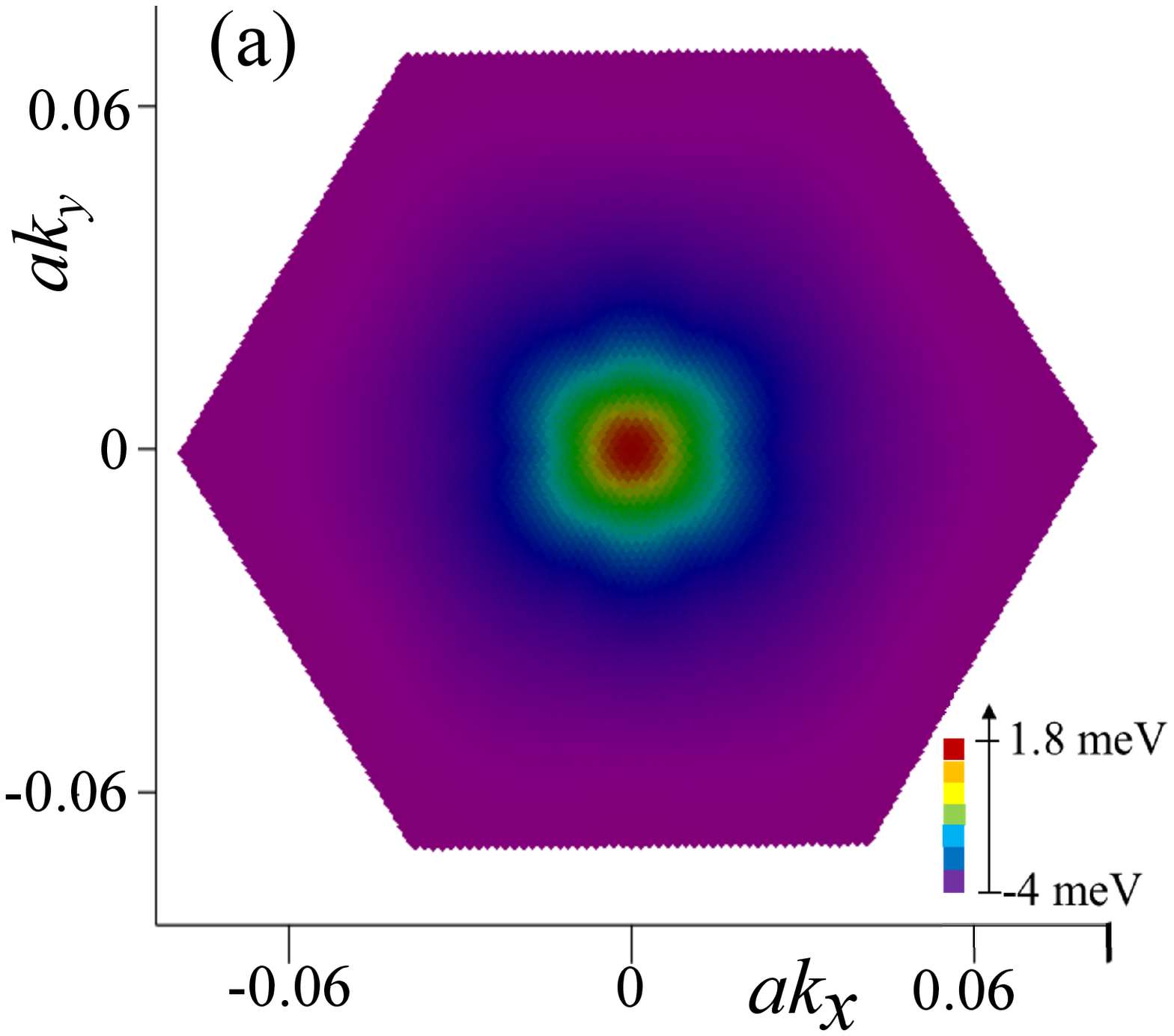}\hspace{5mm}
\includegraphics[width=0.9\columnwidth]{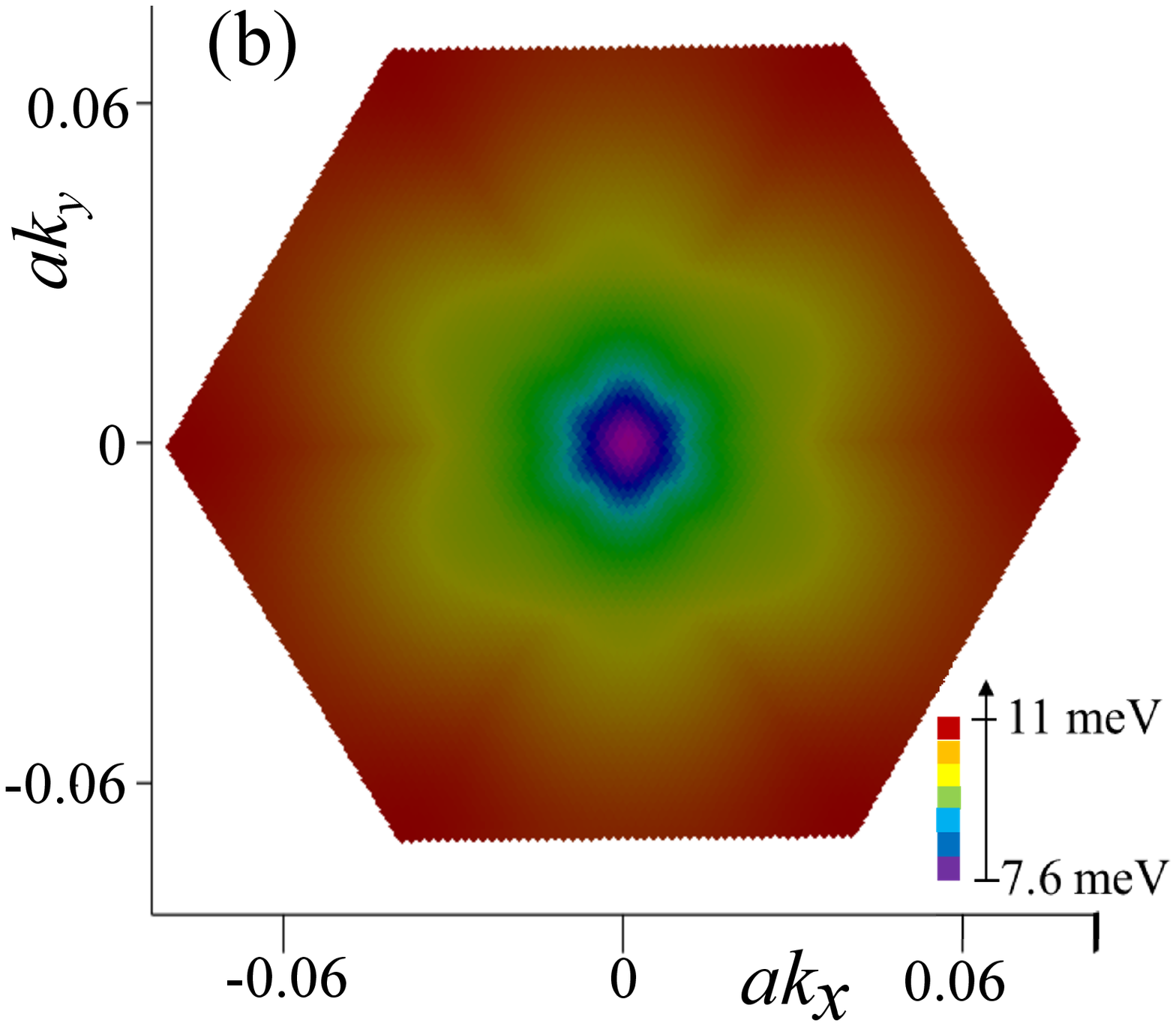}\vspace{5mm}\\
\includegraphics[width=0.9\columnwidth]{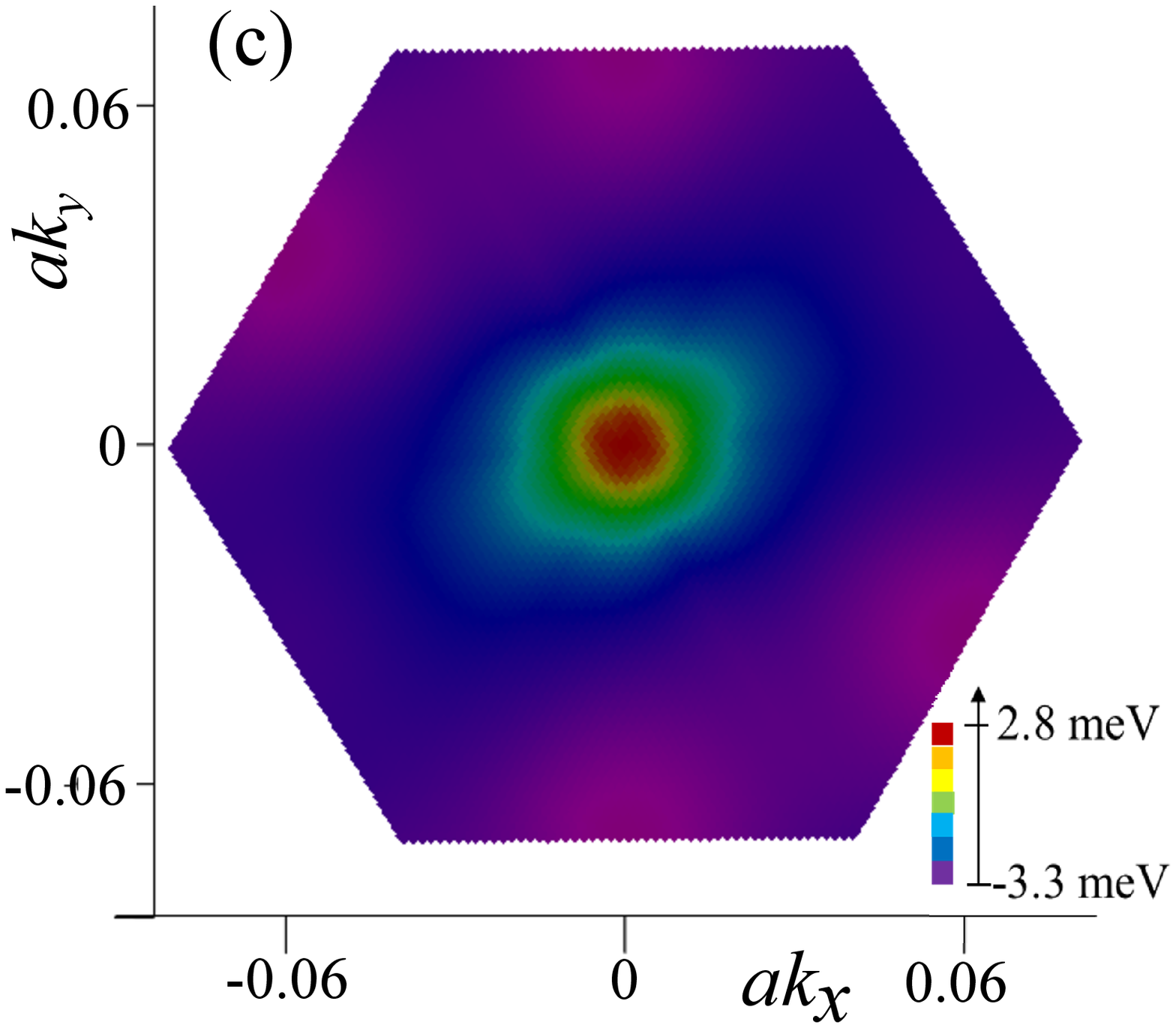}\hspace{5mm}
\includegraphics[width=0.9\columnwidth]{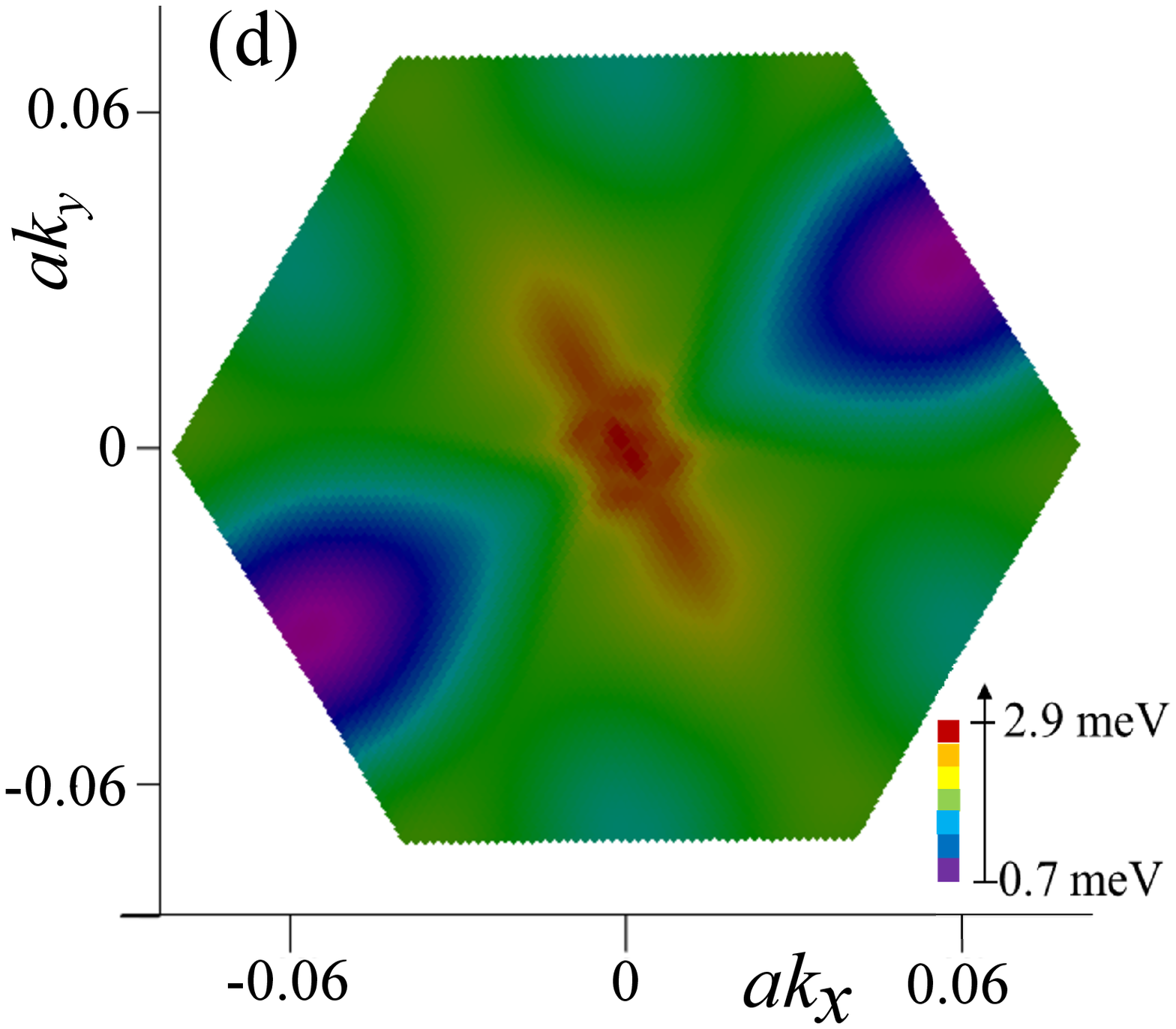}
\caption{Color plots of the mean-field low-energy bands closest to the
Fermi level calculated at the charge neutrality point [panels (a) and (b)] and
at $\nu=-2$ [panels (c) and (d)]; parametrization~II.A. Each plot has its individual color bar. 
Nematicity-induced reduction of the point symmetry group from $C_6$ [(a) and (b)] to $C_2$ [(c) and (d)] is clearly seen.
\label{FigSDWSpecRBZ}}
\end{figure*}

However, the lifting of the spin degeneracy is not the only consequence of
the SDW ordering. The geometrical symmetries of the order parameter affect
the symmetries of the single-electron mean-field spectrum as well. The
plots in
Figs.~\ref{Fig3DSpec}
are not convenient for discussion of this issue, and we will use
Figs.~\ref{FigSDWSpecRBZ}(a)\,--\,(d),
which present individual color plots of the mean-field bands calculated at
different doping levels inside the RBZ [similar data for the non-interacting
case is shown in
Figs.~\ref{FigSpec}(g)\,--\,(j)],
instead.

We start from the charge neutrality point. At zero doping, the low-energy
bands bundle into two groups (four bands per group) of nearly degenerate
bands, see
Figs.~\ref{Fig3DSpec}(a,e).
Such a group will be called a
quartet~\cite{OurtBLGPRB2019}.
The separation between bands within a specific quartet is finite. However,
it is much smaller than the characteristic separation between the quartets
themselves. Because of this near-degeneracy,
it is
sufficient to choose a single band to represent a given quartet. Of four
bands in each quartet, the bands closest to the Fermi level are shown in
Fig.~\ref{FigSDWSpecRBZ}(a)\,--\,(b).
These plots have the same symmetries as those in
Figs.~\ref{FigSpec}(g)\,--\,(j):
they all are symmetric under rotations of $60^{\circ}$ around the $\Gamma$ point,
they also have six mirror symmetry axes, parallel and perpendicular to
$\mathbf{G}_1$,
$\mathbf{G}_2$,
and
$(\mathbf{G}_1+\mathbf{G}_2)$.
The other bands in quartets all have the same symmetries, independent of a
specific parametrization.

Doping reduces the symmetry of the SDW order parameters. As a result, the
symmetry of the mean-field spectrum is also reduced. To illustrate this, in
Figs.~\ref{FigSDWSpecRBZ}(c)\,--\,(d)
we present color plots of two low-energy bands closest to the Fermi level
(one is filled, the other is empty) calculated for parametrization~II.A at
doping
$\nu=-2$.
The spectra now do not exhibit hexagonal symmetry, but they are still
symmetric under a
$180^{\circ}$
rotation around the $\Gamma$ point. There are also two mirror symmetry
axes, parallel and perpendicular to
$(\mathbf{G}_1+\mathbf{G}_2$).
Similar pictures are observed for electron doping and for the two other
parametrizations.

\subsection{Mean-field low-energy spectra: evolution with doping}
\label{subsec::spectra_doping}

We reported previously~\cite{OurtBLGPRB2019}
that the structure of the low-energy single-electron spectrum strongly
depends on the doping level. The study in Ref.~\onlinecite{OurtBLGPRB2019} was performed for a single parametrization (the parametrization employed in
Ref.~\onlinecite{OurtBLGPRB2019}
is a version of parametrization~I).
Below we will extend that analysis by comparing spectra calculated for
different parametrizations. Our main findings are summarized in
Figs.~\ref{Fig3DSpec}(a)\,--\,(h).
These show the spectra inside a reciprocal supercell (centered at the $\Gamma$
point) calculated for parametrizations~I and~II.B at four integer-valued
doping levels
$\nu=0,\,1,\,2\,,3$.
The structures for negative doping levels $n$ are very similar to that for
positive dopings $|n|$.

For fixed doping, the change of parametrization does not introduce
qualitative modifications to the spectrum. However, several quantitative
characteristics are sensitive to the parametrization choice.

\subsubsection{Charge neutrality point}

At the charge
neutrality point, the eight low-energy bands split into two quartets.
Except for a small vicinity of the $\Gamma$ point, the energy
$\Delta_s$
separating the quartets is almost constant everywhere in the RBZ. The
specific value of
$\Delta_s$
depends on the parametrization: for case~I one has
$\Delta_s\approx15$\,meV.
A similar value
($\Delta_s\approx14$\,meV)
was found for parametrization II.A.
At the same time, for the case~II.B this quantity is significantly
larger
$\Delta_s\approx 40$\,meV.

At the $\Gamma$ point, the separation between the quartets is the smallest.
For parametrization~I, see
Fig.~\ref{Fig3DSpec}\,(a),
the splitting between upper and lower quartets is
$\sim 9\times 10^{-5}t$.
Such a splitting is smaller,
but comparable to, the splitting of the non-interacting bands close to the
$\Gamma$ point, as shown in
Fig.~\ref{FigSpec}\,(d).
At the $\Gamma$ point, each quartet consists of two doublets, the splitting
between doublets is about
$8\times 10^{-6}t$.
A similar situation takes place for other two parametrizations, with the only
difference being that the splitting between quartets is one order of magnitude larger
than for parametrization~I. Thus, on a qualitative level, the band
structures at the $\Gamma$ point are similar for all three parametrizations.

The spectra of non-interacting models are four-fold degenerate [(two-fold
spin degeneracy) times (two-fold valley degeneracy)] at the $\Gamma$ point [see
Figs.~\ref{FigSpec}\,(d)-(f)].
Thus, the SDW order partially lifts this degeneracy.

\subsubsection{Doped states}

For $\nu=\pm1$,
each quartet splits into a group of three bands (a triplet) and a single
band (a singlet), with the chemical potential in the (partial) gap between
the triplet and the singlet. At half-filling
($\nu=\pm2$)
each quartet is transformed into two doublets. The chemical potential is
between the doublets. When we have three extra electrons (or extra holes)
per supercell,
$\nu=\pm3$,
the upper (lower) quartet is separated into the doublet and two upper
(lower) singlets. For electron (hole) doping, the chemical potential lies
between the upper (lower) singlets.

As one can see from
Fig.~\ref{Fig3DSpec}(h),
for parametrization~II.B at
$|\nu|\approx3$,
the band warping becomes comparable to the band splitting. This effect is
even more pronounced for parametrization~II.A.
We note that the (approximate) band degeneracy at the $\Gamma$ point
persists for all parametrizations and all levels of doping.

It is instructive to interpret the doping-induced band structure
reconstructions in terms of the minimization of the total energy. At the
charge-neutrality point, splitting of the eight bands into two quartets
acts to lower the total energy, since only four of eight bands are filled.
At half-filling (two extra electrons per supercell), the single-particle
energy is optimized if a filled doublet splinters away from the quartet and
sinks beneath the Fermi level.

When we have only one extra electron per supercell
it is favorable to separate the lower single band (filled) from the
quartet. Finally, when we have three extra electrons per supercell, the
three filled energy bands from the quartet separate from the upper empty
one.

As a result of the spectrum reconstruction, the density of states at the
Fermi energy
$\rho(\varepsilon_F)$
becomes a non-monotonic function of $n$, see
Fig.~\ref{fig::dos}.
One observes that the density of states has a local minimum near or at the
integer value of $\nu$ for all three parametrizations. At the same time,
the dependence of
$\rho(\varepsilon_F)$
on $n$ is sensitive to details of the interlayer hopping. For example, the
minimum at
$\nu = 3$
is very shallow in panel~(b) of
Fig.~\ref{fig::dos},
it is more pronounced in panels~(c), and in panel~(a) the density of state
drops to zero at
$\nu = 3$.
Similar oscillations of the DOS were observed experimentally (for details
see next section).

\section{Discussion}
\label{Discussion}

We argued that doping of the system away from the charge neutrality point reduces
the symmetry of both the order parameters and the electronic spectrum
giving rise to the SDW-driven electronic nematic state. The SDW order
parameters monotonically decrease when doping goes away from zero. On the
other hand, the nematicity demonstrates a different trend. Our calculations
show that for doping
$|\nu|\lesssim0.5$,
the nematicity is virtually absent. At larger doping it starts growing
and achieves maximum at half-filling, that is for $2$ extra electrons or
holes per supercell. With further increase of doping the nematicity
decays, and at
$|\nu|\approx4$
it vanishes together with the SDW order parameters.

\begin{figure}[t]
\centering
\includegraphics[width=0.98\columnwidth]{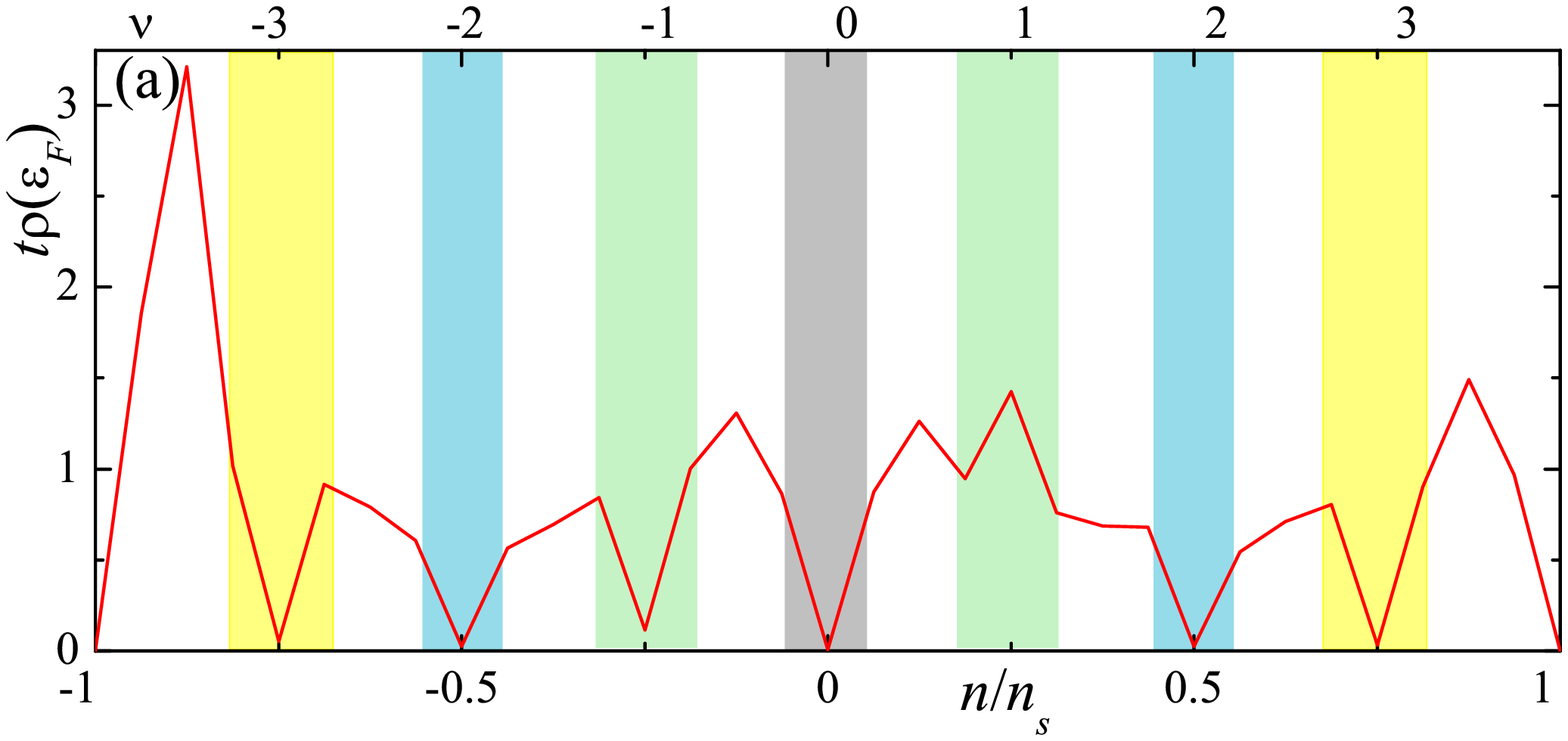}\\\vspace{7mm}
\includegraphics[width=0.98\columnwidth]{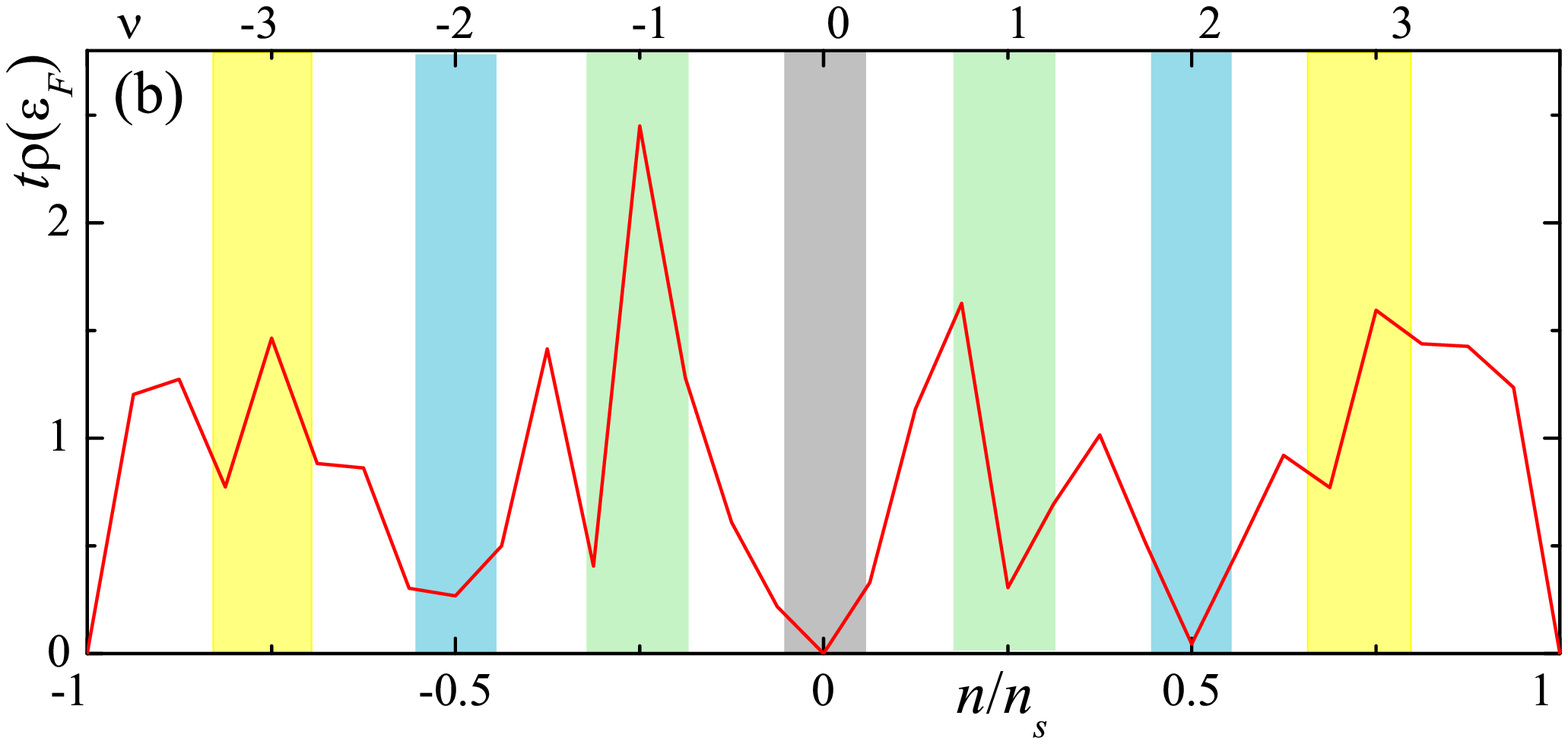}\\\vspace{7mm}
\includegraphics[width=0.98\columnwidth]{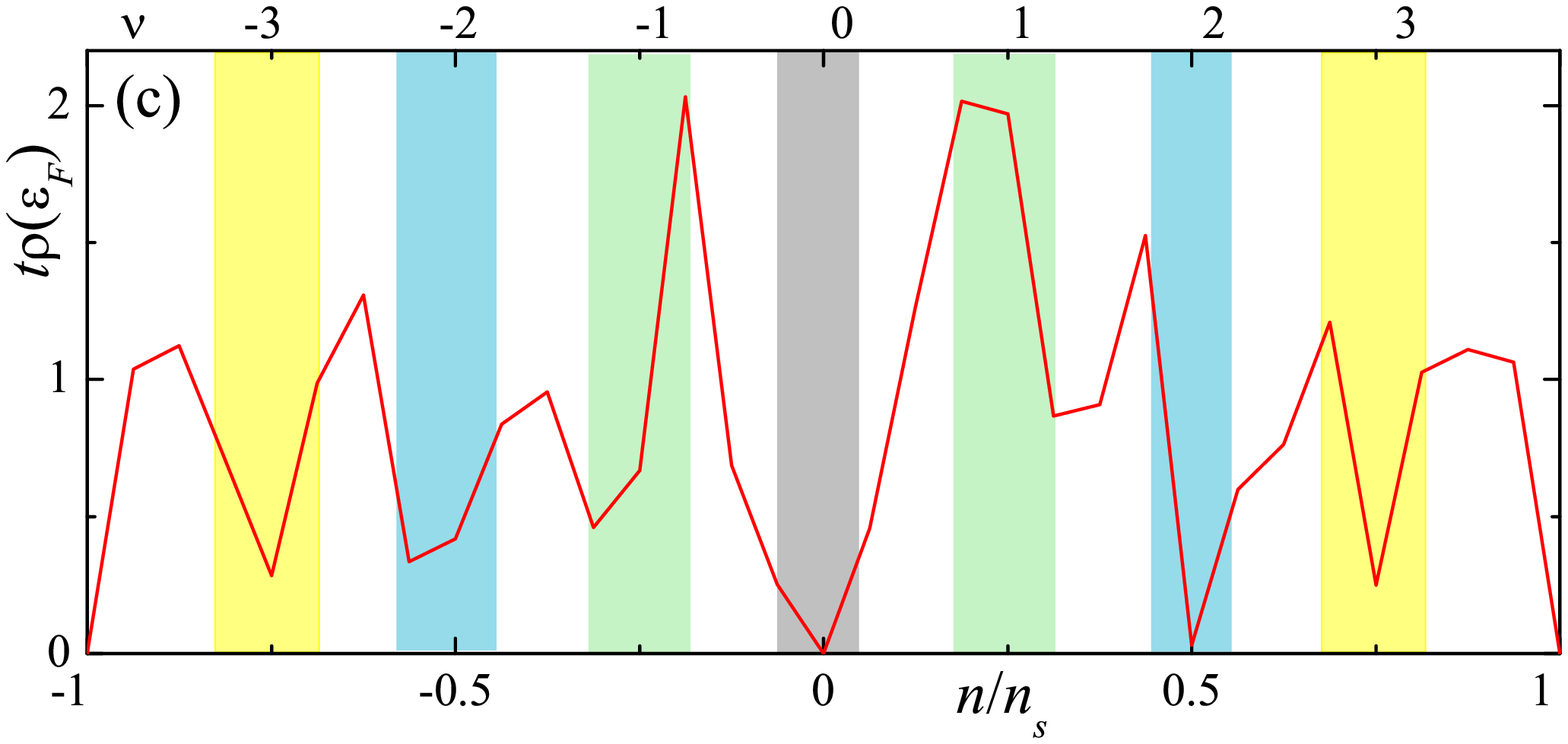}
\caption{The density of states at the Fermi level
$\rho (\varepsilon_F)$
as a function of doping $n$, for different parametrizations: panel~(a)
corresponds to parametrization~I, panels~(b) and~(c) show the results
for the cases~II.A and~II.B. Colored rectangles mark the areas
near integer values of
$\nu = n/(n_s/4)$.
For all parametrizations, the density of states has local minima near or at
integer $\nu$. At the same time, finer details of
$\rho (\varepsilon_F)$
are sensitive to the particulars of the interlayer hopping. Specifically,
the precise locations and the depth of a given minimum vary among the various parametrizations.
\label{fig::dos}
}
\end{figure}
Nematicity reveals itself in the symmetry reduction of both the SDW order
parameters and the electron spectrum.
The reduced symmetry in the order parameters acts to reduce the symmetry of
the charge density and local density of states, see
Fig.~\ref{fig::nematic_DOS}.
``Nematic" features of the local density of states can be detected in STM
measurements, for example, as in
Refs.~\onlinecite{MottNematicNature2019,KerelskyNematicNature2019}.
In these experiments, the bright spots in STM images, centered at the AA
regions of the moir{\'e} superlattice, were uniaxially stretched.
Moreover, the
triangular superlattice was skewed as well.
Reference~\onlinecite{KerelskyNematicNature2019}
reported that the strongest nematicity of STM images was observed near
half-filling, in agreement with our findings. Nematicity of the normal
phase near half-filing ($\nu\approx-2$) was observed in
Ref.~\onlinecite{cao2020nematicity}
by direction-dependent transport measurements. This is in agreement with
our results.
Reference~\onlinecite{cao2020nematicity}
reports also the nematicity of the superconducting phase near half-filling.

Our calculations demonstrate that nematicity of the order parameters and
the energy spectra is very robust to the change of the hopping amplitude
parametrization. This indicates that the nematic state is not an artifact of
some ``lucky" model or parameter choice. Rather, it is an inherent feature
of the MAtBLG.

We observed that the low-energy band structure substantially depends on the
doping level. As a result of the doping-induced spectrum reconstruction,
the density of states at the Fermi level
$\rho(\varepsilon_F)$
passes through minima at (or close to) integer-valued $\nu$'s. Such a
behavior was reported in several experimental
papers~\cite{STM2019,MottNematicNature2019,STMNature2019,
Jiang_twisted_stm2019nature},
see, e.g.,
Fig.~3(a) of
Ref.~\onlinecite{STMNature2019}
or
Fig.~3(e) of
Ref.~\onlinecite{STM2019}.
On the theory side, we reported similar findings in
Ref.~\onlinecite{OurtBLGPRB2019}
for a single specific interlayer parametrization. In the present paper, we extend our previous study considering three more parametrizations, see
Fig.~\ref{fig::dos}.
This is important, because no interlayer tunneling model is universally
accepted, and such an investigation allows us to understand, what physical
properties of the MAtBLG are stable against model variations, and what
properties are fragile and require fine-tuning.

Comparing graphs for $\rho(\varepsilon_F)$
versus $n$ calculated for different interlayer tunneling parametrizations,
we learn an important lesson. On these graphs, the visibility of a specific
minimum is a non-universal quantity, sensitive to the model details. This
was illustrated in Sec.~\ref{subsec::spectra_doping}
with the discussion of the minimum at
$\nu = 3$.
Other minima at odd values of $\nu$ demonstrate a similar non-universality.
The peaks at $\nu = 0, \pm 2$ are not immune to the model modifications either, although to a lesser extent. We believe that the manifestation of this sensitivity might explain the
sample-to-sample variation of the conductivity minima observed
experimentally. Indeed, in
Fig.~1\,(c)
of
Ref.~\onlinecite{MottSCNature2019}
all minima are discernible, in
Fig.~2\,(a)
of
Ref.~\onlinecite{NatureMott2018}
the minimum at
$\nu = -1$
is absent, while the minimum at
$\nu = 1$
is extremely weak.

According to our calculations, the system can be insulating only at the
charge-neutrality point. Specifically, for parametrizations~II.A and~II.B
we see that
$\rho(\varepsilon_F) = 0$
when
$n = 0$
[for parametrization I the gap is very small, cf. Figs.~\ref{Fig3DSpec}(a) and~\ref{Fig3DSpec}(e)].
At other integer-valued $\nu$'s, the mean-field ground state is always
metallic: the bands in the upper and lower quartets are not well separated
in the whole RBZ for any doping levels (see
Fig.~\ref{Fig3DSpec}).
Experimentally~\cite{NatureMott2018,MottSCNature2019},
however, the state at even $\nu$ shows insulating properties. This
discrepancy can be an artifact of the approximation used. First, we
consider only short-range order parameters. Second, for these order
parameters a superlattice periodicity was assumed, that is, no extra
periodicity emerged as in usual antiferromagnets. Removing any of these
constrains will lead to a significant increase of computation costs.

More generally, the doping-induced reconstruction of the spectrum affects
not only
$\rho(\varepsilon_F)$,
but changes the whole curve
$\rho(E)$
versus energy $E$. This dependence was indeed observed in recent
experiments,
Refs.~\onlinecite{STM2019,MottNematicNature2019,KerelskyNematicNature2019}
[see, e.g., sequence of
$\text{d}I/\text{d}V$
curves presented in
Fig.~4(a) of
Ref.~\onlinecite{MottNematicNature2019}].
The band splitting of two quartets, $\Delta_s$, existing at the charge-neutrality point
can be used as a characteristic energy scale of the low-energy band
structure. Our calculations give the values for
$\Delta_s$,
ranging from about $15$\,meV to about $40$\,meV, depending on the interlayer
hopping amplitude parametrization (see
Sec.~\ref{subsec::spectra_doping}).
Such estimates are in agreement with experimental data in
Refs.~\onlinecite{STM2019,STMNature2019,MottNematicNature2019,
KerelskyNematicNature2019,CompressibilityPRL2019,
Jiang_twisted_stm2019nature}

In conclusion, we studied the properties of the magic-angle twisted
bilayer graphene in the doping range from $-4$ to $+4$ electrons per
supercell. A spin density wave is assumed to be the ground state of the
system in the whole doping range. Doping the system away from the charge-neutrality point reduces the symmetry of the order parameters, giving rise to
the SDW-driven electron nematic state. Nematicity is largest near
half-filling ($2$ electrons or holes per supercell). The spatial profile of the
SDW order parameters and nematicity of the electron spectrum are robust to
the change of the interlayer hopping amplitudes parametrization. Our
theoretical results are consistent with several experiments.

\begin{acknowledgments}
This work is partially supported by the JSPS-Russian Foundation for Basic
Research Project No.~19-52-50015, and by the Japan Society for the Promotion of Science (JSPS-RFBR Grant No. JPJSBP120194828).
F.N. is supported in part by: NTT Research, Army Research Office (ARO) (Grant No. W911NF-18-1-0358),
Japan Science and Technology Agency (JST) (via the CREST Grant No. JPMJCR1676),
Japan Society for the Promotion of Science (JSPS) (via the KAKENHI Grant Number JP20H00134),
and Grant No. FQXi-IAF19-06 from the Foundational Questions Institute Fund (FQXi),
a donor advised fund of the Silicon Valley Community Foundation.
We acknowledge the Joint Supercomputer Center of the Russian Academy of Sciences (JSCC RAS) for the computational resources provided.
\end{acknowledgments}

\appendix

\section{Calculation procedure of the SDW order parameters}

Here we present the details of the iteration scheme for calculating the
SDW order parameters. The total Hamiltonian is given by
Eq.~\eqref{H}.
It can be rewritten as
$H=H_0+H_{\text{int}}$,
where
$H_0$
is the single-particle part corresponding to the first term of $H$, while
$H_{\text{int}}$
includes the second and third terms of $H$. The interaction Hamiltonian is
quadrilinear in the terms of electronic creation and annihilation
operators. In the mean-field approximation used here, the following
decoupling is explored:
\begin{eqnarray}
&&\!\!\!\!\!\!d^{\dag}_{\mathbf{n}is\sigma}
d^{\phantom{\dag}}_{\mathbf{n}is\sigma} d^{\dag}_{\mathbf{m}jr\sigma'}
d^{\phantom{\dag}}_{\mathbf{m}jr\sigma'}\!\rightarrow\!
-d^{\dag}_{\mathbf{n}is\sigma}d^{\phantom{\dag}}_{\mathbf{m}jr\sigma'}
\langle
	d^{\dag}_{\mathbf{m}jr\sigma'}
	d^{\phantom{\dag}}_{\mathbf{n}is\sigma}
\rangle-
\nonumber
\\
&&\!\!\!\!\!\!-d^{\dag}_{\mathbf{m}jr\sigma'}
d^{\phantom{\dag}}_{\mathbf{n}is\sigma}\langle
d^{\dag}_{\mathbf{n}is\sigma}d^{\phantom{\dag}}_{\mathbf{m}jr\sigma'}\rangle
\!+\!\langle d^{\dag}_{\mathbf{n}is\sigma}d^{\phantom{\dag}}_{\mathbf{m}jr\sigma'}\rangle\langle d^{\dag}_{\mathbf{m}jr\sigma'}d^{\phantom{\dag}}_{\mathbf{n}is\sigma}\rangle.\nonumber\\
\end{eqnarray}
Assuming that non-zero expectation values are only those shown in
Eqs.~\eqref{Deltania},~\eqref{Anis},
and~\eqref{Mmn}
for the SDW order parameters, we obtain for the mean-field interaction
Hamiltonian:
\begin{eqnarray}
&&H_{\text{int}}^{\text{MF}}
=
-\sum_{\mathbf{n}is}
	\left(
		\Delta_{\mathbf{n}is}^*
		d^{\dag}_{\mathbf{n}is\uparrow}
		d^{\phantom{\dag}}_{\mathbf{n}is\downarrow}
		+ {\rm h.c.}
	\right)
+\sum_{\mathbf{n}is}\frac{|\Delta_{\mathbf{n}is}|^2}{U}\nonumber\\
&&-\sum_{\mathbf{n}i{\ell}\sigma}\left(A^{(\ell)*}_{\mathbf{n}i\sigma}d^{\dag}_{\mathbf{n}+\mathbf{n}_{\ell}i{\cal{A}}\sigma}d^{\phantom{\dag}}_{\mathbf{n}i{\cal{B}}\bar{\sigma}}+{\rm h.c.}\right)
+\sum_{\mathbf{n}i{\ell}\sigma}\frac{|A^{(\ell)}_{\mathbf{n}i\sigma}|^2}{V_{\text{nn}}}\nonumber\\
&&-\sum_{\mathbf{nm}\atop{rs\sigma}}\left(B^{rs*}_{\mathbf{m};\mathbf{n}\sigma}d^{\dag}_{\mathbf{m}1r\sigma}d^{\phantom{\dag}}_{\mathbf{n}2s\bar{\sigma}}+{\rm h.c.}\right)
+\sum_{\mathbf{nm}\atop{rs\sigma}}\frac{|B^{rs}_{\mathbf{m};\mathbf{n}\sigma}|^2}{V^{rs}_{\mathbf{mn}}}\,,\nonumber\\
\end{eqnarray}
where
$V^{rs}_{\mathbf{mn}}
=
V(\mathbf{r}^{1r}_{\mathbf{m}}-\mathbf{r}^{2s}_{\mathbf{n}})$.
The total mean-field Hamiltonian
$H^{\text{MF}}=H_0+H_{\text{int}}^{\text{MF}}$
is quadratic in terms of electron operators, and can be diagonalized.
To proceed with the diagonalization, we switch to the superlattice
quasimomentum representation, as proposed in
Ref.~\onlinecite{PankratovPRB2013}.
To this end, we introduce new electronic operators

\begin{eqnarray}
d^{\phantom{\dag}}_{\mathbf{pG}is\sigma}
=
\frac{1}{\sqrt{{\cal{N}}}}
\sum_{\mathbf{n}}
	\exp{[-i(\mathbf{p}+\mathbf{G})\mathbf{r}_{\mathbf{n}}^{i}]}
	d_{\mathbf{n}is\sigma}\,,
\end{eqnarray}
where
${\cal N}$
is the number of graphene unit cells in the sample in one layer, the
momentum $\mathbf{p}$ lies in the first Brillouin zone of the superlattice,
while
$\mathbf{G}=n\bm{{\cal G}}_1+m\bm{{\cal G}}_2$
is the reciprocal vector of the superlattice lying in the first Brillouin
zone of the $i$th~layer. The number of such vectors $\mathbf{G}$ is equal
to
$N_{\rm sc}$
for each graphene layer. In terms of
$d^{\phantom{\dag}}_{\mathbf{pG}is\sigma}$,
the single-particle Hamiltonian becomes
\begin{eqnarray}
\label{H0}
&&H_0
=-t\sum_{\mathbf{pG}i\sigma}\!\!
\left(
	f_{\mathbf{p}+\mathbf{G}}^{i}d^{\dag}_{\mathbf{pG}i{\cal{A}}\sigma}
	d^{\phantom{\dag}}_{\mathbf{pG}i{\cal{B}}\sigma}+{\rm h.c.}
\right)+
\\
&&\sum_{\mathbf{pG}_1\!\mathbf{G}_2\atop{sr\sigma}}\!\!
\left[
t^{sr}_{\perp}(\mathbf{p}\!+\!\mathbf{G}_1;\mathbf{G}_1\!-\!\mathbf{G}_2)
d^{\dag}_{\mathbf{pG}_11s\sigma}
d^{\phantom{\dag}}_{\mathbf{pG}_22r\sigma}+{\rm h.c.}
\right].
\nonumber
\end{eqnarray}
Here
\begin{eqnarray}
f^1_{\mathbf{p}}&=&1+e^{-i\mathbf{pa}_1}+e^{-i\mathbf{pa}_2}\,,\nonumber\\
f^2_{\mathbf{p}}&=&1+e^{-i\mathbf{pa}'_1}+e^{-i\mathbf{pa}'_2}\,,
\end{eqnarray}
and
\begin{equation}
t^{sr}_{\perp}(\mathbf{p};\mathbf{G})=\frac{1}{N_{\rm sc}}\!\mathop{{\sum}'}_{\mathbf{nm}}\!
e^{-i\mathbf{p}(\mathbf{r}_{\mathbf{n}}^{1}-\mathbf{r}_{\mathbf{m}}^{2})}e^{-i\mathbf{G}\mathbf{r}_{\mathbf{m}}^{2}}
t(\mathbf{r}_{\mathbf{n}}^{1s};\mathbf{r}_{\mathbf{m}}^{2r})\,,
\end{equation}
where the summation symbol with prime
$\sum'_{\bf nm}$
implies that $\mathbf{m}$ runs over sites inside the zero{\it th}
supercell, while $\mathbf{n}$ runs over all sites in the sample. The first
term in
Eq.~\eqref{H0}
corresponds to the intralayer nearest-neighbor hopping, while the second
term describes the interlayer hopping. In terms of operators
$d^{\phantom{\dag}}_{\mathbf{pG}is\sigma}$,
the mean-field interaction Hamiltonian can be written as
\begin{widetext}
\begin{eqnarray}
&&H_{\text{int}}^{\text{MF}}=-\!\!\sum_{\mathbf{pG}_1\!\mathbf{G}_2\atop{is}}\!\!\left[
\left(\frac{1}{N_{\rm sc}}\sum_{\mathbf{n}}\Delta_{\mathbf{n}is}^*e^{-i(\mathbf{G}_1-\mathbf{G}_2)\mathbf{r}_{\mathbf{n}}^{i}}\right)
d^{\dag}_{\mathbf{pG}_1\!is\uparrow}d^{\phantom{\dag}}_{\mathbf{pG}_2\!is\downarrow}+{\rm h.c.}\right]
+\sum_{\mathbf{n}is}\frac{|\Delta_{\mathbf{n}is}|^2}{U}\nonumber-\\
&&-\!\!\sum_{\mathbf{pG}_1\!\mathbf{G}_2\atop{i\sigma}}\!\!\left[
\left(\frac{1}{N_{\rm sc}}\sum_{\mathbf{m\ell}}A^{(\ell)*}_{\mathbf{m}i\sigma}e^{-i(\mathbf{p}+\mathbf{G}_1)\mathbf{r}_{\mathbf{n}_{\ell}}^{i}}e^{-i(\mathbf{G}_1-\mathbf{G}_2)\mathbf{r}_{\mathbf{m}}^{i}}\right)
d^{\dag}_{\mathbf{pG}_1\!i{\cal{A}}\sigma}d^{\phantom{\dag}}_{\mathbf{pG}_2\!i{\cal{B}}\bar{\sigma}}+{\rm h.c.}\right]
+\sum_{\mathbf{n}i{\ell}\sigma}\frac{|A^{(\ell)}_{\mathbf{n}i\sigma}|^2}{V_{\text{nn}}}\nonumber\\
&&-\!\!\sum_{\mathbf{pG}_1\!\mathbf{G}_2\atop{rs\sigma}}\!\!\left[
\left(\frac{1}{N_{\rm sc}}\sum_{\mathbf{nm}}B^{rs*}_{\mathbf{m};\mathbf{n}\sigma}e^{-i(\mathbf{p}+\mathbf{G}_1)(\mathbf{r}^1_{\mathbf{n}}-\mathbf{r}^2_{\mathbf{m}})}
e^{-i(\mathbf{G}_1-\mathbf{G}_2)\mathbf{r}_{\mathbf{m}}^{2}}\right)d^{\dag}_{\mathbf{pG}_1\!1r\sigma}d^{\phantom{\dag}}_{\mathbf{pG}_2\!2s\bar{\sigma}}+{\rm h.c.}\right]
+\sum_{\mathbf{nm}\atop{rs\sigma}}\frac{|B^{rs}_{\mathbf{m};\mathbf{n}\sigma}|^2}{V^{rs}_{\mathbf{mn}}}\,.
\label{HMFint}
\end{eqnarray}
\end{widetext}

Using the operators
$d^{\phantom{\dag}}_{\mathbf{pG}is\sigma}$
we construct the
$N_{R}$-component
vector
\begin{eqnarray}
\label{Psi}
&&\Psi^{\dag}_{\mathbf{p}}=(\psi^{\dag}_{\mathbf{p}1\uparrow},\psi^{\dag}_{\mathbf{p}2\uparrow},\psi^{\dag}_{\mathbf{p}1\downarrow},\psi^{\dag}_{\mathbf{p}2\downarrow}),
\\
\nonumber
&&\psi^{\dag}_{\mathbf{p}i\sigma}
=
(d^{\dag}_{\mathbf{pG}_1\!i{\cal{A}}\sigma},
d^{\dag}_{\mathbf{pG}_1\!i{\cal{B}}\sigma},
\dots,
d^{\dag}_{\mathbf{pG}_{N_{\rm sc}}\!i{\cal{A}}\sigma},
d^{\dag}_{\mathbf{pG}_{N_{\rm sc}}\!i{\cal{B}}\sigma}).
\end{eqnarray}
In terms of this vector, the total mean-field Hamiltonian
$H^{\text{MF}}$
can be written as
$H^{\text{MF}}=\sum_{\mathbf{p}}\Psi^{\dag}_{\mathbf{p}}\hat{H}_{\mathbf{p}}\Psi^{\phantom{\dag}}_{\mathbf{p}}$,
where
$\hat{H}_{\mathbf{p}}$
is the
$N_R\times N_R$
matrix constructed from
$f_{\mathbf{p}+\mathbf{G}}^{i}$,
$t^{sr}_{\perp}(\mathbf{p}+\mathbf{G}_1;\mathbf{G}_1-\mathbf{G}_2)$,
$\Delta_{\mathbf{n}is}$,
$A^{(\ell)}_{\mathbf{n}i\sigma}$,
and
$B^{rs}_{\mathbf{m};\mathbf{n}\sigma}$
according to
Eqs.~\eqref{H0}
and~\eqref{HMFint}.

Since $N_R=8N_{\rm sc}$ with $N_{\rm sc}$ from Eq.~\eqref{Nsc}, one can evaluate $N_R=7,352$ for parametrization~II.B, while for parametrizations~I and~II.A we find
$N_R=22,328$. These numbers are too large for the effective execution of a numerical
procedure which requires multiple diagonalizations of matrices of rank $N_R$.

Fortunately, our task is simplified by the following circumstances.
The main contribution to the order parameter comes from low-energy
single-particle states; consequently, the contributions of the states far
from the Fermi energy can be safely approximated. Beside this, the
Hamiltonian describing these states is particularly simple. Indeed, for
large intralayer kinetic energy, we can neglect both interlayer hoppings
and SDW order parameters. In this limit,
the matrix
$\hat{H}_{\mathbf{p}}$
becomes block-diagonal, with the following
$2\times2$
matrices on its diagonal
\begin{equation}
-t\left(\begin{array}{cc}
0&f_{\mathbf{p}+\mathbf{G}}^{i}\\
f_{\mathbf{p}+\mathbf{G}}^{i*}&0
\end{array}\right).
\end{equation}
The eigenenergies of such a matrix are
$\pm t|f_{\mathbf{p}+\mathbf{G}}^{i}|$.
Both interlayer hopping amplitudes and SDW order parameters are small in
comparison to $t$. As long as we are interested in low-energy features, we
can use the truncated matrix
$\hat{H}'_{\mathbf{p}}$
to calculate the mean-field spectrum. To derive this matrix, we define the
reduced subset of
$\mathbf{G}$
satisfying the inequality
\begin{equation}
t|f_{\mathbf{G}}^{i}|<E_R\,,
\end{equation}
where
$E_R$
is the cutoff energy.
(In all simulations, we use
$E_R \sim 1$\,eV.)
Obviously, the number
$N_1$
of such $\mathbf{G}$'s is an increasing function of
$E_R$.
Also
$N_1<N_{\rm sc}$.
Using this subset of
$\mathbf{G}$'s,
we construct the truncated basis
$\Psi'_{\mathbf{p}}$
and truncated matrix
$\hat{H}'_{\mathbf{p}}$
according to
Eqs.~\eqref{H0}
and~\eqref{HMFint}
with
$\mathbf{G}_1$
and
$\mathbf{G}_2$
belonging to the reduced subset. The rank of the truncated matrix is
$N'_R=8N_1$.

Diagonalization of
$\hat{H}'_{\mathbf{p}}$
gives the wrong result for the eigenenergies
$E^{(S)}_{\mathbf{p}}$
close to
$\pm E_R$.
For this reason we take into account only eigenenergies satisfying the
inequality
$|E^{(S)}_{\mathbf{p}}|<E_0$,
with
$E_0<E_R$.
We use
$E_0=0.2t$,
$E_R=0.4t$
($N'_R=720$)
for parametrization~I and~II.A, and
$E_0=0.5t$,
$E_R=0.7t$
($N'_R=720$)
for parametrization~II.B. Several calculations with smaller and larger
$E_R$
and
$E_0$
show that the results are almost independent on these quantities.

Our goal is to minimize the total energy
$\cal E$
of the system with respect to the order parameters. Since we use the
truncated Hamiltonian, the contribution to the total energy from the
discarded states
$E^{(S)}_{\mathbf{p}}<-E_0$
must be accounted for separately. Since
$E_0\gg\max|\Delta_{\mathbf{n}is}|$,
this can be done perturbatively.

The leading corrections to
$E^{(S)}_{\mathbf{p}}$
are quadratic in
$\Delta_{\mathbf{n}is}$,
$A^{(\ell)}_{\mathbf{n}i\sigma}$,
and
$B^{rs}_{\mathbf{m};\mathbf{n}\sigma}$.
The same is true for the total energy. In our approximation we assume that
the proportionality coefficients are identical for all order parameters and
are equal to
\begin{equation}
\label{eq::susceptibility_approx}
-\frac{1}{V_c(E_0)}=-\frac12\int_{E_0}^{3t}\!\!\!\!dE\,\frac{\rho_0(E)}{E}\,,
\end{equation}
where
$\rho_0(E)$
is the single-layer graphene density of states. Such a correction can be
taken into account by the following replacement in the total mean-field
Hamiltonian
\begin{equation}
\label{Hrepl}
H^{\text{MF}} \rightarrow H'^{\text{MF}}
	\!-\!
	\sum_{\mathbf{n}is}
		\frac{|\Delta_{\mathbf{n}is}|^2}{V_c(E_0)}
	-\!
	\sum_{\mathbf{n}i{\ell}\sigma}
		\frac{|A^{(\ell)}_{\mathbf{n}i\sigma}|^2}{V_c(E_0)}
	-\!
	\sum_{\mathbf{nm}\atop{rs\sigma}}
		\frac{|B^{rs}_{\mathbf{m};\mathbf{n}\sigma}|^2}{V_c(E_0)},
\end{equation}
where
$H'^{\text{MF}}$
is the effective Hamiltonian in the truncated basis.

While the truncation scheme and
Eqs.~(\ref{Hrepl})
and~(\ref{eq::susceptibility_approx})
are an obvious simplification, we want to argue in favor of such an
approach. Working within the mean-field framework, one expects the
expression for the order parameter magnitude $\Delta$ to have the familiar
structure
$$\Delta \sim \Omega_0 \exp(-1/g),$$
where $g$ is the dimensionless interaction constant, and
$\Omega_0$
is the so-called pre-exponential energy scale. Depending on the physical
situation, one estimates
$\Omega_0$
as being of the order of the Debye temperature, or of the order of the bandwidth. However, the intrinsic accuracy of the mean-field approximation
does not allow us to improve our knowledge of
$\Omega_0$
beyond these order-of-magnitude estimates. Keeping this limitation in mind,
we note that
Eqs.~(\ref{Hrepl})
and~(\ref{eq::susceptibility_approx})
approximate contributions to this pre-exponential energy scale. Consequently,
a simplified treatment of these terms is in general agreement with the
accuracy of the mean-field approach.
Also, we need to remember that our simulations are in the
regime
$U < U_c$.
In this limit, two uncoupled single layers of graphene remain in the disordered
state. Thus, the stability of the SDW state relies crucially on the
low-energy band structure, while higher-energy states are of lesser
importance. These reasonings, in addition to numerical checks demonstrating
the insensitivity of final results to specific value of
$N_1$,
lend support to our confidence in the formulated approximation.

For
$E_0=0$
we have
$V_c(0)=2.23t$,
which is equal to the critical Hubbard $U$ for the mean-field transition to
the AFM state of single-layer
graphene~\cite{MF_Uc_sorella1992}.
Thus, the
replacement~\eqref{Hrepl}
is exact for the Hubbard model of the tBLG in the limit of uncoupled
graphene layers.

Our iteration scheme for finding order parameters is the standard one. For a given
$\Delta_{\mathbf{n}is}$,
$A^{(\ell)}_{\mathbf{n}i\sigma}$,
and
$B^{rs}_{\mathbf{m};\mathbf{n}\sigma}$,
we calculate the eigenenergies
$E^{(S)}_{\mathbf{p}}$
and eigenvectors
$\Phi^{(S)}_{\mathbf{p}}$
of the truncated matrix
$\hat{H}'_{\mathbf{p}}$.
Using these quantities we calculate the gradient of the system's energy
according to
\begin{equation}
\label{grad}
\frac{\partial\cal E}{\partial\lambda}
=
\left\langle\frac{\partial H^{\text{MF}}}{\partial\lambda}\right\rangle\,,
\end{equation}
where
$\lambda=\Delta_{\mathbf{n}is}$,
$A^{(\ell)}_{\mathbf{n}i\sigma}$,
or
$B^{rs}_{\mathbf{m};\mathbf{n}\sigma}$.
The new values of the order parameters are calculated according to the
conjugate-gradient method. The averaging in
Eq.~\eqref{grad}
is performed at fixed doping level, where the chemical potential is found
from the condition
\begin{equation}
\frac{n}{n_s}=\sum_{S=1}^{N'_R}\!\int\!\frac{d^2\mathbf{p}}{v_{\text{RBZ}}}\,\theta(\mu-E^{(S)}_{\mathbf{p}})-N'_R/2\,,
\end{equation}
where
$v_{\text{RBZ}}$
is the area of the reduced Brillouin zone.


\bibliographystyle{apsrevlong_no_issn_url}

\end{document}